\documentclass[format=acmsmall,screen=true,review=false,nonacm=true]{acmart}

\pdfoutput=1 
\usepackage[utf8]{inputenc}
\usepackage{amsmath}
\usepackage{listings}
\usepackage{xspace}
\usepackage{xcolor}
\usepackage{tabularx}
\usepackage{subcaption}

\newcommand{\MLIR}{MLIR\xspace}
\newcommand{\LLVM}{LLVM\xspace}
\newcommand{\PDLL}{PDLL\xspace}

\definecolor{orange3}{rgb}{0.425,0.191,0.000}
\definecolor{scarletred3}{rgb}{0.643,0.000,0.000}
\definecolor{green3}{rgb}{0.000,0.405,0.000}
\definecolor{blue3}{rgb}{0.000,0.000,0.704}
\definecolor{aluminium1}{rgb}{0.933,0.933,0.925}
\definecolor{aluminium2}{rgb}{0.827,0.843,0.812}
\definecolor{aluminium3}{rgb}{0.729,0.741,0.714}
\definecolor{aluminium4}{rgb}{0.533,0.541,0.522}
\definecolor{aluminium5}{rgb}{0.333,0.341,0.325}
\definecolor{aluminium6}{rgb}{0.180,0.204,0.212}

\lstdefinelanguage{cpp}{
    basicstyle=\ttfamily\scriptsize,
    keywordstyle=\color{orange3},
    commentstyle=\color{aluminium4},
    keywords=[3]{int64_t, float, double, int, size_t},
    keywordstyle=[3]\color{scarletred3},
    keywords=[4]{template,typename,struct,class},
    keywordstyle=[4]\color{blue3},
    showstringspaces=false,
	breaklines=true,
    breakatwhitespace=true,
    morestring=[b]",
    stringstyle=\color{green3},
}
\lstdefinelanguage{llvm}{
    basicstyle=\ttfamily\scriptsize,
    alsoletter={\%,\#,!},
    keywordsprefix={\%},
    morekeywords={\%},
    keywordstyle=\color{orange3},
    commentstyle=\color{aluminium4},
    keywords=[3]{i64, i32, float, double},
    keywordstyle=[3]\color{scarletred3},
    keywords=[4]{extractvalue,insertvalue,undef},
    keywordstyle=[4]\color{blue3},
    showstringspaces=false,
	breaklines=true,
    breakatwhitespace=true,
    morestring=[b]",
    stringstyle=\color{green3},
    morecomment=[l]{;},
}
\lstdefinelanguage{mlir}{
    basicstyle=\ttfamily\scriptsize,
    alsoletter={\%,\#,!},
    keywordsprefix={\%},
    morekeywords={\%},
    keywordstyle=\color{orange3},
    keywordsprefix=[2]{\@},
    keywordstyle=[2]\bfseries,
    commentstyle=\color{aluminium4},
    otherkeywords={index, f32, i32, f64, i8, vector},
    keywords=[3]{index, f32, memref, vector, i32, f64, i8, affine_map, affine_set, iter_args},
    keywordstyle=[3]\color{scarletred3},
    keywords=[4]{affine,linalg},
    keywordstyle=[4]\color{aluminium6},
    showstringspaces=false,
	breaklines=true,
    breakatwhitespace=true,
    morestring=[b]",
    stringstyle=\color{green3},
    moredelim=[s][\color{blue3}]{\#}{\ },
    moredelim=[s][\color{scarletred3}]{!}{\ },
    morecomment=[l]{//},
}
\lstdefinelanguage{python}{
    basicstyle=\ttfamily\scriptsize,
    alsoletter={},
    keywordsprefix={@},
    morekeywords={def, True, False, with, return, import, from, for, if, in, is},
    keywordstyle=\color{orange3},
    commentstyle=\color{aluminium4},
    keywords=[3]{TensorDef, ScalarDef, AttributeDef},
    keywordstyle=[3]\color{scarletred3},
    keywords=[4]{cast, },
    keywordstyle=[4]\color{aluminium6},
    showstringspaces=false,
	breaklines=true,
    breakatwhitespace=true,
    morestring=[b]',
    stringstyle=\color{green3},
    morecomment=[l]{\#},
}

\title{Composable and Modular Code Generation in \MLIR}
\subtitle{A Structured and Retargetable Approach to Tensor Compiler Construction}

\author{Nicolas Vasilache}
\orcid{0000-0002-4096-3325}
\author{Oleksandr Zinenko}
\orcid{0000-0003-1978-0222}
\author{Aart J.C. Bik}
\orcid{0000-0002-0333-7413}
\author{Mahesh Ravishankar}
\author{Thomas Raoux}
\author{Alexander Belyaev}
\author{Matthias Springer}
\orcid{0000-0001-5077-7131}
\author{Tobias Gysi}
\author{Diego Caballero}
\author{Stephan Herhut}
\author{Stella Laurenzo}
\author{Albert Cohen}
\orcid{0000-0002-8866-5343}
\affiliation{%
  \institution{Google}
  \country{}
}
\authorsaddresses{Google, 1600 Amphitheatre Pkwy, Mountain View, CA 94043, USA}
\renewcommand{\shortauthors}{Vasilache \emph{et al.}}

\date{\today}

\begin{document}

\begin{abstract}
Despite significant investment in software infrastructure, machine learning systems, runtimes and compilers do not compose properly.
We propose a new design aiming at providing unprecedented degrees of modularity, composability and genericity.
This paper discusses a structured approach to the construction of domain-specific code generators for tensor compilers, with the stated goal of improving the productivity of both compiler engineers and end-users.
The approach leverages the natural structure of tensor algebra.
It has been the main driver for the design of progressive lowering paths in \MLIR.
The proposed abstractions and transformations span data structures and control flow with both functional (SSA form) and imperative (side-effecting) semantics.
We discuss the implications of this infrastructure on compiler construction and present preliminary experimental results.
\end{abstract}

\maketitle

\section{Introduction}
Despite significant investment in software infrastructure, Machine Learning (ML) systems are still stuck in a rut~\cite{ml_rut}.
This can largely be traced to a recurring issue: the number and sheer complexity of the steps involved in between programmers' intent and efficient execution on ever more powerful hardware.
Software stacks include technologies as diverse as ML frameworks for distributed training or inference, for server or mobile platforms, as well as user-facing languages for productivity, high-performance communication and compute libraries, and domain-specific compiler implementations and autotuners.

Unfortunately, these stacks are built in silos:
(1) vertical silos follow to top-down, framework-centered considerations: the need to capture the ever-growing user base of ML practitioners encourages framework-specific designs, resulting in duplication of effort; while
(2) horizontal silos follow bottom-up, technology-driven considerations: the lack of performance portability over hardware platforms push for hardware-specific implementations of a fixed set of computational operations and a fixed programming interface.

\emph{None of this is built in a modular fashion; nothing composes.}

Everyone rebuilds similar functionality resulting in incremental improvements at ever-increasing engineering costs. End users do not get access to programmable and portable high-performance building blocks; instead, they have to rely on limited expressiveness interfaces to the underlying implementations. Investments on common infrastructure --- such as the automation of target-specific code generation --- happen very late, making their adoption painful for software engineers with years of experience in a more restricted environment.

Focusing on the performance portability of computational operations in ML frameworks, we explore a better way consisting of multiple cycles combining top-down and bottom-up thinking:
(1) top-down thinking is concerned with making primitives available to the programmer that \emph{gradually decompose} into smaller building blocks with unsurprising, good performance; while
(2) bottom-up thinking is concerned with the creation of primitive building blocks that are well-suited to each hardware architecture and then \emph{gradually compose} them into the larger building blocks that connect to top-down thinking.
We propose a first attempt at embodying such alternating cycles based on compiler technology.
Iterating between top-down and bottom-up thinking opens up multiple opportunities for codesign at the interface between the compiler and programming environments, runtime abstractions, performance tuners and hardware.

From a compiler practitioner's point of view, it is interesting to realize that virtually all modern compiler solutions bottom-out on the \LLVM compiler to produce optimized binaries.
While, \LLVM is a common substrate, it is also a very low-level one.
Users (and ML frameworks) have to make an exclusive choice among the kind of abstractions introduced by either XLA~\cite{XLA}, TVM~\cite{chen2018tvm}, Glow~\cite{rotem2018glow}, Halide~\cite{ragan2013halide}, TACO~\cite{kjolstad2017taco}, polyhedral compilers~\cite{vasilache2019next} or other domain-specific compilers that all eventually produce some flavor of \LLVM IR.
Each one of these compilers comes with its own implementation and does not mix with others: cross-pollination is often reduced to understanding what compiler A did well and retrofit it as new transformations in compiler B.

Stepping back, one problem comes from the lack of an established infrastructure to support such domain-specific compilers.
Another problem comes from jumping the abstraction gap too quickly between user-facing abstractions and levels of compiler Intermediate Representation (IR) on which analyses and transformations occur.
This results in (1) losing information that is available at higher levels of IR and (2) exacerbating phase-ordering issues due to the need to reconstruct high-level semantical information form low-level IR.
This hurts modularity, leaving software engineers without well-understood design patterns for reusable code generator components. 

\MLIR is a new compiler infrastructure that drastically reduces the entry cost to define and introduce new abstraction levels for building domain-specific IRs~\cite{lattner2021mlir}.
It is part of the \LLVM project and follows decades of established practices in production compiler construction.
As such, \MLIR is an ideal solution to the missing infrastructure problem.
Yet, even with the availability of \MLIR, the second problem remains: missing intermediate abstractions and a progressive refinement strategy for the compilation of tensor computations.

This work presents ongoing code generation efforts to build composable abstractions for high-performance code generation in \MLIR:
\begin{enumerate}
\item Composable transformations: generic computations are decomposed hierarchically into smaller tiles, exploiting their algebraic properties and structure.
They may also be fused, and lowered progressively to loops over retargetable vector primitives.
Computations can be instantiated over tensor values (immutable) as well as in-memory buffers (side-effecting).
Such abstractions exist in multiple storage variants such as dense, sparse formats, or quantized representations.
An in-place bufferization pass provides a memory-efficient materialization of programs in tensor form, including transformed programs (tiled, fused, etc.).
\item Programmable operations: on the front-end side, a declarative python-based DSL allows to define operations and their properties, including new operations called ``custom ops''; the latter immediately become first-class IR citizens on which all existing transformations and rewrites are available.
Throughout the transformation and lowering process, we offer a simple ABI definition and interoperability with C, C++ and Python.
\end{enumerate}

As we will see in the following, the abstractions and transformations we propose offer immediate as well as long-term benefits:
\begin{enumerate}
\item High-level, domain-specific information is not discarded prematurely.
Preserving the structure of computational operations allows to transform the IR while avoiding the performance cliffs of numerical libraries (e.g., when a fused operation is lacking a high-performance implementation).
In particular, transformations preserve the ability to lower operations to hardware instructions implementing coarse-grained vector operations, or to numerical libraries --- such as Eigen~\cite{eigen}.
The latter acts as a safety net upon which compiler transformations may add value, opening new avenues for compiler-library co-design.
\item Tiled operations target \emph{subsets} of tensor values or memory buffers. This central notion leverages the natural structure in tensor operations, while remaining completely generic in the actual representation of tensors (values or side-effects, vectors or scalars, dense or sparse, etc.) and in the actual decompositions applied to the computations (e.g., different forms of tiling). This enhances composability while allowing transformations to apply to individual or groups of operations as opposed to loops or control-flow graphs. It also eases the expression of complex transformations and lowering sequences, which in turn facilitates autotuning.
\item The IR remains executable at any intermediate transformation and lowering step. This greatly simplifies debugging, testing and performance evaluation, and somewhat blurs the lines between what is traditionally considered the programmer's and the compiler's responsibilities.
In particular, C and C++ ABI compliance at function boundary eases the integration of domain-specific frameworks by relying on common infrastructure.
It is also an ingredient in the matching of abstract operations into calls to numerical libraries, when available.
\end{enumerate}

The techniques and abstractions described in this paper are used in production at Google, including XLA-based compiler flows~\cite{XLA} for CPU/GPU, and IREE~\cite{iree} for CPU/GPU with an emphasis on mobile and edge computing.
In particular we have successfully replaced uses of the Eigen~\cite{eigen} library in Google-wide production cases.

\section{Overview of the Code Generation Flow}
We leverage the \MLIR compiler infrastructure.
\MLIR drastically reduces the entry cost to define, compose and reuse abstractions for the construction of domain-specific compilers.
It offers a comprehensive collection of solutions to compiler construction challenges by:
(1) standardizing Static Single Assignment (SSA) form representations and data structures,
(2) unifying compiler analyses and transformations across semantic domains through generic programming concepts such as operation traits and interfaces,
(3) providing a declarative system for defining operations with nested regions, and domain-specific type systems, and
(4) providing a wide range of services including documentation, parsing/printing logic, location tracking, multithreaded compilation, pass management, etc.

\MLIR is designed around the principles of parsimony, progressivity and traceability~\cite{lattner2021mlir}. 
The code generation approach presented in this paper has largely contributed to the establishment of these principles and actively leverages them. 
The internal representation in \MLIR is fully extensible, allowing for custom user-defined operations (instructions), attributes and types. 
IR components that are expected to work together are grouped into \emph{dialects}, which can be seen as the intermediate representation analog of dynamic libraries. 
Unlike many earlier compilation flows that have multi-level yet all-or-nothing internal representation, \MLIR affords and encourages the mix of different dialects in a single unit of compilation at any point in the compilation flow. 
For example, a high-level tensor product operation may co-exist with low-level hardware instructions on vector elements in the same function. This provides a great level of modularity, composition and optionality that makes it possible to use the right abstractions for solving a particular problem, \emph{instead of having to solve all problems in a unique  representation}.

\subsection{Bird's Eye View and Motivation of Structured and Retargetable Code Generation}

Code generation approaches for numerical computing have traditionally focused on optimizing the performance of loop nests. 
Associated analyses focus on scalar elements as the body of a loop nest typically computes a single element. 
Such analyses must consider memory dependences and aliasing. 
These approaches have been deeply researched in the past~\cite{AllenKennedy} and have reached a high level of maturity. 
They are well-suited when starting from an input language like C or Fortran where the problem is already specified in terms of loops over data residing in pre-allocated memory.

When focusing on a specific domain (e.g.\ the ML space), we have the luxury of programs defined at a much higher level of abstraction than loops. 
This opens up the opportunity to revisit classical loop optimizations like fusion, tiling or vectorization without the need for complicated analysis and heuristics.
Advantages include reduced complexity and maintenance cost while also scaling naturally to extensions like sparse tensors, that are even more difficult to analyze at the loop level.

\begin{figure}[h!tb]
\centerline{\includegraphics[width=14cm]{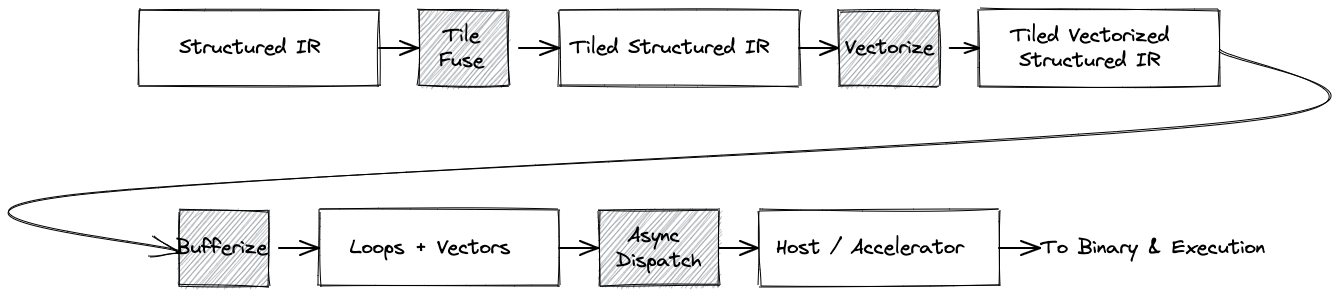}}
\caption{\label{fig:birds-eye-view-of-structured-codegen}Bird's eye view of structured and retargetable code generation.}
\end{figure}

It makes it possible to avoid raising information from lower level representations by means of static analysis, where feasible, and performing optimizations at the highest possible level of abstraction.
We refer to this approach as \emph{structured code generation} since the compiler primarily leverages structural information readily available in the source code. 
Figure~\ref{fig:birds-eye-view-of-structured-codegen} shows a coarse-grained summary structure of the steps and levels of abstraction involved.

The starting point (Structured IR) is composed of tensor algebra operations, organized as a functional program over dense and sparse tensors.

From this level we move to a tiled structured level, which introduces loops by tiling the operations. 
Multiple, gradual tiling steps are possible, and do not necessarily result in loops around scalars. 
Instead, tiling produces loops around structured operations similar to the original ones but on smaller tensors. 
We also perform fusion of tensor operations at this level. 
The final granularity of operations is chosen to make their hardware mapping efficient.
A prototypical example is to tile a matrix multiplication to model the cache hierarchy and then lower the smaller matrix multiplication directly to a superoptimized microkernel in assembly language.

In the next step, we map computations on resulting small tensors to a (retargetable) vector abstraction. 
This mapping exploits high-level knowledge about the operations that we have carefully preserved. 
In particular, it is not required to analyze the loop control flow enclosing the finer-grained tensor operations. 
This step might also include enabling transformations like padding for efficient cache access free of cache-line splitting and vectorization.

What makes \emph{structured code generation} highly composable and reusable is that tiling and fusion transformations are both fully generic in the operations and data types they operate upon. 
These transformations only assume a generic, monotonic (from the point of set inclusion), structural decomposition pattern associated with computations and composite data. 
Both dense and sparse tensor algebra exhibit such blockwise decomposition patterns, and the code generation abstractions and infrastructure generically applies to both.
This is also true of future computations and data structures following the same pattern.

Up until now, computations took place on immutable tensor values. 
We lower this to a representation on side-effecting buffers in the next step. 
This results in a representation with nested loops on vectors and side-effects.
More optimizations on loops and memory accesses happen at this level.

In the final step, we may translate the representation directly to the \lstinline|llvm| dialect of \MLIR for sequential execution on CPU, or offload a GPU kernel, or split up loops into \lstinline|async| blocks for a task parallel runtime, etc.

This flow composes with existing affine analyses and loop optimizations as implemented in \MLIR, and that have been largely explored in the literature.
In fact, the packing and loop peeling transformations in our flow leverage and helped generalize the MLIR affine machinery.

While this flow is but a first stake in the ground, it already demonstrates how to achieve a modular and composable system.
It is also built with optionality in mind: for some operations, skipping levels or even a fully divergent flow are viable options.
This is made possible by our work establishing the \emph{progressive lowering principle}: every step is materialized in the IR and very little load-bearing logic is abstracted away from the user (e.g.\ in the form of complex C++ logic deep inside the compiler implementation of analyses and heuristics).
We expect more operations and data types will be designed and implemented that do not fit the current code generation stack of abstractions.
While this allows to extend the structured and progressive lowering approach to other application domains, we acknowledge that not all computations in ML or HPC will necessarily be covered by the current approach, alone.
This is where composability, modularity and optionality kick in: \emph{not all problems need to be solved in the same abstraction, rather we should use the best abstraction for each different classes of problems}.

\subsection{Short Introduction to MLIR}

\begin{figure}
  \centering
\includegraphics[width=\textwidth]{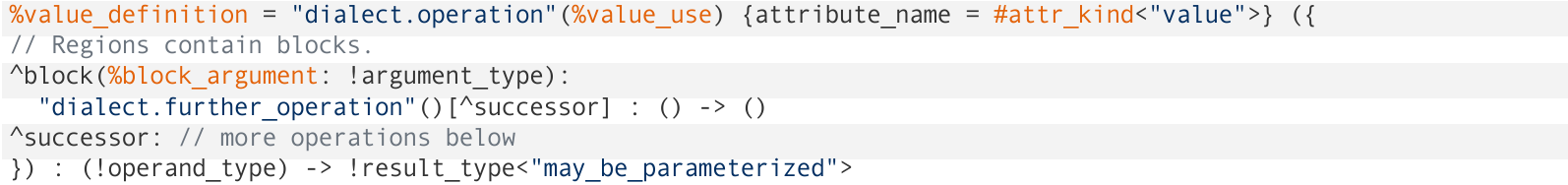}
  \caption{\MLIR concepts in the generic format. \MLIR has an open set of attributes, operations and types. Operations may recursively contain regions of blocks with further operations.}
  \label{fig:mlir_concepts}
\end{figure}

The \MLIR infrastructure builds on the success of \LLVM~IR while providing unprecedented extensibility.
\MLIR has an open, easily extensible set of instructions, called \emph{operations} that typically represent the dynamic semantics of the program.
Operations can represent anything from hardware instructions, or even hardware itself, to building blocks for machine learning models such as layers or blocks thereof.
They define and use values, which represent units of immutable data in SSA form.
The compile-time knowledge about values is captured in \emph{types}, and the knowledge about operations is captured in \emph{attributes}.
Attribute and type systems are similarly open and extensible.
IR objects can be logically grouped together in libraries, called \emph{dialects}.
The \MLIR~IR has a recursive structure where operations may have additional \emph{regions} containing a graph of (basic) \emph{blocks}, which in turn contain further operations. Figure~\ref{fig:mlir_concepts} illustrates key \MLIR concepts.

In addition to common components such as the compiler pass infrastructure, \MLIR provides tools to manage its extensibility, many of which evolved or were specifically designed to support the code generation flow presented in this document. In particular, \MLIR features attribute, operation and type \emph{interfaces} similar to object-oriented programming languages allowing one to work with abstract properties rather than (fixed) lists of supported concepts. Interfaces can be implemented separately from operations, and mixed in using \MLIR's registration mechanism, thus fully separating IR concepts from transformations.

\subsection{Dialects Relevant to Code Generation}
\label{subsec:dialects-relevant-to-code-generation}

The domain-specific abstractions we design and implement comprise representations for the following dialects, listed in increasing level of abstraction.
Following our modularity and optionality design principles, any of these dialects can be mixed with others or simply bypassed if it does not provide a useful abstraction for a particular case.

\subsubsection{\lstinline|vector| Dialect}
\label{sec:vector-dialect}

This dialect provides a fixed-rank \emph{n-D} vector type, e.g., \lstinline|vector<4x3x8xf32>|, as well as operations that form an intuitive and retargetable \emph{vector programming model} that conceptually extends the traditional 1-D vector instructions to arbitrary rank.
Such operations can decompose progressively into lower-rank variants of themselves. They further lower to \LLVM vector instructions (e.g.\ \lstinline|shufflevector|) when the backend heuristics are robust enough to generate near-peak assembly or bypass that level and directly target hardware-specific intrinsics (e.g.\ \lstinline{gpu.subgroup_mma_compute_matrix} 2-D vector instructions or \lstinline{amx.tile_mulf} 2-D tile instructions).

\subsubsection{\lstinline{gpu} Dialect}

The \lstinline{gpu} dialect defines the retargetable \emph{GPU programming model}.
It features abstractions common to SIMT platforms, such as the host/device code separation, workitem/group (thread/block) execution model, communication and synchronization primitives, etc.
This dialect can be produced from the \lstinline{vector} dialect and can itself be lowered to platform-specific dialects such as \lstinline{nvvm}, \lstinline{rocdl} or \lstinline{spirv}.
It is only listed to illustrate the overall retargetability of our approach and is not discussed further in the paper.

\subsubsection{\lstinline|memref| Dialect}

The \lstinline|memref| dialect introduces the \lstinline|memref| data type, which is the main representation for \emph{n-D} memory buffers in \MLIR and the entry point to the side-effecting memory-based operations, and the operations to manage buffer allocation, aliasing (memref \emph{views}) and access.
Unlike traditional pointers, \lstinline|memref|s are multi-dimensional buffers with explicit layout that allows for decoupling the indexing scheme from the underlying storage: \lstinline|memref<10x10xf32, strides: [1,10]>| affords column-major access while having row-major storage.
The \lstinline|memref| data type also provides an unsurprising ABI to interoperate with external C code, useful to interact with libraries.

\subsubsection{\lstinline|tensor| Dialect}
\label{subsubsec:tensor-dialect}
The tensor dialect operates on an abstract n-D tensor type for which we have not yet decided on a representation in memory. 
During the compilation, sufficiently small tensors of static sizes may be placed directly in (vector) registers while larger or dynamically-sized tensors are put into memory storage thanks to the bufferization process.
Tensor values are \emph{immutable} and subject to def-use SSA semantics, operations on tensors are often free of side-effects.
This allows classical compiler transformations such as peephole optimizations, constant subexpression and dead code elimination, or loop-invariant code motion to apply seamlessly to tensor operations regardless of their underlying complexity.
\emph{Since tensor values are immutable, they cannot be written into. Instead, ``value insertion'' operations create new tensors with a value or a subset thereof replaced.}
\footnote{This is analogous to the design of \texttt{struct} in LLVM IR:
\texttt{\%1 = insertvalue \{f64, f32, i32\} \%0, f32 42.0, 1} defines a new value \texttt{\%1} that holds the same elements as \texttt{\%0} except for the element at position \texttt{1} that now holds \texttt{42.0}.
}

\subsubsection{\lstinline|scf| Dialect}
The \emph{structured control flow} \lstinline|scf| dialect provides operations that represent looping and conditionals (e.g.\ regular \lstinline|scf.for| and \lstinline|scf.while| loops without early exit as well as an \lstinline|scf.if| conditional construct) and \emph{embeds them into the SSA+regions form} of \MLIR.
This is structured at a higher-level of abstraction than a control flow graph. Notably, \lstinline|scf| loop operations may yield SSA values and compose well with other operations and dialects with either memory-based side-effecting semantics or SSA-based side-effect-free semantics.

\subsubsection{\lstinline|linalg| Dialect}
The linalg dialect provides higher-level compute primitives that can operate on both \lstinline|tensor| and \lstinline|memref| containers. These primitives can decompose into versions of themselves operating on structured \emph{subsets} of the original input data and producing similarly structured subset of their results. They also capture program invariants and structural information, such as independency of certain parts of computation or reduction patterns, that decreases the need for expensive analyses prior to transformation.

\subsubsection{\lstinline{sparse_tensor} Dialect}
The sparse tensor dialect provides the types and transformations required to make sparse tensor types first-class citizens within the MLIR compiler infrastructure. The dialect bridges high-level \lstinline|linalg| that operates on sparse tensors with lower-level operations on the actual sparse storage schemes that save memory and avoid performing redundant work.

\subsection{Lower-level Dialects: Producing LLVM IR and Binaries}
\begin{figure}[h!tb]
\centerline{\includegraphics[width=14cm]{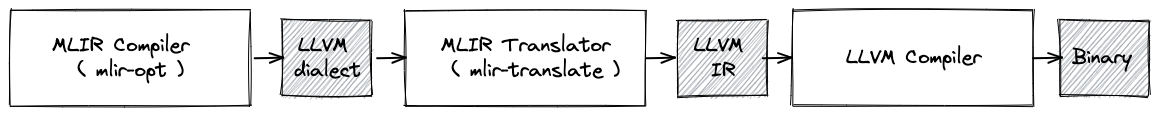}}
\centerline{\includegraphics[width=14cm]{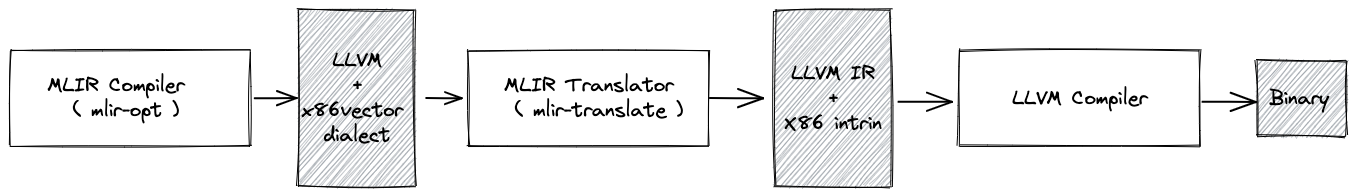}}
\caption{Simple visual description of the MLIR compiler flow: (top) \texttt{llvm} dialect only, (bottom) \texttt{llvm} and \texttt{x86vector} dialects, the latter containing hardware-specific ``intrinsic'' operations.}
\label{fig:simple-mlir-flow}
\end{figure}

At the end of the transformation process, \MLIR produces low-level dialects common to multiple compilation paths.
The \lstinline|llvm| dialect closely mirrors \LLVM IR and is the output we consider in this paper.
The \MLIR module using this dialect can be \emph{translated} to \LLVM IR before being handed off to the \LLVM compiler to produce machine code.
Figure~\ref{fig:simple-mlir-flow}(top) summarizes the tool flow.

Similarly to the rest of \MLIR, this dialect can be mixed with other ones.
In particular, it reuses built-in \MLIR types such as integer (\lstinline{i32}) or floating-point (\lstinline{f32}) scalars.
A detailed example of such a dialect mix is discussed in Appendix~\ref{sec:appendix-llvm-mix-example}.

While our flow mostly relies on the \LLVM compiler to perform common middle-end and back-end optimizations, some performance-critical scenarios require stronger guarantees of specific hardware instructions being emitted.
Therefore \MLIR provides a handful of low-level platform-specific dialects: \texttt{nvvm}, \texttt{rocdl}, \texttt{x86vector}, \texttt{arm\_neon}, \texttt{arm\_sve}, \texttt{amx}, etc.
These dialects partly mirror the corresponding sets of \LLVM IR intrinsic functions, which themselves typically map to hardware instructions.
Beyond making these instructions first-class operations and providing, these dialects also define slightly higher-level operations that make use of \MLIR's extensible type system and other capabilities.
For example, the
\begin{lstlisting}[language=mlir]
arm_neon.2d.sdot : vector<4x4xi8>, vector<4x4xi8> to vector<4xi32>
\end{lstlisting}
operation is naturally expressed on a \MLIR multidimensional vector type.
Before converting to \LLVM IR, it is first lowered to
\begin{lstlisting}[language=mlir]
arm_neon.intr.sdot : vector<16xi8>, vector<16xi8> to vector<4xi32>
\end{lstlisting}
that operates on flattened 1-D vectors to match \LLVM's convention.
The full example is provided in Appendix~\ref{sec:appendix-arm-neon-example}.

\section{Transformations}
\label{sec:transformations}

\begin{figure}[h!tb]
    \noindent
    \includegraphics[width=\textwidth]{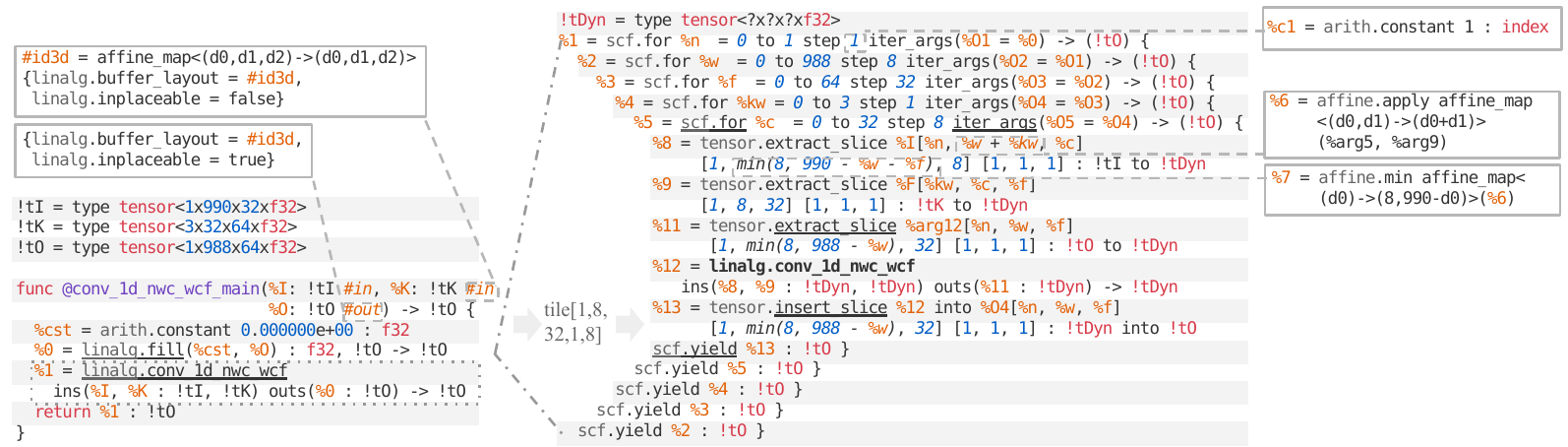}
    \caption{Tiling a convolution on tensors introduces loops with secondary induction variables, pseudo-IR. Parts in italic are simplified for clarity and expanded in callouts. Underscored parts refer to new concepts: (left) operations on immutable tensors, (right) secondary induction variables and tensor slicing.}
    \label{fig:lifecycle_tiling}
\end{figure}

We follow the IR step by step as it is transformed, considering a \lstinline|linalg.conv_1d_nwc_wcf| operation and its lowering to a tiled, padded and vectorized form.
The input IR is shown in Figure~\ref{fig:lifecycle_tiling} (left).

This level of abstraction operates on immutable SSA values: new tensor values are created from existing ones.
Memory locations only occur as annotations at function boundary to
specify how these tensors will materialize into memory, in a subsequent lowering step (more on this later).

The index notation corresponding to \lstinline|linalg.conv_1d_nwc_wcf|
is given by a $5$-D rectangular iteration domain operating on $3$-D tensors along with the expression:\footnote{The operation also allows specifying static sizes and strides that we omit here for simplicity}
\begin{equation*}
O[n,w,f] = I[n,w+k_w,c] . K[k_w,c,f]
\end{equation*}
The iteration domain is implicit in the operation description and is such that the iterators span the entire data of the operands. 
In this example, this is given by the inequalities 
$$
0\le n<O.0, \quad 0\le w<O.1, \quad 0\le f<O.2, \quad 0\le k_w<K.0, \quad 0\le c<K.1
$$
where $O.d$ denotes the size of the $d$-th dimension of $O$.
The derivation for these quantities follows the same rules as Tensor Comprehensions~\cite{vasilache2019next}. 
In the dense case, they can be derived with successive applications of the Fourier-Motzkin elimination procedure~\cite{schrijver86}. 

\subsection{Tiling}

Tiling the operation introduces \lstinline|scf.for| loops as well as subset operations (\lstinline|tensor.extract_slice| and \lstinline|tensor.insert_slice|) to access the tiled data Figure~\ref{fig:lifecycle_tiling} (right).
The tiled form of the operation is itself a \lstinline|linalg.conv_1d_nwc_wcf| operating on the tiled subsets.
The derivation of dense subsets is obtained by computing the image of the iteration domain by the indexing function for each tensor.
Non-dense iteration domains and subsets require IR extensions and inspector-executor~\cite{kjolstad2017taco}
code generation that are outside the scope of this paper.

Back to our example, we chose tile sizes of \lstinline|1x8x32x1x8|.
While these sizes are static, some divisions are not integral and the boundary tiles are subject to full/partial tile classification.
As a result, there is no \emph{single static tensor type} that is valid for every loop iteration; the tiled tensor type \lstinline|!tDyn| must be relaxed to a
dynamically shaped tensor.\footnote{Note that our compilation flow supports full dynamism, although for illustration purposes we focus on static inputs.}.
The dynamic tile sizes needed to access the tile data slices are \%8, \%9 and \%11.
Separate canonicalizations later kick in to further refine the types that can be determined to be partially static.

The \lstinline|scf.for| loops introduced by this ``tiling on tensors'' transformation 
perform iterative yields of the full tensor value that is produced at each iteration of
the loop nest. The yielding form of \lstinline|scf.for| combined with subset operations 
generalize the insertion into a structure to a more dynamic type (see Section~\ref{subsubsec:tensor-dialect}).
Since tensor values are immutable, new values are produced by each \lstinline|tensor.insert_slice| and \lstinline|scf.yield|. 
It is the responsibility of the bufferization process to avoid superfluous
allocations and copies.

\subsection{Padding Values and Packing}
\label{subsec:padding-values-and-packing}

When tiling is applied, the content of a tile can often become more dynamic to account for boundary effects.
This hampers vectorization which requires static sizes.
There are multiple mitigating options:
\begin{enumerate}
    \item Tiling may trigger multi-level loop peeling (or versioning) to isolate the statically known constant part of the problem in a main loop, followed by cleanup loops for the boundaries. 
    The cleanup loops still exhibit dynamic behavior but they can always be tiled by $1$ and
    further reduce to a dimension of size $1$ that can be vectorized in a finer-grained form.
    \item An alternative is to pad the dynamic tile to a larger known static size.
    The value used for padding must be the neutral for the consuming operation.
    This introduces extra copies and more computations at the boundary but all tiles now become full.
    \item A third alternative is to turn to a representation that involves explicit masking.
    This is work in progress and outside the scope of this paper.
\end{enumerate}

\begin{figure}[h!tb]
    \noindent
    \includegraphics[width=\textwidth]{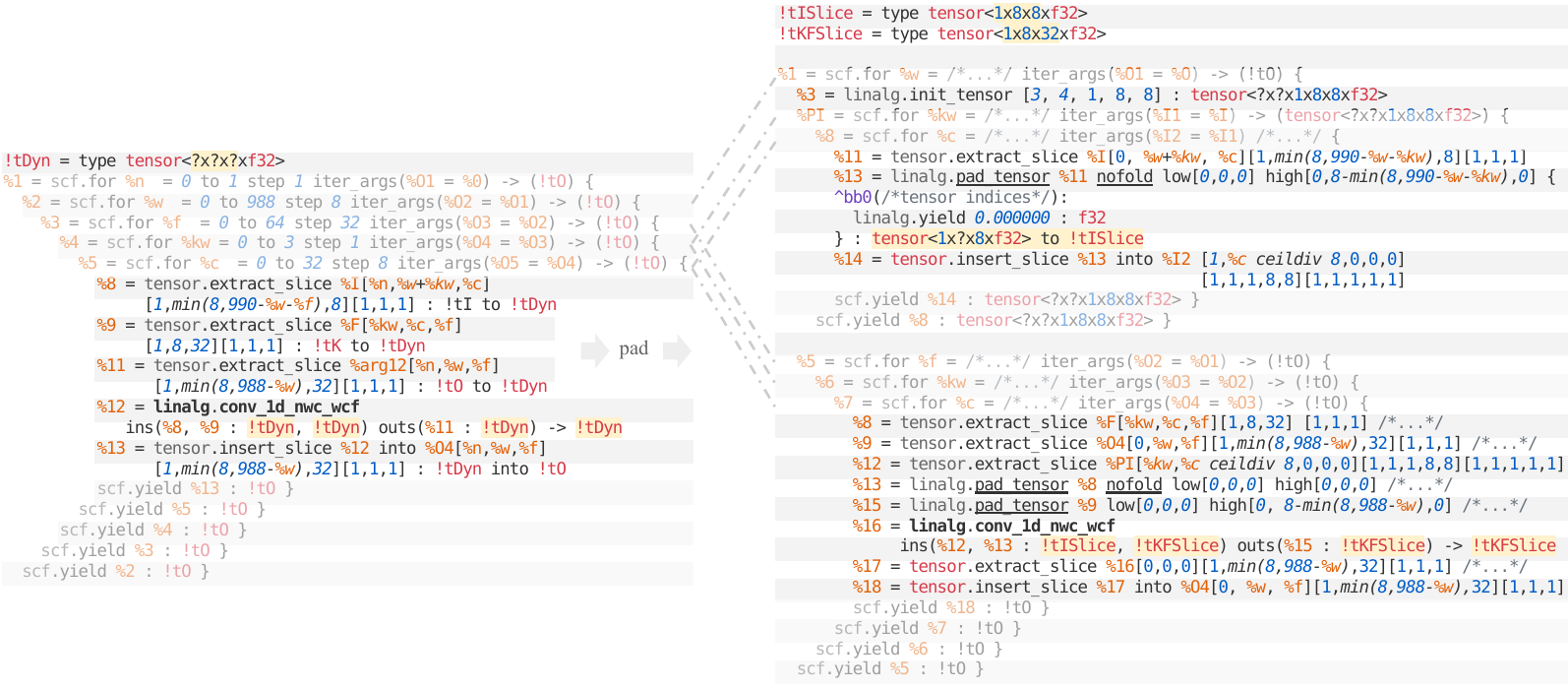}
    \caption{Padding a tiled operation to obtain fixed-size tensors (highlighted), pseudo-IR. Parts in italic are simplified for brevity. Constants in roman font are attributes, in italic are \texttt{arith.constant} operation results. The \texttt{nofold} padding persists in the IR even without type change and may be later placed in a fast buffer. Trailing type annotations sometimes omitted for brevity.}
    \label{fig:lifecycle_padding}
\end{figure}

When the computation does not entail enough temporal locality, peeling is almost always the 
better choice.
As soon as some temporal locality is available, the copy required for padding can be amortized.
Padding then also serves the purpose of aligning memory accesses in the padded buffer, once bufferization has occurred.
This is particularly important to avoid \emph{cache line splitting}, i.e., partially clobbering cache lines and inducing extraneous data transfers through the cache hierarchy; it is often required to reach the highest levels of performance.
In other words, value padding also implies address padding.\footnote{Additional masking may in turn facilitate address padding.}

Padding is materialized by the introduction of \lstinline{tensor.pad} operations, and its size is obtained by subtracting the dynamic tile size from the static tile size. 
All elements in the padded region are set to the constant value \lstinline[mathescape]!%cst!. 
This makes all operands of the tiled convolution statically shaped.

When the tile shape is already known to be static, value padding is not required for vectorization. In such cases, \lstinline|linalg.pad| would
simply fold away.
We additionally support a \emph{nofold} attribute to force padding to 
occur in such cases where address alignment is essential to avoid cache line split.

Operations that exhibit temporal reuse of the data may additionally
benefit from hoisting the padding operation out of the tile loops and storing the padded tiles in a higher-dimensional packed tensor. This allows both
amortizing the cost of copying as well as physically laying out tiles
contiguously in an intermediate buffer. This results in a smaller distance in memory between tiles that are reused within a short time-span
and reduces TLB misses.

The amount of hoisting is configurable per tensor and exhibits various
trade-offs between memory consumption, cost of copy and benefits to the 
computation primitive.

In the example of Figure~\ref{fig:lifecycle_padding}, the input tensor padding is hoisted by $3$ loops. 
This introduces an additional tile loop nest to precompute the padded tiles and insert them into the packed tensor of type \lstinline|tensor<?x?x1x8x8xf32>| containing all padded tiles.
Inside the original tile loop nest, the padding is replaced by an access to the packed tensor \lstinline|%12= tensor.extract_slice %PI...|.

Similarly to the tiling case, the sizes of packed tensor are obtained by computing the image of the iteration domain by a function of the enclosing loop variables and the tensor's indexing function.

\subsection{Vectorization}
\label{subsec:vectorization}

\begin{figure}[h!tb]
    \noindent
    \includegraphics[width=\textwidth]{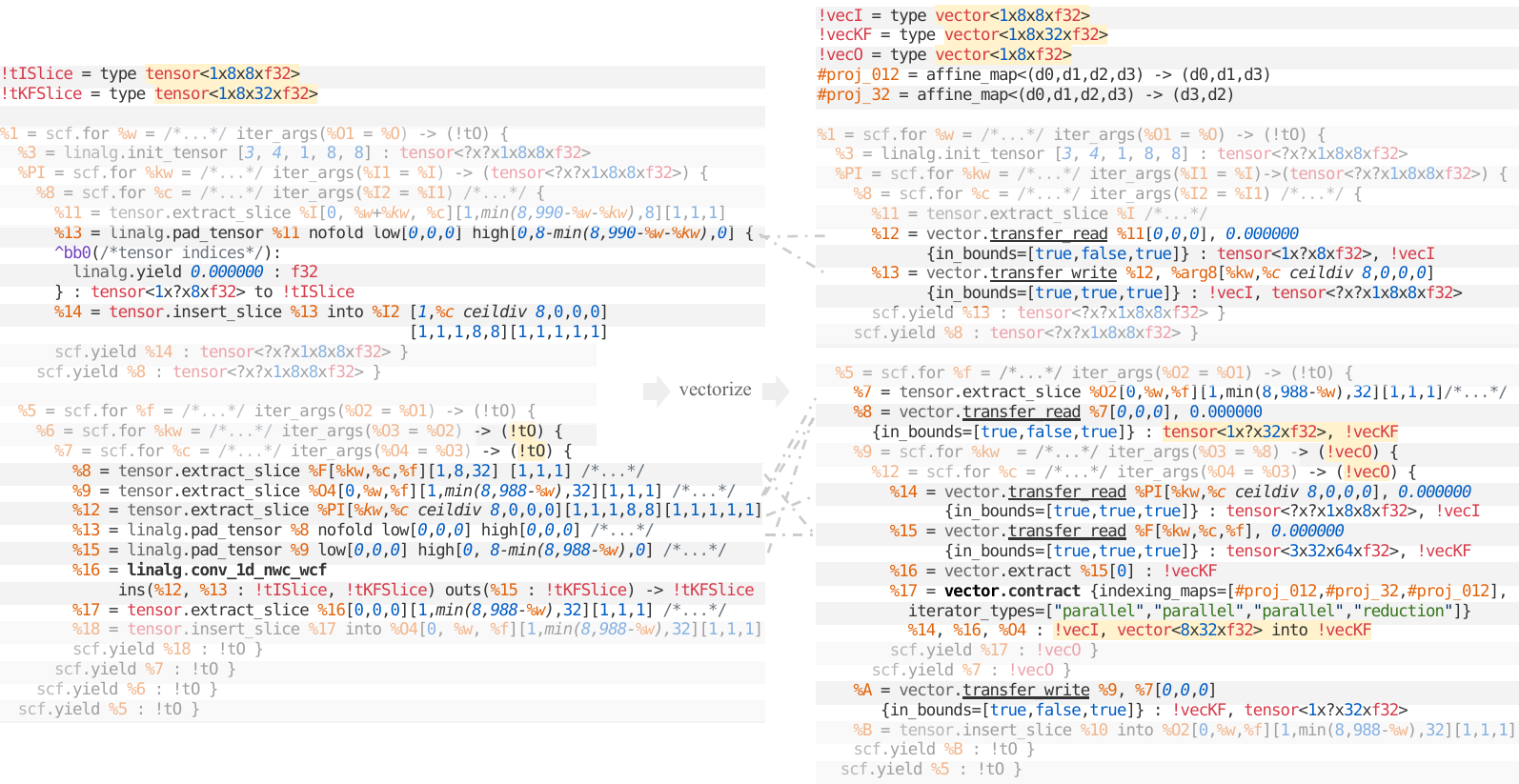}
    \caption{Operations on tensors of fixed sizes can be directly vectorized, pseudo-IR. Parts in italic are simplified for brevity as in Figure~\ref{fig:lifecycle_padding}. Vector values are immutable. They can be read from and written into tensors. Out-of-bounds access is allowed; reads replicate the scalar operand, writes are ignored.}
    \label{fig:lifecycle_vectorization}
\end{figure}

After tiling and padding, the convolution operands are statically shaped and are in a good state for vectorization, see Figure~\ref{fig:lifecycle_vectorization} (left).
In the current IR, only $2$ types of operations need to be
vectorized: \lstinline|tensor.pad| and 
\lstinline|linalg.conv1d_nwc_wcf|.

The vectorization of \lstinline|tensor.pad| is implemented with a simple one-off pattern that reduces it to a pair of \lstinline|vector.transfer_read|, \lstinline|vector.transfer_write| operations. 
The \lstinline{vector.transfer} operations are often referred to as a \emph{Swiss army knife} for bridging the gap between memory and vectors. 
In particular, they carry enough information to encode various multi-dimensional vector memory read and write patterns (e.g., broadcasted, permuted, masked, padded accesses).
They can be easily retargeted to specifics of current and future memory subsystems and vector ISAs.
With the existence of such an operation, the vectorization of 
\lstinline|tensor.pad| is comparatively a very small and progressive lowering step.\footnote{Especially if one thinks about all the load, store, indexing and fill operations that result from such a construct.}
The vector dialect additionally provides a first-class representation for high-intensity operations.
These operations are a key building block for high-performance codegen.
Figure~\ref{fig:lifecycle_vectorization} (right) illustrates one such operation, \lstinline{vector.contract}, in action.

The vectorization of \lstinline|linalg| operations follows a recipe that introduces a \lstinline{vector.transfer_read} for each operand, performs the computation in vector form and commits it back to the proper tensor or buffer via a \lstinline{vector.transfer_write}.
The \lstinline{vector.transfer} operations are indexed following the indexing expressions of the 
\lstinline|linalg| operation.
This behavior is generic for the data movement part of all \lstinline|linalg| operations.

The vectorization of the computation part is subject to variations.
Under the hood, every \lstinline|linalg| operation has a body region that expresses the scalar form of
the computation \footnote{The body may be printed explicitly (when expressed in \lstinline|linalg.generic| form)
or simply elided when expressed in "named" form (such as \lstinline|linalg.conv_xxx|).}.
The body vectorization depends on the type of indexings the parent \lstinline|linalg.generic| performs:
\begin{enumerate}
\item In the simplest case of pointwise operations (indexings are all identity), every operation in the body is simply written as a pointwise vector variant.
\item Lower dimensional tensor operands can be \lstinline|vector.broadcast| into higher-dimensional vectors where needed and reduce to the previous case.
\item Permutations in indexing expressions are handled with \lstinline|vector.transpose| operations.
\item Reduction dimensions lower to a first-class \lstinline{vector.contract} or \lstinline{vector.multi_reduction} depending on further analysis of the body.
\item Sliding window patterns such as convolutions are handled specially by unrolling along 
certain dimensions and extracting slices that further reduce to \lstinline{vector.contract} or
\lstinline{vector.fma}. This simple strategy delivers high performance
while capturing strided and dilated convolutions.
\end{enumerate}

In our running example of Figure~\ref{fig:lifecycle_vectorization} (right), the dimension corresponding to the \lstinline|%kw| loop would be unrolled. For the purpose of this illustration we have used a 
tile size of $1$ which does not further unroll. Note the 
\lstinline|%16 = vector.extract %15[0] : !vecK| operation which is a degenerate form of unrolled slice
extraction for size $1$.
Additional canonicalization and folding patterns occur that simplify chains of \lstinline|vector.transfer| operations, as well as move loop-independent instructions out of loops
(e.g.\ \lstinline|%8 = vector.transfer_read|).
Also note that the loops \lstinline|%9 = scf.for ...| and \lstinline|%12 = scf.for ...| both
yield vector values without inserting or extracting from a tensor.
This will guarantee no roundtrip to memory after bufferization.

All these transformations are implemented by following SSA def-use chains and legal by design (see Section~\ref{subsec:discussion} for a discussion of this principle).

\subsection{Bufferization}
\label{subsec:tensor-bufferization}

Bufferization is the process of materializing \texttt{tensor} values into memory (\texttt{memref}). 
It is necessary to make tensor programs concretely executable with a source
of data residing in memory.
In our current compilation pipeline, it is one of the last steps.

Tensors in \MLIR are immutable. An operation that produces a new tensor value (maybe from another input tensor) is conceptually an entirely new tensor. Unlike with \texttt{memref}s, there is no concept of updating/writing to a tensor in-place. To achieve good performance, it is essential to:

\begin{itemize}
    \item Allocate as little memory as possible.
    \item Copy as little memory as possible.
\end{itemize}

Buffers should be reused and updated in-place whenever possible or risk large performance penalty when program transformations result in unexpected allocation and copying.

\begin{figure}[h!tb]
    \includegraphics[width=\textwidth]{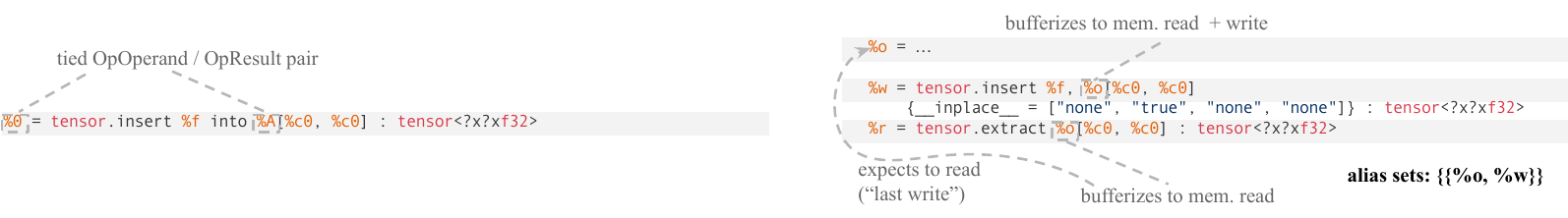}
    \caption{Left-hand side: output tensor arguments, tied with the result of an operation, in destination-passing style. Right-hand side: example of a read-after-write conflict.}
    \label{fig:bufferization_notation_conflict}
\end{figure}

\paragraph{Read-after-Write Conflicts}
Allocating a new buffer for every memory write is always safe, but wastes memory and introduces unnecessary copies. 
On the other hand, reusing a buffer and writing to it in-place can result in invalid bufferization if the original data at the overwritten memory location must be read at a later point of time. 
When performing transformations, one must be careful to preserve program semantics exposed by dependencies~\cite{AllenKennedy}.
The right-hand side of Figure~\ref{fig:bufferization_notation_conflict} illustrates a potential \emph{Read-after-Write (RaW) conflict} that prevents in-place bufferization. 
The problem of efficient bufferization is related to register coalescing, the register allocation sub-task associated with the elimination of register-to-register moves.

\paragraph{Destination-Passing Style}
The current heuristic we are offering for bufferization is well-suited for operations that are in \emph{destination-passing style}. 
In such operations, one of the tensor arguments is tied with the resulting tensor for in-place bufferization.
Such a tensor argument is called an \emph{output} tensor, see the left-hand side of Figure~\ref{fig:bufferization_notation_conflict}. 
Intuitively, output tensors are similar to C++ \emph{output parameters} that are passed as non-const references and used for returning the result of a computation. 
Except these ties between an output tensor (argument) and the operation's result serve as a bufferization constraint with no observable impact on the functional semantics; in particular, output tensors still appear as immutable.
During bufferization, only output tensors are considered when looking for a buffer to write the result of an operation into.

\begin{figure}[h!tb]
    \includegraphics[width=\textwidth]{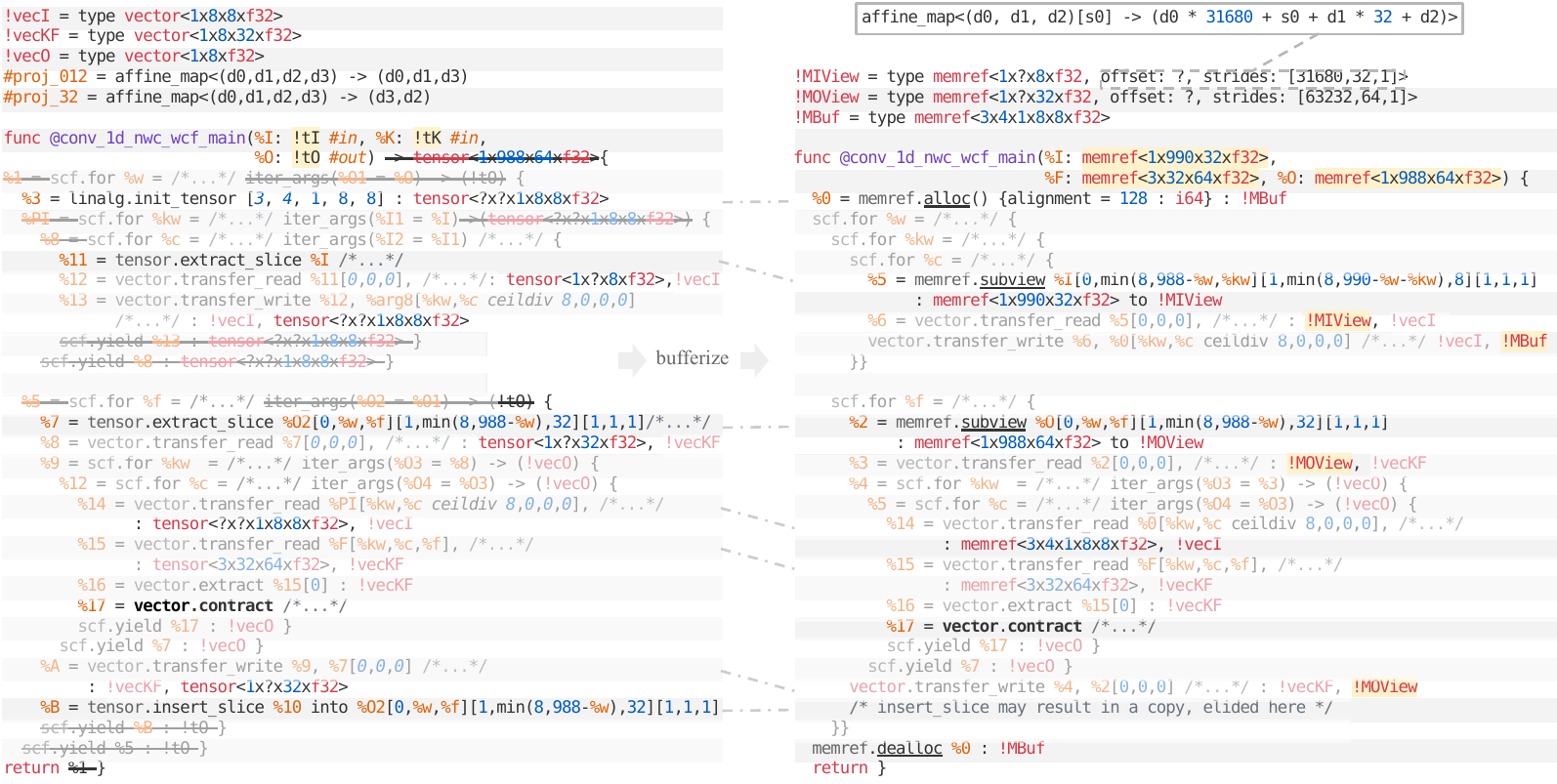}
    \caption{Bufferization assigns tensor values to buffers, taking into account function-level annotations \texttt{\#in}, \texttt{\#out} from Figure~\ref{fig:lifecycle_tiling}. Data flow is replaced by side effects, unnecessary values are crossed out on the left. Temporary buffers may be allocated to ensure contiguous access patterns. ``Computational payload'' dialects such as \texttt{linalg} and \texttt{vector} are designed to support both \texttt{tensor} and \texttt{memref} (buffer) containers.}
    \label{fig:lifecycle_bufferization}
\end{figure}

The rationale derives \emph{from first principles} when composing structured operations with \lstinline{scf.for}, the natural target for lowering multidimensional tensor operations.
Since \lstinline{scf.for} yields a value as a result, its nested region must yield fully defined tensors rather than arbitrary subsets.
Since nested operations typically apply to tensor subsets --- often resulting from \lstinline{linalg} tiling transformations --- a pair of matching \lstinline{extract_slice}/\lstinline{insert_slice} operations are typically injected.
These, and the associated \lstinline{scf.yield} operation, naturally consume their tensor argument (i.e., there cannot be any subsequent uses of it), which makes them ideal candidates for in-place bufferization. 
A comprehensive example is shown in Figure~\ref{fig:lifecycle_bufferization}.

This heuristic design appears to work well on the kind of IR we are dealing with when operating on the \texttt{linalg} dialect:
\begin{itemize}
    \item We saw that tiling produces outer loops iterating on tiled subsets. Operations to manage these subsets, such as \lstinline{extract_slice}, \lstinline{insert_slice}, are naturally in destination-passing style.
    \item Padding, packing, vectorization, and other transformations also produce operations with a destination-passing style semantics, on full tensors or subsets.
    \item \lstinline{linalg.generic} itself is designed as a destination-passing style operation. This includes \lstinline{linalg.matmul} and any other operation that reduces to \lstinline{linalg.generic}.
\end{itemize}

To illustrate this rationale, consider \texttt{tensor.insert} as an example of an operation in destination-passing style. 
A tensor result of an operation may have one or multiple \emph{potentially aliasing OpOperands} in the bufferization framework. 
For example, the only potentially aliasing OpOperand of \texttt{\%0} in the example is \texttt{\%A} (Figure~\ref{fig:bufferization_notation_conflict}, left-hand side), meaning that after bufferization:
\begin{itemize}
    \item \emph{buffer(}\texttt{\%0}\emph{)} = \emph{buffer(}\texttt{\%A}\emph{)}
    \item Or: \emph{buffer(}\texttt{\%0}\emph{)} is a newly allocated buffer.
\end{itemize}
No other operands are taken into account when choosing a buffer. 
Operations that do not have a potentially aliasing OpOperand for their tensor results always allocate a new buffer. 
For example, \texttt{tensor.generate} always allocates after bufferization.

This bufferization design is a heuristic: operations with no natural candidate output tensor (e.g.\ the sum of two tensors) allocate and copy by default.
This simplifies the bufferization problem, reducing it to an analysis of use-def chains.
The tradeoff is that upstream compilation passes are responsible of rewriting the IR in destination-passing style.
We believe a global copy elimination problem could be formalized on top of destination-passing style, offering the best of both worlds in terms of allowing passes to optimize bufferization at a global scale, while still enabling a robust, in-place bufferization path for the important special case of refining structured operations.

\paragraph{Analysis of Tensor SSA Use-Def Chains}
During bufferization, each operation with tensor semantics is replaced with an operation with memref semantics. 
Before modifying any IR, an analysis decides for each tensor OpOperand \texttt{\%t} whether \emph{buffer(}\texttt{\%t}\emph{)} (\emph{in-place bufferization}) or a copy thereof (\emph{out-of-place bufferization}), denoted by \emph{copy(buffer(}\texttt{\%t}\emph{))}, should be used with the new memref operation. 
The analysis simulates \emph{a future in-place bufferization of the OpOperand} and checks if a RaW conflict can be found under this assumption. 
If not, the analysis greedily commits to this in-place bufferization decision. 
Furthermore, the analysis stores the fact that the OpOperand and its potentially aliasing OpResult are now \emph{known to alias}, by merging their alias sets.

The search for RaW conflicts is based on a traversal of tensor SSA use-def chains. 
When simulating the in-place bufferization of an OpOperand \texttt{\%o}, the analysis looks for all uses of this SSA value (and its aliases) that bufferize to a memory read.
For each read, it walks the IR back the \emph{last write} of the tensor; i.e., the value that defines the contents of the tensor (skipping over operation like \texttt{tensor.cast} or \texttt{tensor.extract\_slice} that do not bufferize to a memory write). 
This is the value that the operation is expected to read. 
If an interleaved ``future write'' of an aliasing buffer exists, then there would be a RaW conflict.

The in-place bufferization of \texttt{\%o} of the \texttt{tensor.insert} (Figure~\ref{fig:bufferization_notation_conflict}, right-hand side) is a RaW conflict (and would thus be an invalid bufferization) because the operation is in-between the definition of \texttt{\%o} and the \texttt{tensor.extract} operation, which is expected to read the original value of \texttt{\%o} and not \texttt{\%w}.

\paragraph{Extension Points}
The bufferization framework is customizable and can be adapted for different use cases via two main extension mechanisms.

First, the order in which operations are analyzed is a heuristic and affects the order in which RaW conflicts are detected. 
There are often multiple out-of-place bufferization candidates that can prevent a RaW conflict. 
Only when a RaW conflict is no longer avoidable, the analysis decides to bufferize an OpOperand out-of-place. Therefore, operations (and their OpOperands) that are analyzed earlier are less likely to bufferize out-of-place. By analyzing a certain group of operations first, the analysis can be steered towards bufferizing these operations in-place and trying to avoid RaW conflicts by bufferizing other operations out-of-place.

Second, operations can refine the analysis by specifying conidtions that should not be treated as RaW conflicts. 
For example, a \texttt{tensor.insert\_slice} operation only affect a subset of the entire buffer. 
This informatio can be used to make better bufferization decisions around matching \texttt{tensor.extract\_slice}/\texttt{tensor.insert\_slice} pairs.
This paves the way for more advanced analyses based on intersection and difference, as well as partial reuse, of buffer regions.

\subsection{Progressive Lowering of Multidimensional Vector Operations Towards LLVM}
\label{subsec:rewrites-and-foldings}

\begin{figure}[h!tb]
    \noindent
    \includegraphics[width=\textwidth]{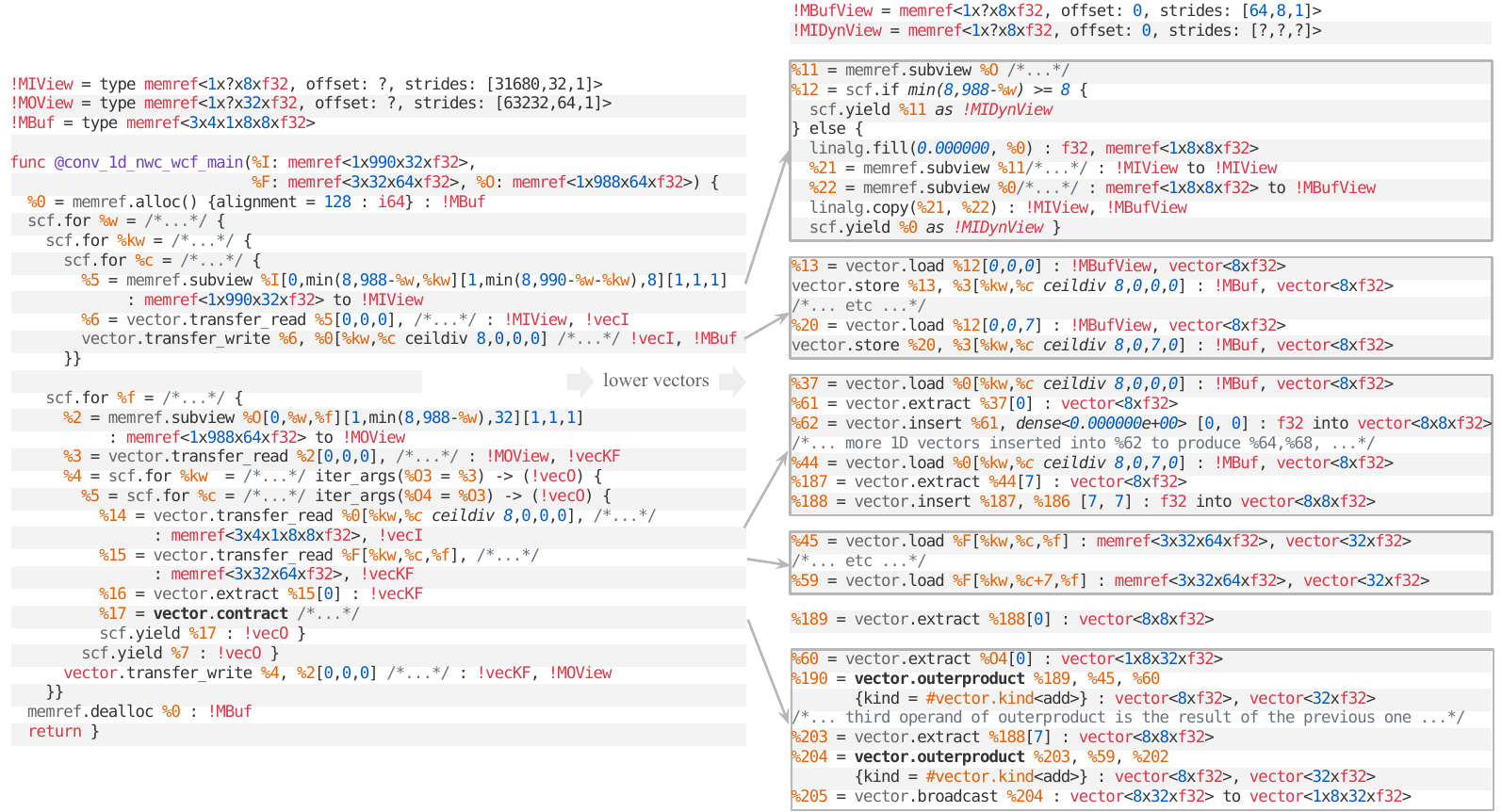}
    \caption{The \texttt{vector} dialect can be lowered progressively to simpler operations on 1-D vectors. Illustrated on lowering contractions to outer products, with parts in italic simplified for brevity and repetitive parts omitted. Lower-level vector operations require constant indices and are produced by unrolling the outer dimensions.}
    \label{fig:lifecycle_lower_vectors}
\end{figure}

At this time, the IR has reached a level of abstraction consisting of loops around buffers containing
multi-dimensional vectors and operations on those. This is now close to the C + vectors paradigm of
LLVM, with the exception that we operate on multi-dimensional vectors, whereas LLVM only has 1-D vectors.

In the simplest case, multi-dimensional \lstinline|vector.transfer| operations lower to multiple $1$-D \lstinline|vector.load| and \lstinline|vector.store| operations.
When supported by hardware, they can also lower to n-D DMA operations.
In more complex cases, transfer operations lower to a combination of broadcasts, tranpositions and masked scatter/gather.
In the particular case where the \lstinline|vector.transfer| cannot be determined to be in-bounds, 
one must resort to an additional separation between full and partial transfer, akin to the full and 
partial tile separation at the tile level.
This is illustrated in Figure~\ref{fig:lifecycle_lower_vectors} (right) in the \lstinline|else| block
around the \lstinline|linalg.copy(%21, %22)| operation.

\begin{figure}[h!tb]
\includegraphics[width=\textwidth]{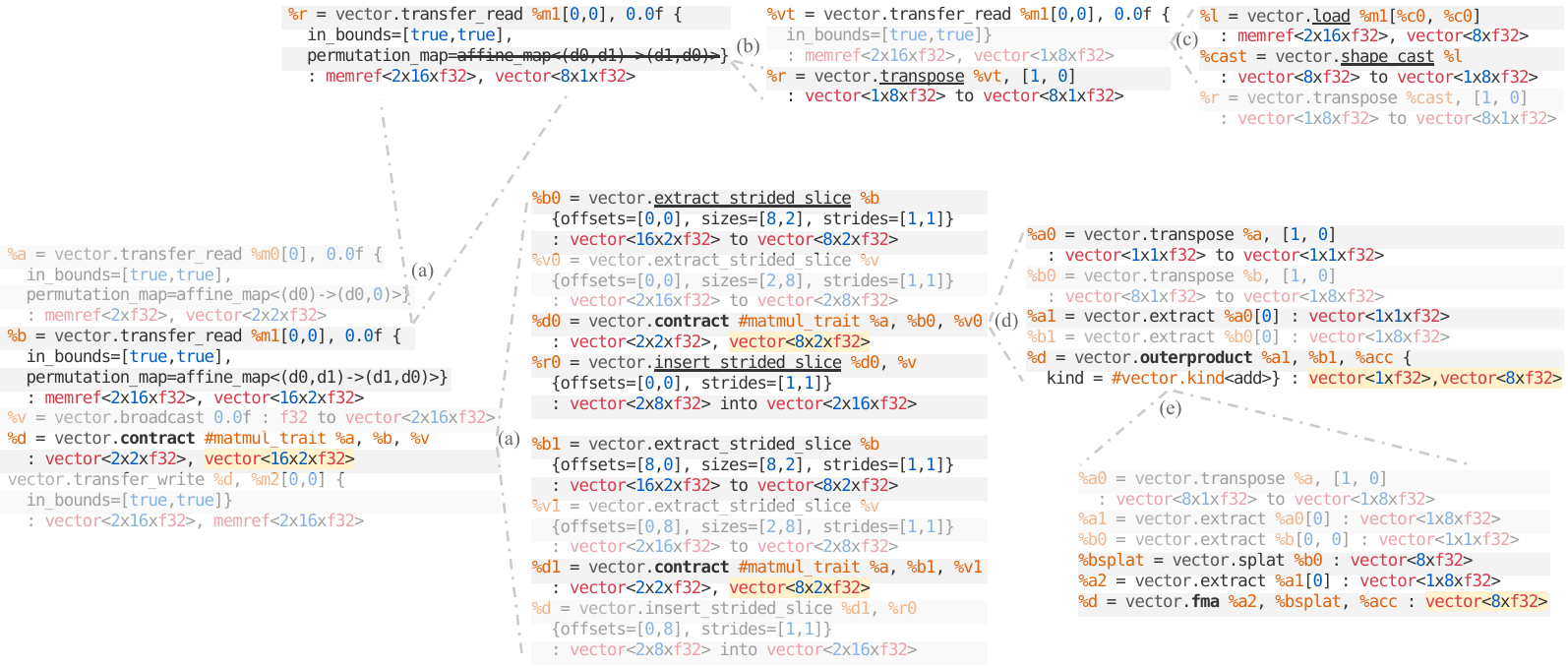}
\caption{Progressive lowering of the \texttt{vector} dialect operations representing matrix product: (a) vector unrolling with target shape $2\times8\times2$ introduces vector slice manipulation; (b) the transfer permutation is materialized as a \texttt{transpose} operation; (c) 1-D transfers become plain loads with shape adaptation; (d) contractions rewrite as outer products (other options are possible), which in turn lower to (e) fused multiply-add instructions.}
\label{fig:vector_deeper}
\end{figure}

Our pervasive use of n-D vector types effectively shielded an "unroll-and-jammed vector form", known to be efficient on vector hardware, from intermediate compilation phases that could interfere with later vectorization. At this point, this form is ready for progressive lowering into 1-D operations with almost 1-1 mapping to LLVM IR.

Let us illustrate the progressive lowering of operations on n-D vectors to lower-dimensional equivalents closer to hardware instructions. Starting from the vectorized matrix product code on the left of the Figure~\ref{fig:vector_deeper}. This IR is (mostly) retargetable since it is using higher-level \lstinline{transfer} and \lstinline{contract} operations that do not usually correspond to the available hardware instructions.\footnote{Even with \texttt{transfer} operations, vector sizes become hardware-specific and other aspects such as special ISA dimensions and the number of available registers come into play, but the representation is still easy to transform.} We first apply vector ``unrolling'' as in Figure~\ref{fig:vector_deeper}(a). The goal of this transformation is twofold: (1) it breaks down vector operations to sizes known to be well supported by the target, e.g., mapping to AMX instructions, and (2) preemtively handling non-power of $2$ sizes into power of $2$ compositions, e.g., \lstinline{vector<12xf32>} into 3 \lstinline{vector<4xf32>} to avoid suboptimal backend code generation~\cite{OuterproductSpilling}. The resulting IR is still partially retargetable as \lstinline{transfer} and \lstinline{contract} operations are still present and need to be lowered to a closer-to-hardware representation using one of the available schemes.

After unrolling, \lstinline|vector.extract_strided_slice| and \lstinline|vector.insert_strided_slice| extract and insert
slices of a vector into a larger vector. 
They are conceptually the same as their \lstinline{tensor} dialect counterparts.
When unrolling operations folding patterns can lead to insert and extract operations cancelling each other if the target shapes match.
Other peephole optimization patterns, applied along with the unrolling transformation,
track subsets through broadcast and transpose operations, all the way to transfer operations
where the may fold into memory reindexings.

Higher-level \lstinline|vector.transfer_read| usually cannot be directly lowered to a load instructions and are handled progressively:
first, by materializing transpositions as in Figure~\ref{fig:vector_deeper}(b), then, by creating 1-D loads and broadcasts as in Figure~\ref{fig:vector_deeper}(c).
Transpositions can in turn be implemented elementwise, with LLVM's \lstinline{shuffle} instruction or using dedicated intrinsics, depending on the configuration.

Similarly, \lstinline|vector.contract| can be lowered to outer products, inner (dot) products or LLVM IR matrix intrinsics.
In this example, it is lowered to outer products in Figure~\ref{fig:vector_deeper}(d)
to enable further mapping to SIMD fused multiply-add instructions in Figure~\ref{fig:vector_deeper}(e).
Each stage of the progressive lowering is accompanied by foldings and peephole optimizations that reduce the amount of IR to handle and enable additional transformations.
As the result, the fully lowered vector IR operates on \lstinline{vector<8xf32>}, supported for example by AVX2, and is quite compact. The resulting code for our example has several dozen operations and is provided in Appendix~\ref{sec:appendix-lowered-vector}. These operations are now ready for lowering to the LLVM dialect and further translation to LLVM IR.

\subsection{Discussion}
\label{subsec:discussion}

The levels of IR we introduced allow for developing an expressive set of essential transformations. These transformations are legal by design, in the sense that their legality and applicability derive from the operation's properties and structure.
We refer to this philosophy as \emph{transformations-oriented IR design}. 

\subsubsection{Transformation-Oriented IR Design}
\label{subsubsec:transformation-oriented-ir-design}

Traditional compiler analyses and transformations for numerical computing~\cite{AllenKennedy} revolve around trade-offs and questions related to:
\begin{itemize}
\item Legality, i.e.\ what transformations can be applied without changing the observed program semantics?
Legality conditions are often checked through static analyses. They may be performed upfront or on-demand, and their results may be updated as the IR is transformed.  
For example, dominance analysis produces necessary conditions for code motion: uses should remain dominated by definitions.
\item Applicability, i.e.\ how complex is the IR matching process for finding the place where to apply a transformation? How complex does the IR become after applying the transformation?
Applicability also encompasses considerations related to how much information has potentially been lost, whether the IR remains analyzable and whether subsequent transformations continue to be easily applicable.
\item Profitability, i.e.\ what are the transformations deemed beneficial for a given metric?
Profitability is often determined by heuristics or performance models. For example, polyhedral compilers often focus on finding an objective function to minimize (universal or target-specific) \cite{vasilache2019next}, while auto-tuners may rely on a learned performance model to accelerate search \cite{zheng2020ansor}.
\end{itemize}

It is of central importance to control the abstractions on which the transformation legality, applicability and  profitability questions relate to.
The finer-grained the IR, the more general and canonical the representation, but also the more intractable the analyses and transformations.
Indeed, canonicalization to some flavor of \LLVM IR consisting of a Static Single Assignment (SSA) form Control Flow Graph (CFG) has proven invaluable in enabling the reuse of common infrastructure for middle-end and back-end compilers.
But lowering abstractions and domain knowledge too quickly reduces the amount of structure available to derive transformations from.
While manipulating loops is a net gain compared to control flow graphs for a certain class of transformations, important information is still lost (e.g.\ parallel and reduction multi-loop semantics, loop-level data flow and memory footprint, substitution of a full loop nest with an external implementation).
It induces non-trivial phase ordering issues: loop fusion to enhance temporal locality may alter the ability to recognize an efficient BLAS-2 or BLAS-3 implementation in a numerical library.
Workarounds introduce constraints on the compiler that interfere with other decisions and passes, which is a long-standing and well-known problem in the compiler community~\cite{Click95}.
This work seeks to alleviate the issue by designing higher-level IR components that are more conducive to transformations.

\subsubsection{Top-down: Orchestrating Transformations}
A higher-level IR facilitate the declarative specification of transformations:
most transformations target individual operations in the IR, rather than large multi-operation constructs such as loops, making it easy to specify transformation targets.
Notably, tiling, fusion and unrolling apply to high-level operations rather than loops.\footnote{Some transformations such as explicit distribution or software pipelining remain naturally attached to loops.}
Loops and other constructs may be produced as a result of transformations, but they rarely need to be targeted by further high-level transformations.
On the other hand, target specifications can be arbitrarily complex yet readable if expressed, e.g., using the pattern-matching infrastructure available in \MLIR such as the PDL dialect.
Encoding operational knowledge directly in the IR allows us to design transformations that are legal by design. The fact that our \emph{fundamental units of IR rewrites} encompass more IR variety than loops, together with generic properties of data types (values or side-effects, dense or sparse, etc.) offers additional genericity and extensibility benefits, and applicability at both high level (e.g.\ \lstinline|linalg.generic|, \lstinline|vector.contract|) and low level (numerous canonicalizations on scalar and vector operations, folding to yield values across loops, etc.).

Furthermore, in presence of a meta-programming dialect suitable for IR manipulation, it becomes possible to express the transformations entirely declaratively as yet another \MLIR dialect.
Optimizing transformations can be then stored, analyzed and transformed, and shipped \emph{separately} from the main compiler.
It provides a way to specify transformations and the units of IR they manipulate and produce, while enabling local pattern rewrites almost everywhere.
This is paramount to quickly deliver performance improvements when new hardware becomes available without having to update the entire compilation stack.

This declarative approach also facilitates the design of custom passes by selecting specific rewrite rules. This allows mixing transformations, canonicalizations, constant folding and other enabling rewrites in a single transformation. The result is a system where pass fusion~\cite{Click95} is simple to achieve and alleviates phase ordering issues.
Indeed, our structured and retargetable code generation approach is deliberate about extending the notion of passes with \emph{more flexible and
controlled application of rewrite rules}.
The reader familiar with the program synthesis literature may think of these transformations as resembling TASO~\cite{TASO}, while additionally encompassing tiling, fusion, interchange, padding, packing, vectorization and the ability to decompose operations into smaller operations.

Specifying transformations as patterns expressed in an IR also supports search over possible transformations.
A sufficiently advanced search mechanism can analyze the declaratively specified patterns for which the transformation applies and build the search space from mutually exclusive choices.
Patterns can also be defined in a parametric form (e.g., tile sizes are not hardcoded in the pattern) and instantiated at transformation time.
The search procedure then consists in trying different sequences of compatible transformations with different parameters, resulting in different \emph{compilation strategies}.

The multi-level nature of MLIR makes it possible to build higher-level dialects to define IR transformations superimposed on the existing infrastructure and handled by progressively lowering.
For example, the transformation sequence and some of the parameters can be reified into a new ``strategy'' operation that gets lowered into primitive transformation operations with the lowering also specified declaratively.
Just as with any other dialect, IR modules using such meta-programming dialects can be created programmatically from any language.
The textual or binary IR format can be used to communicate between the front-end language, in which the transformation is written, and the compiler infrastructure with loose coupling.
The expressiveness of such transformations is similar to that of RISE/ELEVATE~\cite{rise} but without restrictions to the specification language.

\subsubsection{Bottom-Up Discussion: A Thick Blanket of High-Performance Patterns}
The example discussed in Section~\ref{subsec:rewrites-and-foldings} is just a tiny slice of the various rewrite patterns we have been developing.
Applying sets of orthogonal low-level pattern rewrites results in desirable larger scale behavior.
Application of these patterns is often bottom-up and are akin to assembling low-level building blocks into larger building blocks.
The patterns of interest encompass canonicalization, folding and other enabling patterns, as well as vector lowering patterns from higher-dimensional vector forms to hardware specific vector forms. 
For example, decomposing a 2-D \lstinline{vector.contract} into unrolled 1-D outer product microkernels for x86 CPUs supporting broadcast-load vector instructions.
Individual patterns at this level have few parameterization options, often limited to a choice among a few variants (e.g.\ lower a higher-dimensional vector reduction by using horizontal reductions or transpositions).
The application of individual patterns is often unsurprising and the performance of the resulting IR is quite predictable prior to their application.
These types of patterns bear resemblance to lower-level ones in the \LLVM back-end but operate on higher-level retargetable IR.

At this time we have started exploring the application of search techniques to build parametric and composable pattern sets. One  direction of interest is to aggressively search for tuples of
$$
(\textit{patterns}, \textit{parameters}, \textit{op-dag matching constraints}, \textit{problem sizes}).
$$
that reach a high fraction of peak hardware performance, given a fixed set of \LLVM compiler flags.
Thanks to our hierarchical IR design, we believe any such tuple can be saved and applied, with a low degree of surprise.
We expect this part of the system to make heavy use of the upcoming \PDLL  infrastructure~\cite{PDLL} to succinctly encode the search results and ship it to production.
Future considerations will include how such patterns should be managed, updated, curated and prioritized.

\section{Single Thread CPU Experiments}

We evaluate our code generation framework on a small set of kernels of high relevance to the machine learning community. 
All benchmarks measure single threaded CPU performance and compare to the peak performance of the machine.

\subsection{Preamble: Driving Experiments}
\MLIR provides a set of bindings for Python that support IR creation and manipulation at a generic level.\footnote{\url{https://mlir.llvm.org/docs/Bindings/Python}}
The infrastructure described in this paper aims to facilitate multi-level metaprogramming and drove the design of these bindings.

\begin{figure}[h!tb]
\centering
\includegraphics[width=\textwidth]{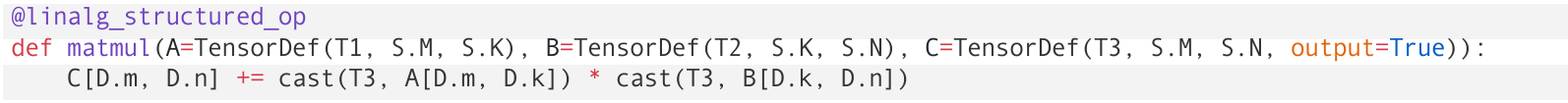}
\caption{Example: type-polymorphic matrix multiplication in OpDSL}
\label{fig:opdslexample}

\medskip
\centering
\includegraphics[width=\textwidth]{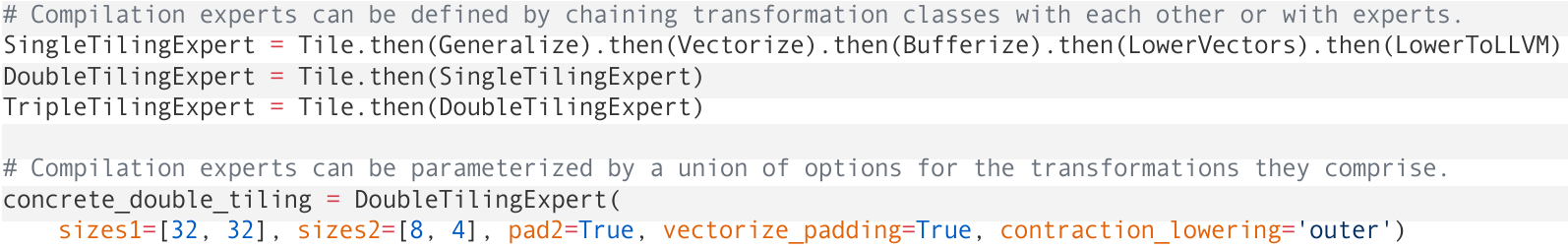}
\caption{Example: defining tiling transformations in Python}
\label{fig:tilingexample}
\end{figure}

We additionally provide a custom domain-specific language (DSL) embedded into Python, referred to as OpDSL. 
The purpose of OpDSL is to shift the API paradigm from constructing a compiler IR to expressing computation in a concise, human-readable and mathematically compelling form, that has shown successful with Tensor Comprehensions~\cite{vasilache2019next}.
Figure~\ref{fig:opdslexample} illustrates a type-polymorphic matrix multiplication in OpDSL.

The flow leverages and extends the minimal \MLIR execution engine for JIT-compilation and execution.
Structured data objects processed by the flow are exposed in Python as objects compatible with the Python buffer protocol.
They can therefore be converted to and from NumPy arrays, which are further convertible to framework-specific data types.
More details are given in Appendix~\ref{sec:appendix-python}.

In addition, we provide a test and benchmarking harness in Python to automate the measurement of compilation- and run-time as well as performance numbers such as $\mathrm{GFLOP}/s$ for computation and $\mathrm{GB}/s$ for memory traffic. 
The harness also wraps around the compilation and execution with multiple transformation strategies as described below.

\subsection{Preamble: Driving Transformations}
The transformations described in the previous sections are provided as configurable and callable ``transformation objects'' in Python.
These objects can be applied to \MLIR modules and perform the specified transformations on them.
Under the hood, the transformation results in a custom pass pipeline, consisting of the specified transformation pass with options set up as well as a handful of properly configured enabling/cleanup passes.
The pass pipeline is then run on the module.

The transformations currently available in Python are listed in Table~\ref{tbl:list-of-transformations}.
Certain transformations are applied in combination with tiling.
As discussed previously, the general notion of multi-dimensional subset --- applied to tensors, vectors, memrefs --- is load-bearing in our approach.
Transformations initially introduced on classical loops are generalized to multi-dimensional structured operations with the explicit goal of preserving structure.
Transformations may not be applicable until some loops are materialized by tiling as a means of explicitly and selectively discarding some parts of the structure.
In turn this improves the composition of transformations written as patterns: the matching conditions do not trigger until prerequisites are met.
For example, interchange applies to \emph{the loops produced by tiling} around the target structured operation.
Similarly, peeling only occurs on partial tiles.
Peeling and padding (Section~\ref{subsec:padding-values-and-packing}) are used as a means to ensure the main operations become fixed-shape and more easily amenable to vectorization (Section~\ref{subsec:vectorization}).

\begin{table}[h!tb]
    \centering
    \footnotesize
    \begin{tabularx}{\textwidth}{l|X}
    \toprule
    \textbf{Transformation} & \textbf{Options} \\\midrule
    \textsc{Tile} & \texttt{sizes} --- array of tile sizes \\
    ~ & \texttt{interchange} --- order of loops after tiling \\
    ~ & \texttt{pad} --- whether to pad partial tiles\\
    ~ & \texttt{pack\_paddings} --- non-removable padding for arrays\\
    ~ & \texttt{hoist\_paddings} --- number of loops from which to hoist padding for arrays\\
    ~ & \texttt{peel} --- loops from which to peel off partial tiles\\
    ~ & \texttt{scalarize\_dyn\_dims} --- whether to emit scalar code for non-vectorizable (dynamic) dimensions\\
    \hline
    \textsc{Vectorize} & \texttt{vectorize\_padding} --- whether to vectorize pad operations\\
    \hline
    \textsc{PipelineOneParentLoop} & \texttt{parent\_loop\_num} --- which parent loop to pipeline\\
    ~ & \texttt{II} --- Iteration Interval \\
    ~ & \texttt{read\_latency} --- Latency of a read operation \\
    \hline
    \textsc{UnrollOneParentLoop} & \texttt{parent\_loop\_num} --- which parent loop to unroll\\
    ~ & \texttt{unroll\_amount} --- by how many iterations to unroll the loop \\
    \hline
    \textsc{UnrollOneVectorOp} & \texttt{source\_shape} --- source shape of the vector op to unroll\\
    ~ & \texttt{source\_shape} --- target shape to unroll the vector op to \\
    \hline
    \textsc{Bufferize} & \emph{none}\\
    \hline
    \textsc{Sparsify} & \emph{none}\\
    \hline
    \textsc{LowerVectors} & \texttt{contraction\_lowering} --- how to lower vector contractions (outer/inner product, LLVM matrix intrinsics)\\
    ~ & \texttt{multi\_reduction\_lowering} --- how to lower multidimensional reductions (inner or outer parallel)\\
    ~ & \texttt{transpose\_lowering} --- how to lower transpositions (elementwise, flat, vector shuffle, target-specific)\\
    \hline
    \textsc{LowerToLLVM} & \emph{none}\\
    \bottomrule
    \end{tabularx}
    \smallskip
    \caption{\label{tbl:list-of-transformations}Transformations (and their options) that are currently available in Python. Some loop transformations apply only to loops produced by tiling and must be combined with tiling (e.g., peeling or interchange).}

  \medskip
  \footnotesize
  \centering
  \begin{tabularx}{\textwidth}{X|ccccccc}
    \toprule
     \textbf{Benchmark}             & L1 @ 12.8KB & L2 @ 20\% & L2 @ 40\% & L2 @ 90\% & L3 @ 40\% & L3 @ 80\% & DRAM \\
    \midrule
        copy (1 read + 1 write / B) &     289.5   &    89.3   &    83.9     &    54.8   &   25.7    &   17.2    &  12.2 \\
    \hline
   \begin{tabular}{@{}l@{}}extrapolated \\ (2 reads + 1 write / B)\end{tabular} &    434.25  &    134    &   125.8   &    82.2   &   38.5    &   25.8    &  18.3 \\
    \bottomrule
  \end{tabularx}
  \smallskip
  \caption{Measured and extrapolated single core copy performance in $\mathrm{GB}/s$.}
  \label{tbl:measured-bandwidths}
\end{table}

The transformations listed in Table~\ref{tbl:list-of-transformations} latch on a structured unit of the IR, with additional information and constraints.
E.g.\ for the \lstinline|UnrollOneVectorOp| case, the structured unit is \lstinline|vector.contract|, with additional constraints on its current shape, and providing the target shape after transformation.
Note that this direct control of multi-dimensional structured operations is not available in IRs that treat loops as the unit handle for transformations~\cite{ragan2013halide, chen2018tvm, vasilache2019next}.

To complete the environment to control and run transformation and code generation experiments, we provide a computation chaining API to define transformation sequences, see Figure~\ref{fig:tilingexample}.
% Together with Python metaprogramming, harnesses and interoperability capabilities, this provides a flexible, all-batteries-included environment to devise, control and run experiments.

\subsection{Experimental Setup}

Experiments are run on an Intel Xeon Gold 6154 CPU @ $3.00\mathrm{GHz}$.
This is a processor with dual-issue AVX512 fma instructions and 32KB of L1D, 32KB of L1I, 1MB L2 cache per core and a unified 25MB L3 shared by 18 cores.
Single-threaded single precision computations peak at $192\mathrm{GFLOP}/s$ 
(2 \lstinline|fma| operations per cycle, 2 operations (\lstinline|mul| + \lstinline|add|) each on 16 \lstinline|f32|).
The theoretical L1 bandwidth of a single core is $384\mathrm{GB}/s$ (assuming 1 load and 1 store instruction per cycle, each one of 64B).

Measurements are nearly bare-metal, in privileged mode, following scientific computing~\cite{benchmark2015} and LLVM benchmarking recommendations~\cite{LLVM_benchmarking}.
In particular, we disable turbo boost, address space randomization and the SMT pair of the core we run on.
We also run in a specific shielded \lstinline|cpuset| consisting of a single core and migrate all 
processes away from the core we execute on.

We further measure a peak memory bandwidths given in Figure~\ref{tbl:measured-bandwidths} with a simple contiguous copy benchmark.
We report the highest measured L1 throughput in $\mathrm{GB}/s$ which we find by fine-tuning, this maximum occurs for $12.8\mathrm{KB}$ read buffer size (i.e.\ $25.6\mathrm{KB}$ total buffer sizes, or approximately 80\% of L1 capacity).
Allocations occur at 64B boundary to guarantee the absence of cache line splitting given the sizes and transformations considered 
for this peak bandwidth analysis.
% Other than 64B-aligned allocations, we do not control allocation to e.g. guarantee good associativity behavior.
Since the hardware can issue 2 loads and 1 store per cycle, we also extrapolate the ideal scaling of the 
measured bandwidth to the actual load/store mix of a given scenario (i.e.\ by mechanically scaling the peak bandwidth by up to 50\% over the one measured with the copy benchmark).
Depending on the benchmark, the roofline~\cite{roofline} is either the measured copy bandwidth (e.g.\ for transpose or reduction) or the extrapolated bandwidth (e.g.\ for depthwise convolution that perform multiple reads for 1 write).
                            
In the following, all experiments consist of single threaded execution times measured on our benchmark system. 
We perform 100 measurements and plot the median. Black error bars show the 25\% and 75\% quantiles to quantify the measurement variance.
We report steady-state performance after dropping warm-up iterations due to compulsory cache misses and other overheads.\footnote{Note that with AVX-512, due to throttling issues, significant warmup may be required.}
Such overheads are relevant in larger-scale experiments in which fusion would be required to keep the L1 hot.
We compare the benchmark results to reference implementations to guarantee correctness.

\subsection{Benchmarks}

We evaluate the effectiveness of the strategies developed within our infrastructure on a range of kernels that dominate the execution time of machine learning workloads. All kernels execute a single tensor algebraic operation. Our results highlight the raw operator performance, independently of fusion and layout optimization opportunities across multiple operations.

We distinguish between memory-bound and compute-bound kernels. The memory-bound kernels move and reorder data to match the access patterns of the more compute intense operations. We benchmark the following memory-bound kernels:
\begin{itemize}
    \item Copy performance is an essential performance metric. The \emph{Copy2D} benchmark operates on $2$-D tensors that store contiguous data. This setup allows us more flexibility in targeting tiling, vectorization and unrolling than a flat $1$-D buffer but is otherwise equivalent.
    \item Transposition is a ubiquitous operation that comes in different shapes and sizes. The \emph{Transpose2D} benchmark implements a $2$-D transpose. It is the common denominator of higher dimensional transpose operations since a $n>2$-D transpose can be rewritten as iterated $2$-D transposes at various locations within the tensor. For instance a $4$-D $i,j,k,l\xrightarrow{} k,j,l,i$ transposition can be rewritten as $j\times k$ transposes of $(i,\ldots,l\xrightarrow{}\ldots,l,i)$. 
    This is always possible while keeping the fastest varying dimensions contiguous for both input and output tensors. 
    \item Reductions are data aggregation operations and an important algorithmic motif. Here we focus on the bandwidth-bound reductions associated with matrix-vector products or similar operations occurring in data analytics and neural networks. 
    The \emph{ColRed2D} and \emph{RowRed2D} benchmarks reduce the rows or columns of a $2$-D tensor to a $1$-D tensor, respectively. 
\end{itemize}

Compute-bound kernels have significant reuse and exhibit much higher computational than memory bandwidth needs. 
Their execution time is thus limited by the compute throughput rather than the memory bandwidth.
We benchmark the following compute-bound kernels:
\begin{itemize}
\item Matrix multiplication is ubiquitous in numerical computing. 
  The \emph{Matmul} benchmark implements plain matrix multiplication.
\item Convolutions ($1$-D and $2-D$ with strided and dilated variants) dominate
  the execution time of many machine learning models. In this paper we focus on 
  the so-called \emph{NHWC} format but other formats are trivial to generate with
  our OpDSL approach.
\end{itemize}
Lastly, we discuss the performance of depthwise convolutions for sizes relevant to the popular MobileNet~\cite{MobileNetV2} model.
\emph{DepthwiseConv2D} is a \emph{NHWC} format kernel whose compute to communication ratio presents challenging and interesting problems. 

For each benchmark, we manually derive up to 5 expert compiler strategies using the transformations of Table~\ref{tbl:list-of-transformations}. 
In each case, we run a few manual experiments to devise good register tile sizes such that the performance of L1-resident kernels is high.
We then fix the tile sizes and select the best performing strategy among the 5 in each case.
This is akin to a fixed expert-driven heuristic.

Systematic autotuning and search space exploration is an area of active investigation that is expected to bring significant improvements.
We discuss preliminary results with autotuning in Section~\ref{subsec:prelim-tuning-results}.

\subsection{Performance of Memory-Bandwidth-Bound Kernels}

A bandwidth-bound kernel may be limited by different levels of the memory hierarchy, depending on problem size and access pattern. 
We run the bandwidth-bound kernels on different problem sizes and analyze their performance for all three cache hierarchy levels. 
In the particular case of the \lstinline|copy| benchmark, we run a small search over different $2$-D vector sizes, load/store interleaving, and loop unrolling specific for every problem size.
The measured bandwidth then becomes our L1 bandwidth measuring stick.

\subsubsection{L1 Bandwidths}
\begin{figure}[h!tb]
\centering
\minipage[t]{\textwidth}
\includegraphics[width=0.98\textwidth]{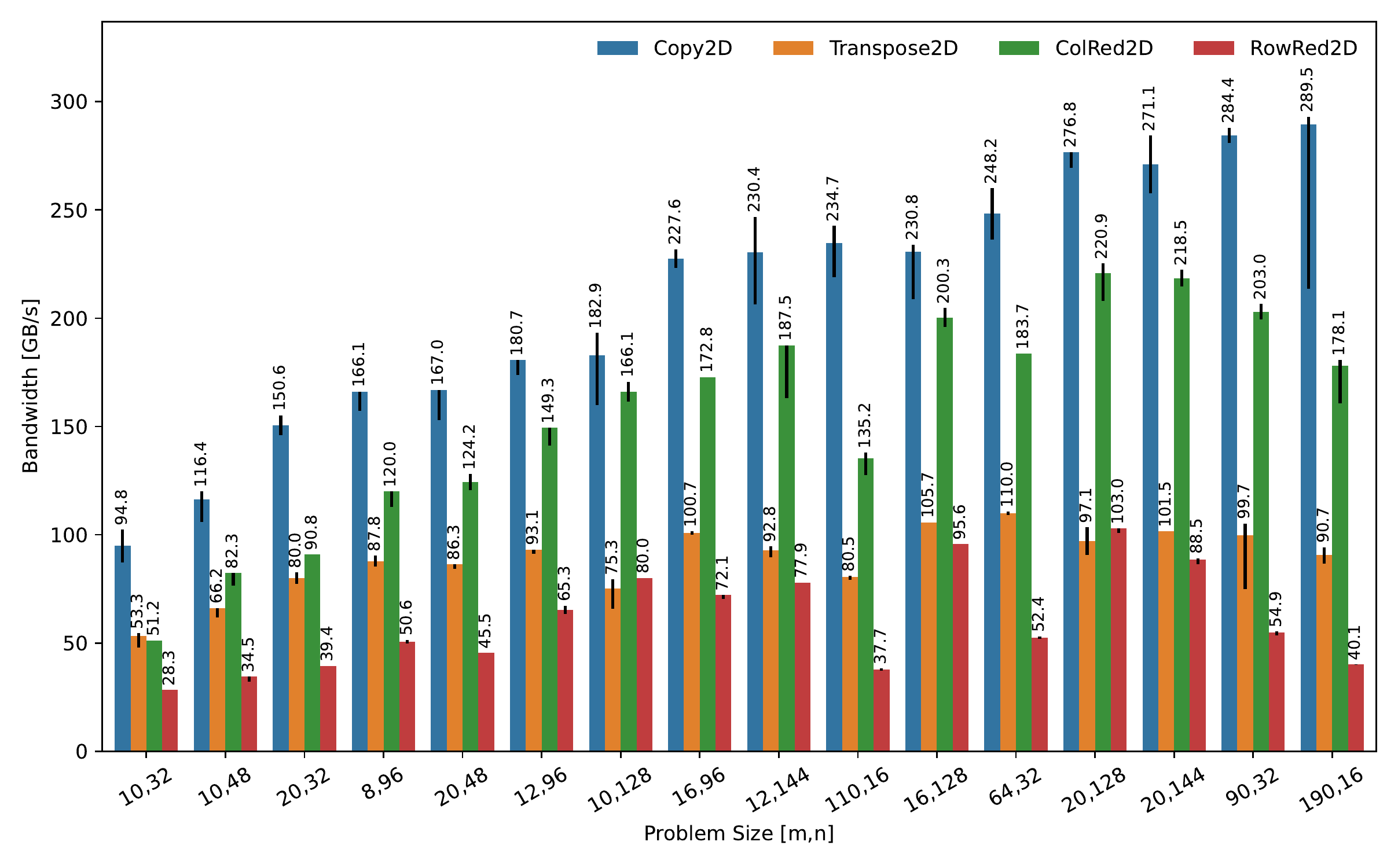}
\endminipage
\caption{\label{fig:l1-bandwidth}Memory bandwidth for bandwidth-bound kernels on problem sizes fitting L1 cache. Theoretical copy peak bandwidth is 384 GB/s (289 GB/s measured). Transpose performance is limited by \lstinline|xmm| loads and \lstinline|ymm| shuffles, more work is needed to get good \lstinline|zmm| assembly (see Section~\ref{subsubsec:transpose-discussion}).}
\end{figure}

\autoref{fig:l1-bandwidth} shows the achieved memory bandwidth for all bandwidth-bound kernels on problem sizes fitting L1 cache. 
We observe the highest bandwidths for the \emph{Copy2D} kernel that performs exactly $1$ \lstinline|vector.load| and $1$ \lstinline|vector.load| per 64B of data. 
This kernel does not perform any computation nor rearranges data. 
Despite executing in a tight loop with data resident in L1 cache, the benchmarks show that offsetting latency requires a sufficiently large problem size. 
We start seeing L1 bandwidth performance in the 200 GB/s zone only at around 4KB of read data (8KB total, i.e.\ 25\% L1 capacity).
A 200GB/s L1 bandwidth is sustained only in the 8KB--14KB range of read data (16KB--28KB total, i.e.\ 50-87\% L1 capacity).
We observe the maximal bandwidth of 289 GB/s at roughly 75\% L1 cache utilization. 
Larger problem sizes achieve lower bandwidths, presumably, due to conflict misses.
Lastly, variance starts increasing greatly around 80\% of L1 capacity.

\emph{Transpose2D} rearranges the data while moving.
This results in significantly more complex instruction sequences than a \emph{Copy2D}'s simple $1$ load $1$ store pattern.
We achieve bandwidths of 30-60\% of \emph{Copy2D} and up to 109 GB/s L1 bandwidth for the most favorable size.
We discuss the characteristics of this kernel more deeply in Section~\ref{subsubsec:transpose-discussion} as well as opportunities for improvement.

\emph{ColRed2D} and \emph{RowRed2D} read more data (a full $2$-D tensor) than they write (a $1$-D tensor).
This is beneficial on our test system that can perform $2$ reads and $1$ write per cycle (see Table~\ref{tbl:measured-bandwidths}).
At the same, time they perform computations.
In particular \emph{RowRed2D} performs an expensive horizontal reduction along the vectorization dimension. 
While \emph{ColRed2D} achieves high bandwidths of up to 212 GB/s, we observe only up to 99 GB/s for \emph{RowRed2D}.
The lower performance is particularly pronounced for problem sizes with a short reduction dimension. 
This is due to the mapping of horizontal reduction on Intel processors and the notorious cost of crossing \lstinline|ymm| and \lstinline|xmm| boundaries.
The resulting performance is not better than combining \emph{Transpose2D} and \emph{ColRed2D}.
Finer tuning and better AVX-512 patterns are expected to improve the situation in the future.

Overall, rearranging data or performing computation during data movement immediately reduces the achieved L1 cache bandwidth.

\subsubsection{L2 Bandwidths}
\begin{figure}[h!tb]
\centering
\minipage[t]{\textwidth}
\includegraphics[width=0.98\textwidth]{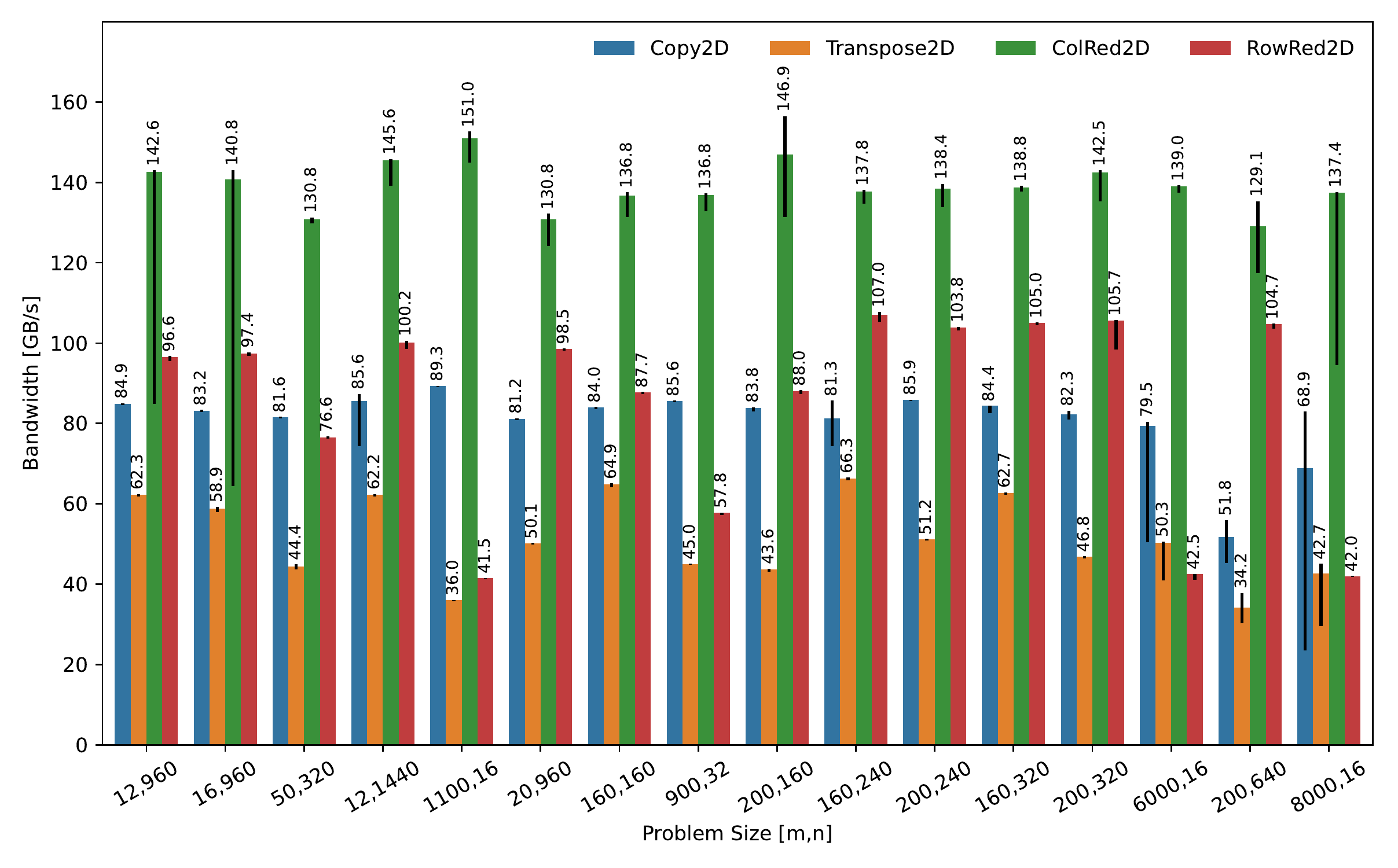}
\endminipage
\caption{\label{fig:l2-bandwidth}Memory bandwidth for bandwidth-bound kernels on problem sizes fitting the L2 cache. 
Measured peak copy bandwidth is $83.9 \mathrm{GB}/s$ ($1$ read and $1$ write per byte). 
Reductions can go past this because they perform an amortized $2$ reads and a fraction of a write per iteration. 
Despite the extra compute, dropping most writes is still a net win.}
\end{figure}

\autoref{fig:l2-bandwidth} shows the achieved memory bandwidths for problem sizes fitting L2 cache. 
\emph{ColRed2D} and \emph{RowRed2D} achieve the highest bandwidths due to their favorable read-to-write ratio. 
We observe bandwidths of up to 125 GB/s. 
\emph{RowRed2D} achieves similarly high bandwidth but also falls behind for problem sizes with small vector dimension, due to the slow reduction of the last vector in the row. 
The difference between \emph{Copy2D} and \emph{Transpose2D} is less pronounced since the cost for rearranging the data partly overlap with the slower data movement.

\subsubsection{L3 Bandwidths}
\begin{figure}[h!tb]
\centering
\minipage[t]{\textwidth}
\includegraphics[width=0.98\textwidth]{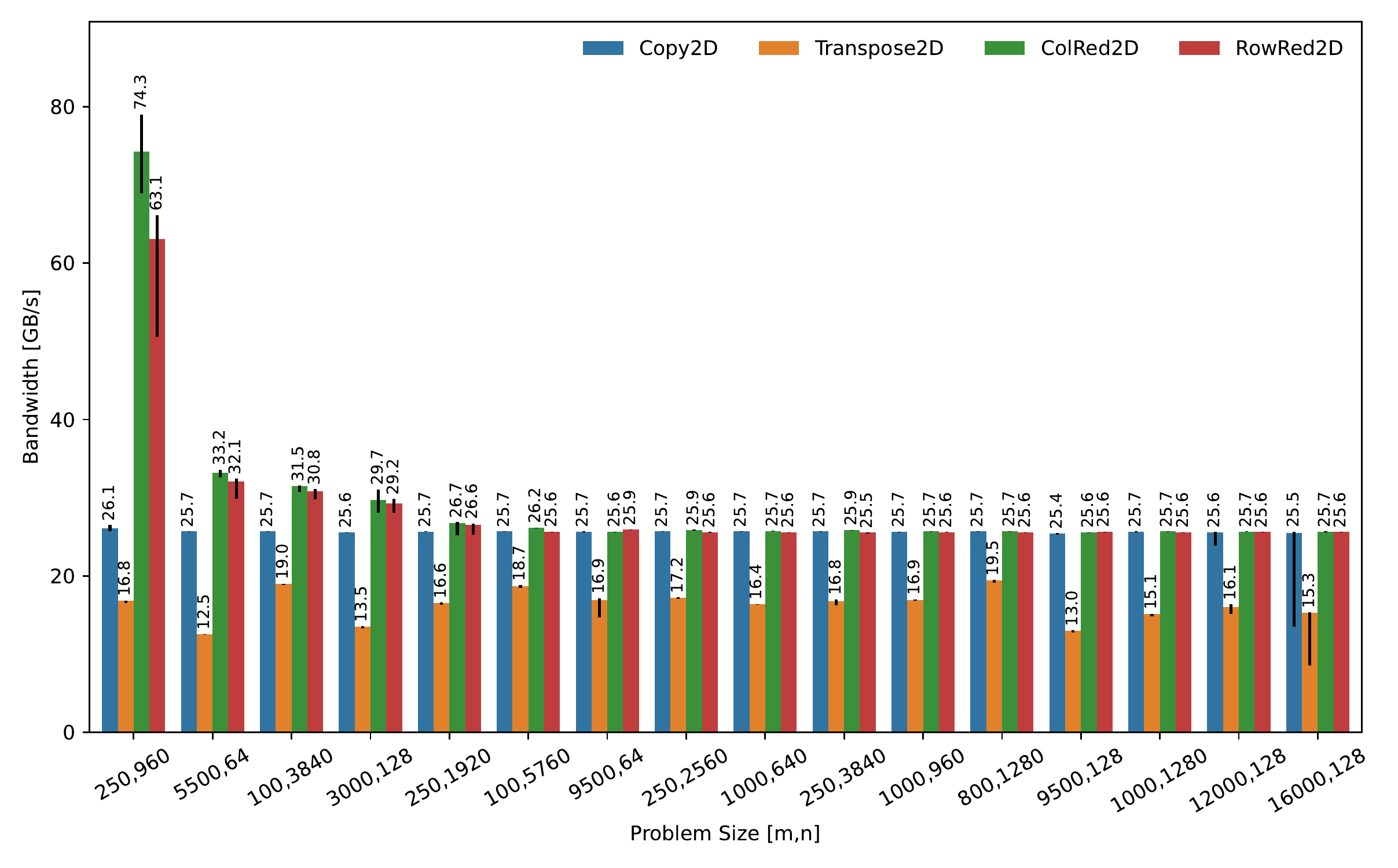}
\endminipage
\caption{\label{fig:l3-bandwidth}Memory bandwidth for bandwidth-bound kernels on problem sizes fitting L3 cache. 
Measured copy peak is 25.7 GB/s (L3 @40\%). 
Note that reductions of 250x960xf32 still fit in L2 because the write buffer is much smaller 
(i.e. \lstinline|250xf32| for row reduction and \lstinline|960xf32| for columns reduction.}
\end{figure}

\autoref{fig:l3-bandwidth} shows the achieved memory bandwidths for problem sizes fitting L3 cache. 
All benchmarks except for \emph{Transpose2D} peak at an achieved memory bandwidth of 26 GB/s. 
We attribute this performance difference to the L3 cache latency of that cannot be hidden for the transpose access patterns.

\subsubsection{Transpose Discussion}
\label{subsubsec:transpose-discussion}
In this section, we dive deeper into the specific performance of the transpose
patterns we generate and discuss opportunities for improvements.

First, to get to the best transposed version, the Intel Reference Optimization Manual~\cite{IntelOptimizationManual} recommends using the \lstinline|vblendps| instruction (example 15.19).
The vector dialect provides a dedicated \lstinline|vector.transpose| operation as well as multiple lowering strategies when going to LLVM. 
Since we emit LLVMIR, we do not have direct control over register allocation and instruction selection.
To offset this lack of control, hardware-specific vector dialects (e.g. the \lstinline|x86vector| dialect) provide
access to intrinsics to match clang intrinsics such as \lstinline{_mm256_blend_ps}.
Unfortunately, some of these clang intrinsics (including \lstinline{_mm256_blend_ps}) are not backed by a real intrinsic 
implementation and do not provide direct access to the corresponding ISA.
Instead, they lower immediately to generic LLVM \lstinline|shufflevector| instructions!
LLVM then relies on peephole optimization and especially \emph{SelectionDAG}; but these contain ``very little
arch-specific shuffle combines''~\cite{VBlendPsLimitation}.
In such a case, we were not able to compile to the desired \emph{vblendps} operation. 
Instead, we had to originally settle to a pure shuffle-based implementation\footnote{Note that such a pure shuffle-based 
implementation is similar to the one used in the
Eigen~\cite{eigen} library (more specifically, the \lstinline|ptranspose(PacketBlock<Packet8f,8>& kernel)| routine).}.

To circumvent this issue, we also provide an \lstinline{InlineAsmOp} and specific asm-based intrinsics that encode
the intended instructions (e.g. an \lstinline{mm256BlendPsAsm} lowering helper that is \emph{guaranteed} to emit the \lstinline{vblendps} instruction where we intend it to)\footnote{
We provide such specialized avx2 lowerings under the \lstinline|transpose\_lowering=avx2| for \lstinline|vector| transpose
of size \lstinline|8x8|.}.
This provides a nice \lstinline|30-40\%| performance boost on small transpose sizes over the version
generated by LLVM for an 8x8 transpose implemented only with intrinsics that lower to \lstinline|shufflevector|.

The Intel Reference Optimization Manual~\cite{IntelOptimizationManual} further recommends using specific permutation instructions from memory to further reduce shuffle issue port pressure.
These are not yet available as first-class citizens in our lowering but LLVM's peephole optimization was able to recover the
pattern in the very specific case of a \lstinline|4x8| tiling of the transpose.
While this does not make use of \lstinline|avx512| instructions and is thus potentially 2x slower 
than we could hope for, the LLVM  \lstinline|4x8| version performs best, by far.
We summarize our findings in Table~\ref{tbl:comparison-transpose}.

\begin{table}[h!tb]
  \small
  \centering
  \begin{tabularx}{\textwidth}{X|cccc}
    \toprule
    \textbf{Size}     & \textbf{Tile8x8Shuffle}  & \textbf{Tile16x16Shuffle} & \textbf{Tile8x8AVX2} & \textbf{Tile4x8Shuffle}  \\
    \midrule
    \lstinline|16x16| &          24.1     &       22.5       &     41.8    &      55.3     \\
    \hline
    \lstinline|32x32| &          29       &       27         &     64      &      95       \\
    \bottomrule
  \end{tabularx}
  \smallskip
  \caption{Median (p50) measured performance of a \lstinline|vector.transpose| lowering strategies (GB/s).
   The "natural" \lstinline|16x16| native AVX512 lowering with \lstinline|shufflevector| performs significanly
   worse than a custom \lstinline|8x8| AVX2 lowering mixing \lstinline|vblendps| and \lstinline|shufflevector|.
   The less obvious \lstinline|4x8| tiling and \lstinline|shufflevector| performs significantly better despite
   using \lstinline|xmm| loads.}
  \label{tbl:comparison-transpose}
\end{table}

Based on this understanding, we conducted additional experiments using \lstinline|UnrollOneVectorOp| 
to try and keep a \lstinline|16x16| \lstinline|vector.transpose| shape while forcing \lstinline|xmm| and \lstinline|ymm| loads. 
This experiments resulted in a performance degradation, likely due to the \lstinline|shufflevector| that we do not yet canonicalize. 
More work will be needed to get to the bottom of  this. 

The performance has room for future improvements as we devise a better-suited AVX512 version and more finely tune tile sizes and codegen strategies.
We expect a better AVX512 solution will likely involve insight from prior work on Knight's Landing~\cite{IPDPS2013}.
For now, the measured efficiency is between 30\% and 55\% for a fully isolated $2$-D transpose operation.
This has to be put in additional context.
The copy kernels we studied in the previous section perform exactly $1$ 64B load and $1$ 64B store for each operation.
The transpose kernels are significantly more complex.
They involve 16B and 32B loads but still reach a comparatively high performance, given the small loads and additional shuffle instructions.

Lastly, an additional argument to consider is that transpositions are by themselves rarely isolated in a vacuum.
Instead, they often compose with other operations (e.g. matmul), in which case their cost is often amortized.

\subsection{Performance of Compute-Bound Kernels}

Let us now review the performance of matrix multiplication and convolutions.

\subsubsection{Matrix Multiplication}

Attaining a high matrix multiplication throughput is essential in numerical computations. 
We measure the performance for plain matrix multiplication and transposed variants:
\begin{equation*}
\begin{split}
C[m,n] = A[m,k]B[k,n] & \qquad(AB) \\
C[m,n] = A[k,m]B[k,n] & \qquad(A^TB) \\
C[m,n] = A[m,k]B[n,k] & \qquad(AB^T)
\end{split}
\end{equation*}

\begin{figure}[h!tb]
\centering
\minipage[t]{\textwidth}
\includegraphics[width=0.98\textwidth]{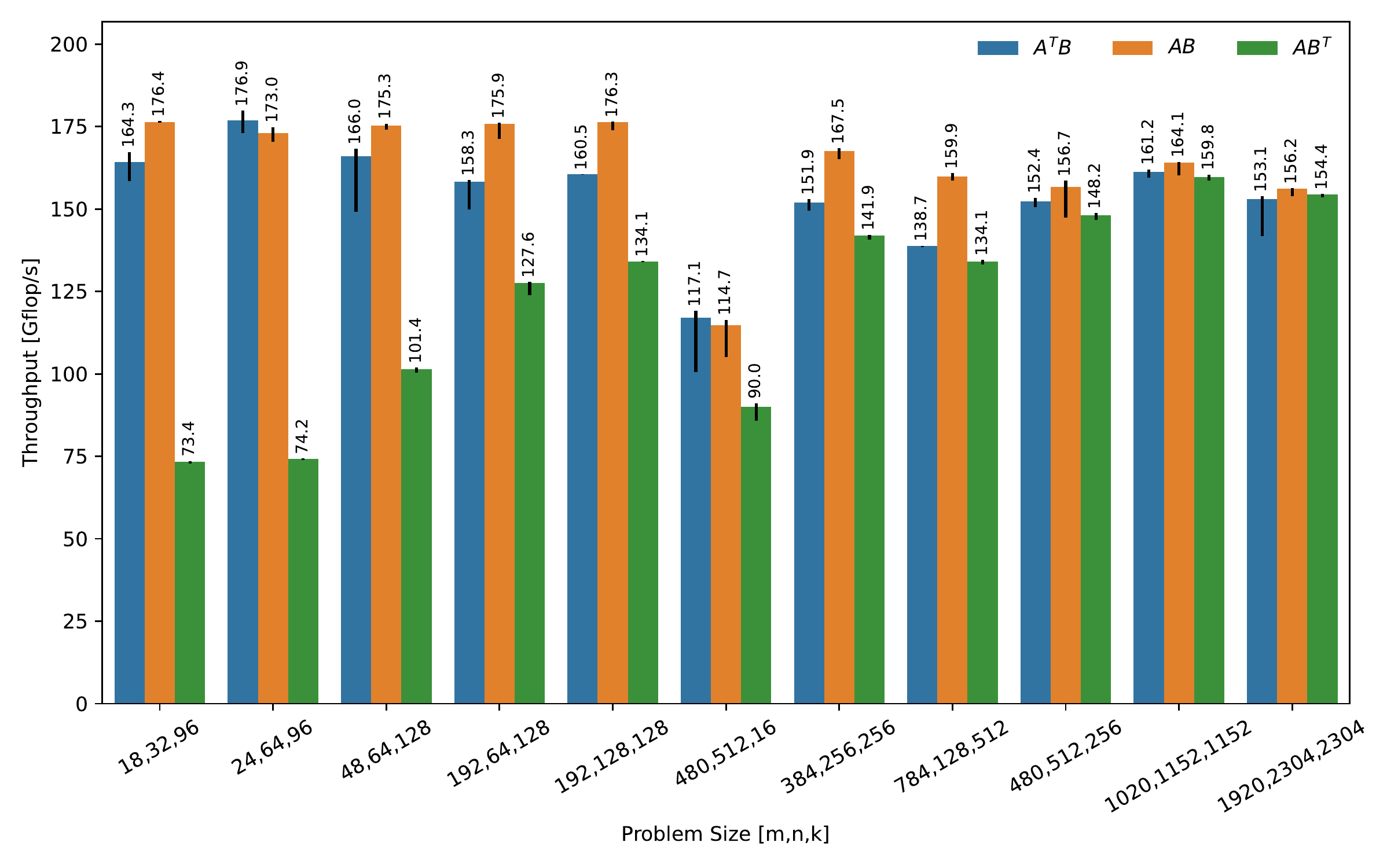}
\endminipage
\caption{\label{fig:matmul}Matrix multiplication compute throughput for different storage layouts and problem sizes (best of 5 fixed strategies). The theoretical peak is $192 \mathrm{GFLOP}/s$, fine peformance tuning is left for future work.}
\end{figure}

Figure~\ref{fig:matmul} illustrates the performance of matrix multiplication for various sizes. 
We reach a 92\% efficiency in the \lstinline|AB| kernel case, which demonstrates that our codegen approach emits close to peak arithmetic intensity kernels.
It is worth noting that in the low-latency regime cases, layout has a significant impact on performance.
In particular, the $AB^T$ has a significantly lower performance due to the layout of reduction dimension along the fastest varying memory dimension (i.e.\ $C[m,n] = A[m,k]B[n,k]$). 
This is a similar horizontal reduction issue as for \emph{RowReduction2D}.
At low-latency sizes, the cost of a transpose is prohibitive and performance suffers. 
As we reach larger sizes, the tranpose becomes beneficial and no more performance difference is observable. 

Larger sizes use a fixed tiling of \lstinline|288x128x512| but are not specifically tuned.
Additionally, we do not yet emit prefetch instructions or try to pipeline data movements with computation. 
These transformations along with deeper tuning is left for future work.
For a preview discussion of achievable performance combining autotuning with our infrastructure, see Section~\ref{subsec:prelim-tuning-results}.

\subsubsection{Convolution}

Convolution folds an input tensor with a multi-dimensional kernel following a \emph{sliding window} pattern.
%\ntv{note, I dropped the "image" terminology because in 1-D these are typically signals (e.g. sound)}.
We specify a convolution operator in terms of its input image and kernel dimensions following standard practice in the ML literature:
\begin{itemize}
 \item $H$: height of the image,
 \item $W$: width of the image,
 \item $N$: batch number of the image (input and output only),
 \item $C$: input channels of the image,
 \item $F$: output filters of the image (kernel only),
\end{itemize}
Additional parameters are the kernel widths $K_h$ and $K_w$, the strides $S_w$ and $S_h$, and the dilations $D_w$ and $D_h$. We measure the performance for $1$-D and $2$-D convolution in \lstinline|NHWC| format:
\begin{equation*}
\begin{array}{llll}
O[n,w,f] &=& I[n,w\times S_w+k_w\times D_w,c]\cdot K[k_w,c,f] & (1-D) \\
O[n,h,w,f] &=& I[n,h\times S_h+k_h\times D_h,w\times S_w+k_w\times D_w,c]\cdot K[k_h,k_w,c,f] & (2-D)
\end{array}
\end{equation*}
where the stride ($S_h$ and $S_w$) and the dilations ($D_h$ and $D_w$) parameters control the input image access pattern. 

\begin{figure}[h!tb]
\centering
\minipage[t]{\textwidth}
\includegraphics[width=0.98\textwidth]{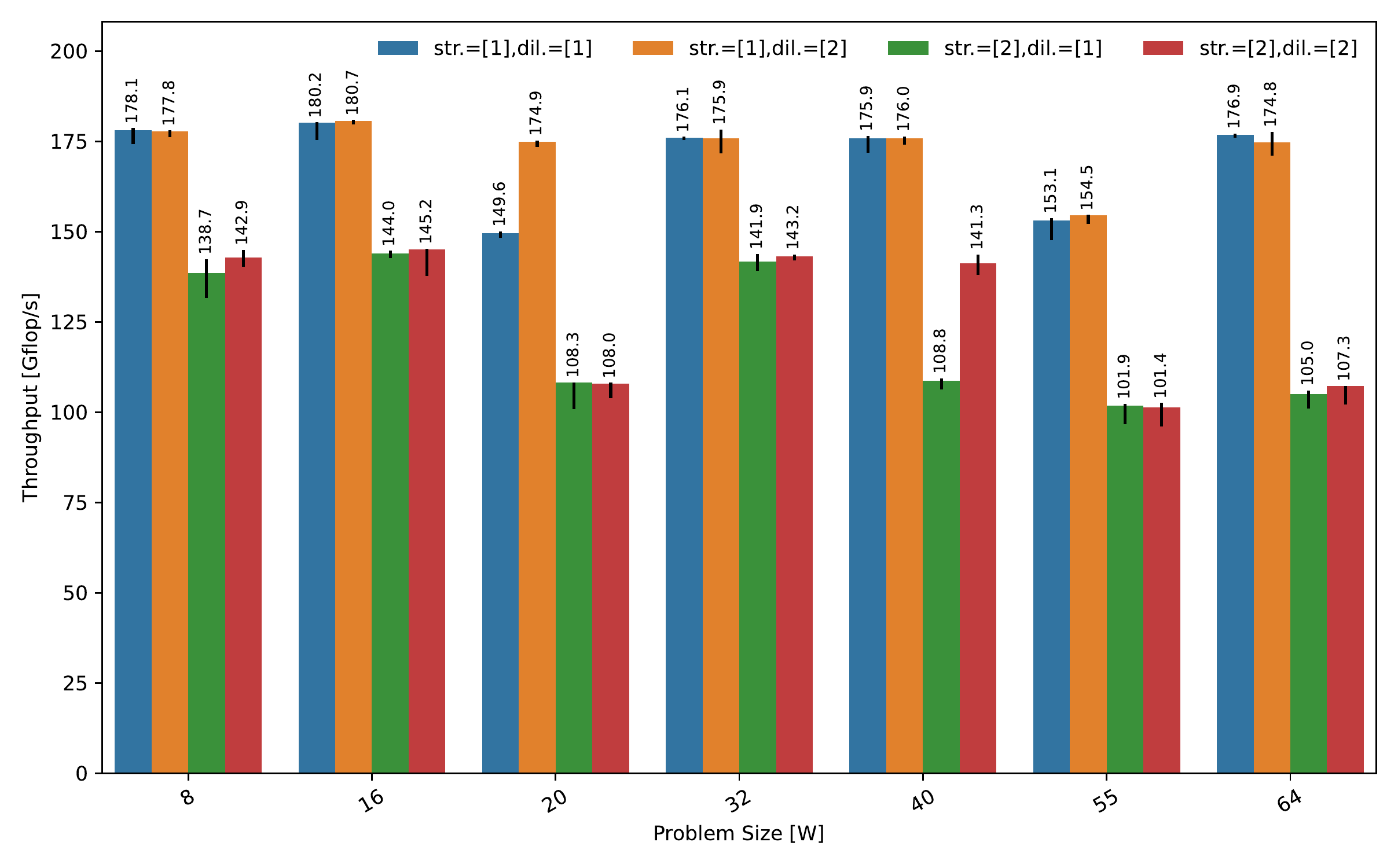}
\endminipage
\caption{\label{fig:conv-1d}L1-resident $1$-D convolution for different strides, dilations, and problem sizes ($(N,C,F,K_w)$ fixed to $(1,32,64,3)$). Theoretical peak is $192 \mathrm{GFLOP}/s$, fine tuning of performance is left for future work.}
\end{figure}

\autoref{fig:conv-1d} shows the compute throughput for the $1$-D (left) and the $2$-D (right) convolution after tiling once. 
All problem sizes not shown in the plot are constant ($(N,C,F,K_w)=(1,32,64,3)$ for the $1$-D case and $(N,C,F,K_w)=(1,32,64,3,3)$ for the $2$-D case).
The reader familiar with the literature may remark that we are providing performance results for the more difficult \emph{batch size 1} configuration.
The particularity of the problem is that it contains one fewer dimension of parallelism.

We measure high performance for both the $1$-D and the $2$-D convolution when the stride is $1$, with a peak at around $96\%$ of theoretical hardware peak.
We observe a slowdown for non-unit strides.
The slowdown is more pronounced in the $1$-D case where the data size is quite small and the computation do not make up for the loss in access pattern efficiency.
The inefficiencies start to dissipate above input size $40$ and almost disappear in the $2$-D case.
Note that sizes $20$ and $55$ are not perfect multiples of the good vector sizes and require multi-dimensional loop peeling, resulting in lower performance (padding is not beneficial at such small sizes)\footnote{Specific size tuning to avoid peeling along some dimensions is expected to further improve the performance in such cases.}.

\begin{figure}[h!tb]
\centering
\minipage[t]{\textwidth}
\includegraphics[width=0.98\textwidth]{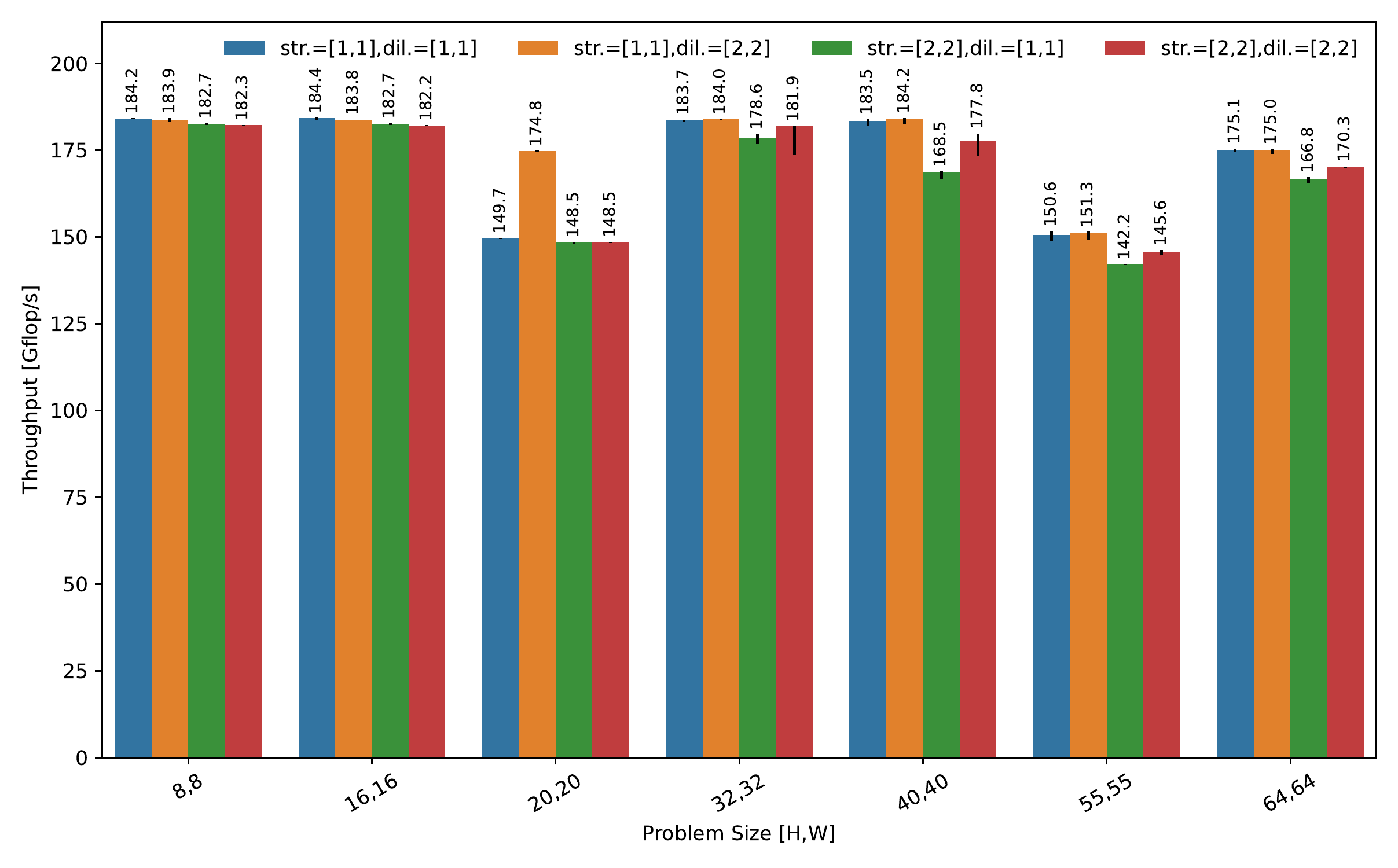}
\endminipage
\caption{\label{fig:conv-2d} L1-resident $2$-D convolution for different strides, dilations, and problem sizes ($(N,C,F,K_h,K_w)$ fixed to $(1,32,64,3,3)$). 
Theoretical peak is $192 \mathrm{GFLOP}/s$, fine tuning of performance is left for future work.}
\end{figure}

Since this is a heavily compute-bound kernel with multiple parameters, we only focus on the performance of the vectorized kernel for small sizes resident in L1.
Thanks to our modular and composable approach, larger sizes can be built from smaller sizes, following previously discussed padding, packing and peeling strategies (see Section~\ref{subsec:padding-values-and-packing}).
In particular, the compiler implements a single $1$-D vectorization strategy that exhibits enough arithmetic intensity and is reused in the $2$-D case.
To achieve this, the $2$-D strategy simply tiles the $H$ and $K_h$ dimensions by $1$ then folds and canonicalizes it further.
The $2$-D case reduces to the $1$-D case where high-intensity vectorization patterns apply.

\subsection{Depthwise Convolution}

Depthwise convolution is a compute-efficient twist on convolution mainly oriented at mobile inference.
$1$-D and $2$-D depthwise convolution in \lstinline|NHWC| format are expressed as:
\begin{equation*}
\begin{array}{llll}
O[n,w,c] &=& I[n,w\times S_w+k_w\times D_w,c]\cdot K[k_w,c] & (1-D) \\
O[n,h,w,c] &=& I[n,h\times S_h+k_h\times D_h,w\times S_w+k_w\times D_w,c]\cdot K[k_h,k_w,c] & (2-D)
\end{array}
\end{equation*}
where the strides ($S_h$ and $S_w$) and the dilations ($D_h$ and $D_w$) control the input image access pattern. 
Compared to classical convolutions, the volume of the computation kernel is reduced by $1$ dimension (typically a $16\times$ to $512\times$ reduction in both compute and data volume).
The computation is also altered to only reduce across the window dimension(s) (and not along the channel dimension).
The depthwise convolution kernel has a low arithmetic intensity ($\frac{K_w}{\mathrm{sizeof}(\mathrm{element\_type})}$ $\mathrm{FLOP}/s$ per byte in the $1$-D case and $\frac{K_h.K_w}{\mathrm{sizeof}(\mathrm{element\_type})}$ $\mathrm{FLOP}/s$ per byte in the $2$-D case).
As a result, depthwise convolution is memory bandwidth-bound on modern server processors.

Since the kernel is memory-bound, it is important to measure the volume of data properly.
In the following experiments, we evaluate the total data volume as follows ($K_h$ and $H$ are $1$ in the $1$-D case):
\begin{equation*}
\begin{array}{lll}
\mathrm{Total}_{\mathrm{size}} &=& \mathrm{Out}_{\mathrm{size}} + \mathrm{Ker}_{\mathrm{size}} + \frac{\mathrm{In}_{\mathrm{size}}}{\Pi_i \gcd(\mathrm{stride}_i, \mathrm{dilation}_i)} \\
           &=& N\cdot C\cdot 
                \left(H\cdot W + \frac{H_{\mathrm{in}} \cdot W_{\mathrm{in}}}{\Pi_i \gcd(\mathrm{stride}_i, \mathrm{dilation}_i)}\right) + K_h\cdot K_w\cdot C 
\end{array}
\end{equation*}
The conservative reasoning follows:
\begin{enumerate}
    \item Every element of the convolution kernel is read.
    \item Every element of the output is written but is not considered read; the data is simply overwritten.
    \item Only a subset of the input is read, depending on the values of the stride and dilations quantities. This is captured by a scaling \emph{adjustment} factor.
\end{enumerate}
The scaling adjustment factor is obtained by the intersection of the stride lattice and the dilation lattice, within the limits of the input boundary.
The gcd is a good approximation when the size of the kernel is greater than the stride.
This guarantees we only compute the input points that participate in the derivation of at least $1$ output element.
This is a conservative measure: any element that is unused but falls in a common cache line with a used element is not counted; still the data will be moved by the hardware.

\begin{figure}[h!tb]
\centering
\minipage[t]{\textwidth}
\includegraphics[width=0.98\textwidth]{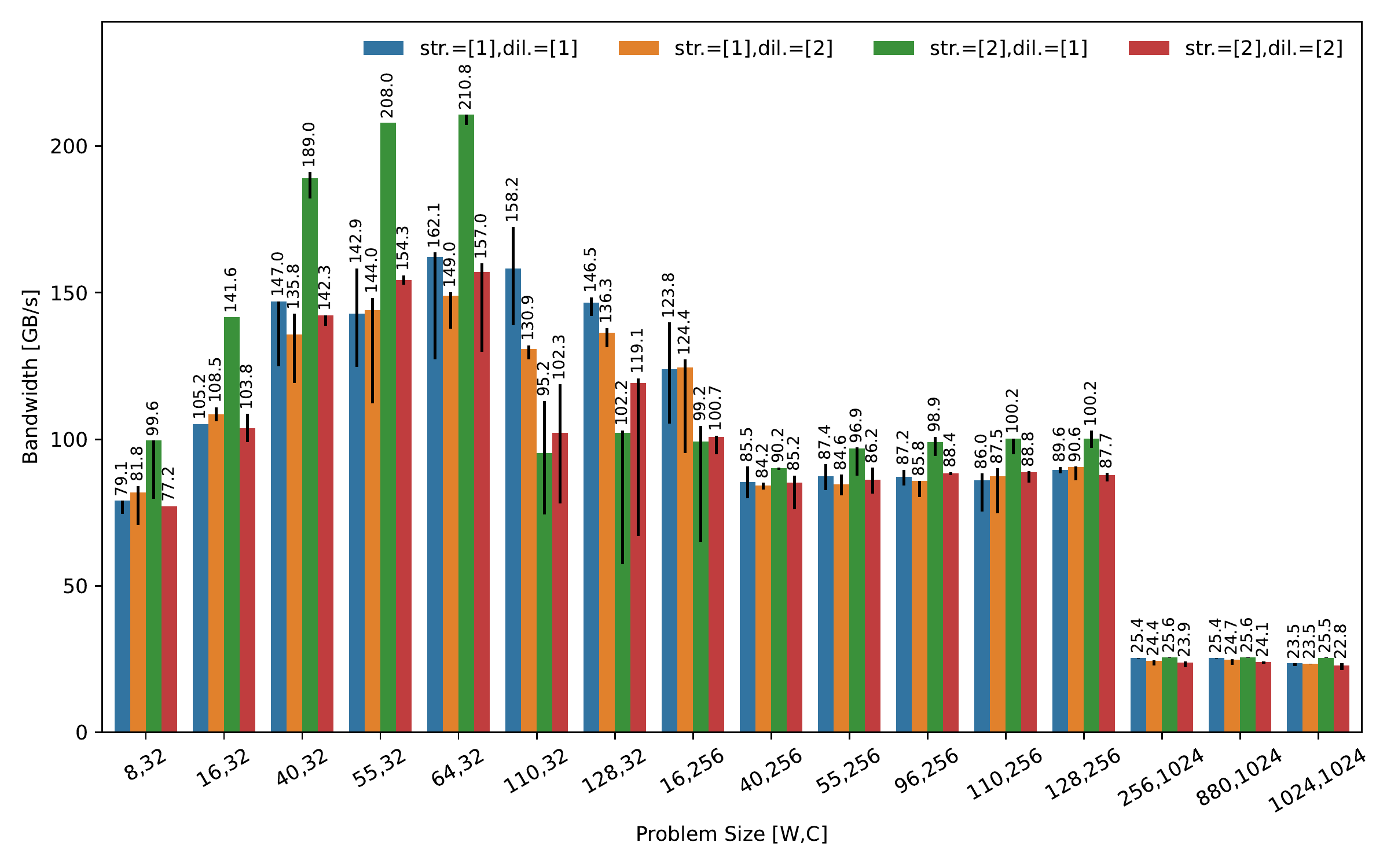}
\endminipage
\caption{\label{fig:depthwise-conv-1d}1-D depthwise convolution for different strides, dilations, and problem sizes. Measured copy peak $289 \mathrm{GB}/s$ (L1), $89.3 \mathrm{GB}/s$ (L2), $25.7 \mathrm{GB}/s$ (L3 @40\%).}
\end{figure}

\begin{figure}[h!tb]
\centering
\minipage[t]{\textwidth}
\includegraphics[width=0.98\textwidth]{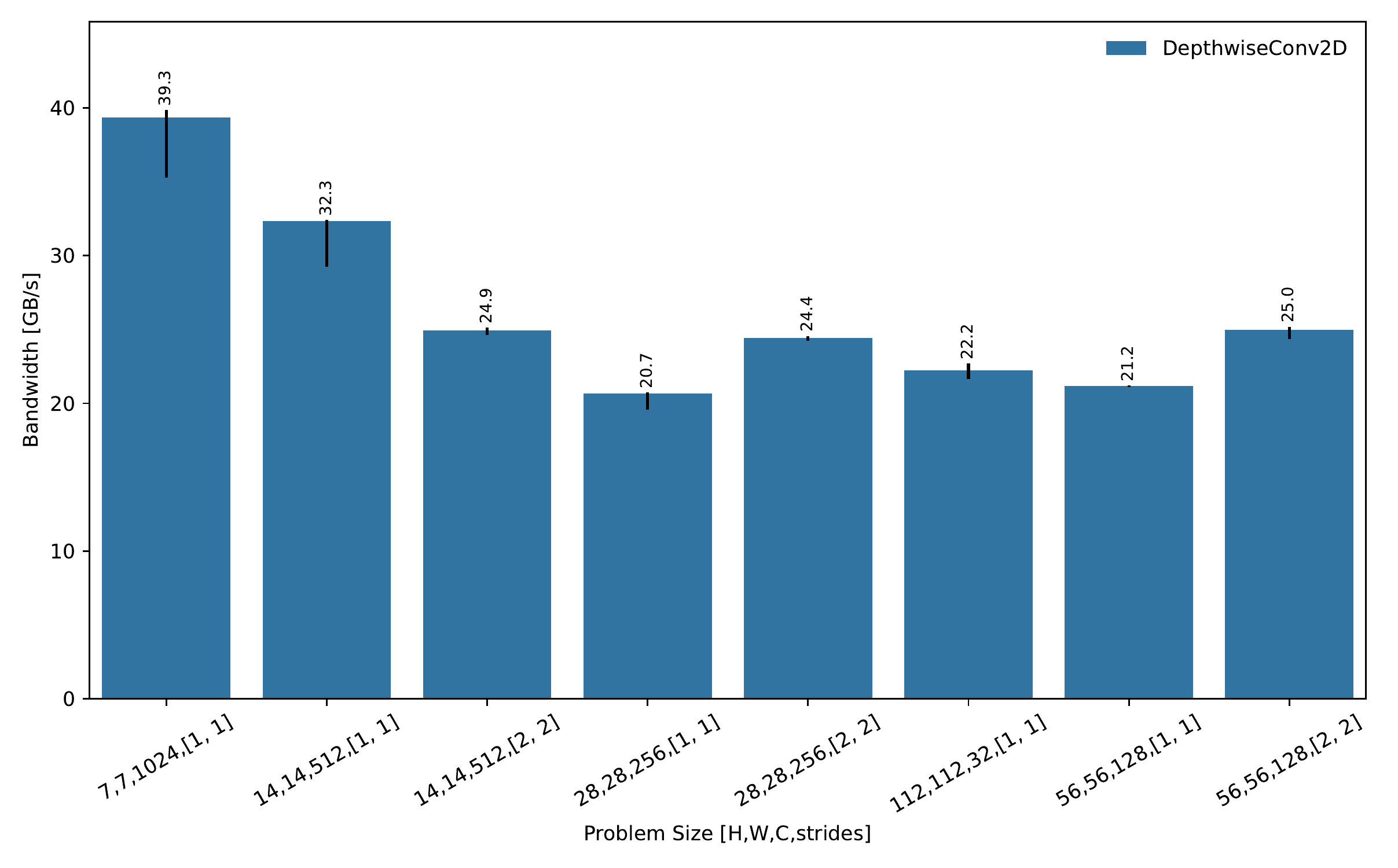}
\endminipage
\caption{\label{fig:depthwise-conv-2d}2-D depthwise convolution for different strides, dilations, and problem sizes. Measured copy peak $54.8 \mathrm{GB}/s$ (L2 @90\%), $25.7 \mathrm{GB}/s$ (L3 @40\%). The \lstinline|7x7x1024| problem takes \lstinline|16x16x1024xf32=1.05M| (input and output tensors). This slightly exceeds L2. Other sizes are deep in L3 territory and require parallelism to increase bandwidth (outside the scope of this paper).}
\end{figure}

Figures~\ref{fig:depthwise-conv-1d} and~\ref{fig:depthwise-conv-2d} show the memory bandwidth for the $1$-D and $2$-D depthwise convolution.
All problem sizes not shown in the plot are constant ($(N,C,K_w)=(1,32,3)$ for the $1$-D case and $(N,C,K_h,K_w)=(1,32,3,3)$ for the $2$-D case).
In the $1$-D case, we measure $45-75\%$ L1 bandwidth for the smaller problem sizes compared to \emph{Copy2D}.
As size increases, we see performance close to the L2 and L3 bandwidth of \emph{Copy2D}, which signals strong data reuse.
For the $2$-D case, we limit the exploration to sizes related to MobileNet layers with batch size $1$. 
We again see performance within $10-20\%$ of the peak L3 bandwidth for comparable sizes, suggesting good data reuse.

Lastly, note that our approach scales easily to other kernel sizes, strides and dilations than the traditional $1$ or $2$ values that are 
often the only supported by library implementations (and often not in all combinations).

\subsection{A Brief Evaluation of Sparse Code Generation}

We end this experimental evaluation with an example of sparse tensor code generation. As stated earlier, the \lstinline{sparse_tensor} dialect introduces the concept of sparse compilation into MLIR. The dialect makes sparse tensor types first class citizens by bridging high-level operations on these sparse tensors types with lower-level operations on the actual sparse storage schemes. To this end, the sparse tensor dialect introduces a tensor encoding attribute that allows specifying the formats of the Tensor Algebra Compiler (TACO)~\cite{kjolstad2017taco}. 

\begin{figure}[h!tb]
    \centering
    \includegraphics[width=\textwidth]{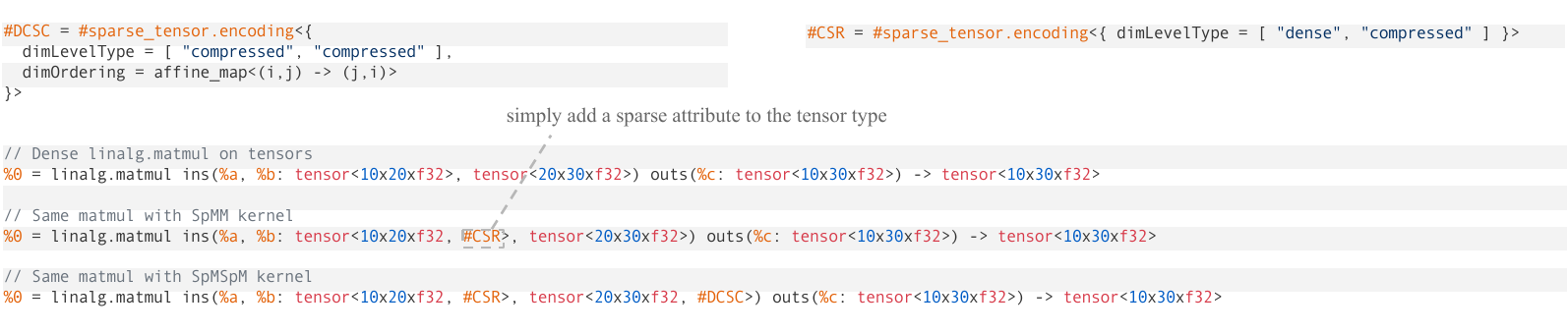}
    \caption{Definition of sparse tensor layouts and their usage in a matrix multiplication.}
    \label{fig:sparse_layouts}

    \medskip
    \centering
    \includegraphics[width=\textwidth]{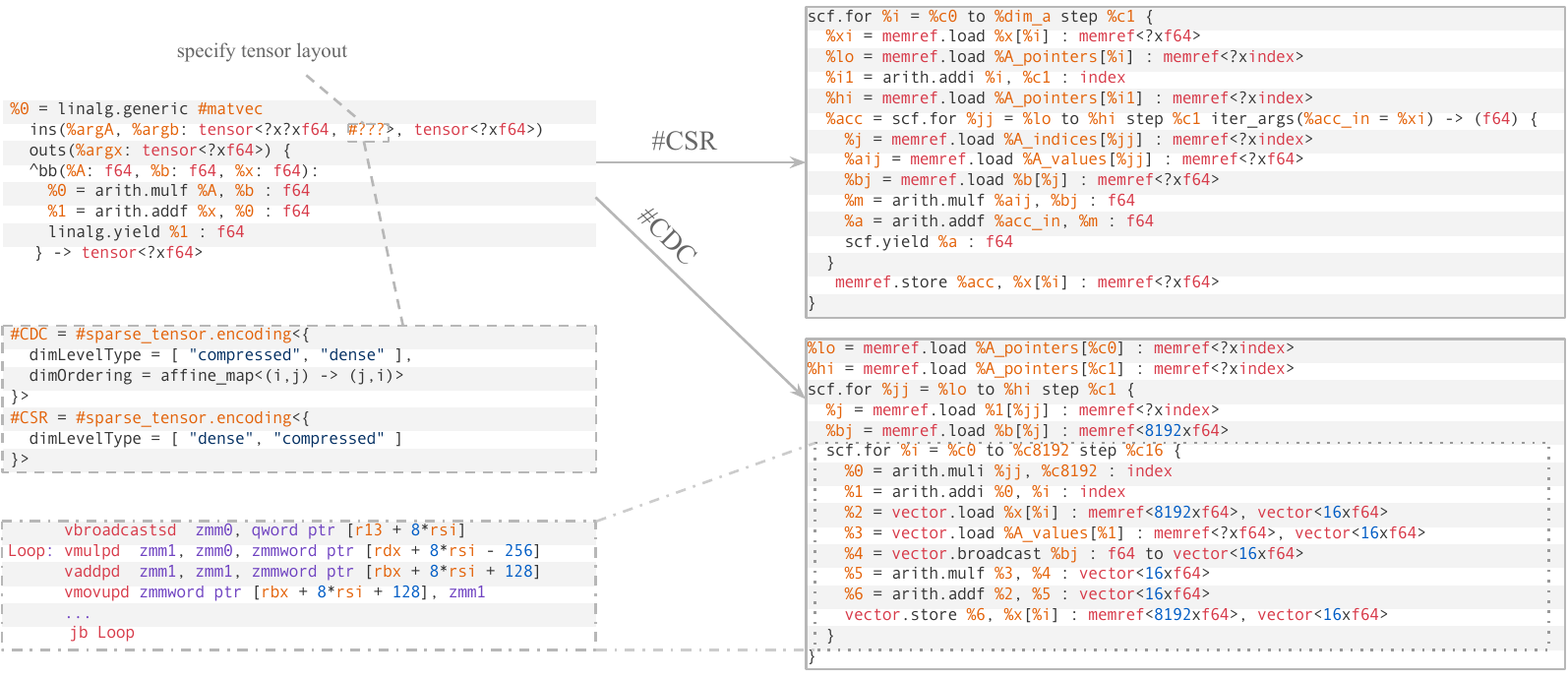}
    \caption{Example: lowering of matrix-vector multiplication with sparse input matrix.}
    \label{fig:sparse_lowering}
\end{figure}

Figure~\ref{fig:sparse_layouts} illustrates how to use the the well-known Doubly Compressed Sparse Column (DCSC) storage format for sparse matrices in a matrix multiplication. A sparse tensor is specified by simply placing the encoding attribute inside the tensor type. This way of defining sparse tensor formats is quite powerful. 
For a single $d$-dimensional tensor, the dimension level formats and ordering alone give rise to $2^d \cdot d!$ different formats. 
This allows annotating a sparse kernel in a such a way that eventually results in the generation of effective sparse code. Even the output tensor could be made sparse to save memory when storing the result.

A set of \lstinline{sparse_tensor} dialect rewrite rules take care of lowering the kernel to sparse storage schemes and imperative constructs that only store and iterate over the nonzero elements to perform the matrix multiplication. Such an approach to automatic sparse code generation was first proposed in~\cite{bik96,biktms} in the context of sparse linear algebra, and later generalized to sparse tensor algebra in~\cite{kjolstad2017taco}. Even though the details of these rewrite rules are outside the scope of this paper, they follow a similar high-level modular and composable philosophy as discussed in Section~\ref{sec:transformations}, with variations in the transformations due to the use of sparse index sets and slightly more complicated bufferization due to the compound nature of sparse storage schemes. The rewrite rules similarly interoperate with \lstinline|linalg|, \lstinline|tensor|, \lstinline|memref|, \lstinline|scf|, and \lstinline|vector| abstractions, thereby progressive lowering a sparsity agnostic definition of a kernel into a form that fully exploits the sparsity of tensors as well as all performance features of the target architecture.

Consider, for example, a matrix times vector computation \texttt{x = Ab} for a general sparse matrix \texttt{A} (Figure~\ref{fig:sparse_lowering}). When \texttt{A} is in Compressed Sparse Row (CSR) format, nested \lstinline|scf| loops are used to iterate over the outer dense dimension, and, by means of an indirection, just the nonzero elements of the compressed inner dimension. 

\begin{table}[h!tb]
  \small
  \centering
  \begin{tabularx}{\textwidth}{X|rr|rr}
    \toprule
    \textbf{Size} & TACO random    & MLIR random   & TACO column        & MLIR column \\
    \midrule
    \lstinline|16,384 x 16,384| &  3.32 &  2.93 &  1.96 &  1.58 \\
    \lstinline|32,768 x 32,768| & 12.71 & 12.59 &  8.12 &  6.90 \\
    \lstinline|65,536 x 65,536| & 51.49 & 50.48 & 32.21 & 30.48 \\
    \bottomrule
  \end{tabularx}
  \smallskip
  \caption{Sparse kernel runtimes (in ms) on an Intel Xeon W2135 $3.7\mathrm{GHz}$.}
  \label{tbl:sparse-matvec}
\end{table}

Now suppose that the sparse matrix \texttt{A} is actually a structured sparse matrix, where most columns are empty, but when nonzero, the columns are dense, i.e.\ mostly filled. In that case, \texttt{CDC} makes more sense, favoring column-wise access over the default row-wise storage, and using compression for the outermost dimension (now columns) only. Specializing for this format for a sparse $8192 \times 8192$ matrix and a vector length of 16, the sparse rewriting rules compose well with vectorization to eventually yield sparse vector code that skips over empty columns, but performs the innermost dense update using vectors. After more progressive lowering within MLIR, eventually an LLVM IR representation is handed off to the backend compiler which, when targeting AVX512, generates an innermost loop that performs the vector updates in full SIMD.

Table~\ref{tbl:sparse-matvec} shows the runtime (in ms) on an Intel Xeon W2135 $3.7\mathrm{GHz}$, comparing sparse computations generated by both TACO and MLIR, evaluated on sparse matrices with a random uniform density of about 1\% (random) and matrices with 1\% of the columns filled (column). For TACO, we used the default settings of the built-in benchmark, but excluded compilation time for a fair comparison. The table clearly shows that the sparse compiler component of MLIR is on-par with a state-of-art compiler like TACO, with additional benefits when the transformation of MLIR's sparse compiler are composed with other optimizations, such as vectorization.

\subsection{Preliminary Autotuning Results}
\label{subsec:prelim-tuning-results}

Our collaborators at Nod.AI have been using our modular and composable codegen strategy and tuning it for specific use cases. 
They have built an autotuner using CompilerGym~\cite{CompilerGym}, a compiler tuning environment based on the OpenAI Gym~\cite{OpenAIGym}.
They report fully-automatic preliminary results on recent Alder Lake CPUs~\cite{NodeAi-tuner} on which they outperform both Intel's MKL library and the TVM autotuner on a variety of matrix multiplication sizes relevant to the machine learning community.
When integrated within the IREE~\cite{iree} compiler-based runtime system, and further tuned for a BERT~\cite{Bert} model on NVIDIA A100 GPUs, they report the fastest end-to-end PyTorch-based~\cite{PyTorch} inference numbers~\cite{NodeAi-runtime}.

\section{Related Work}

We chose to focus this related work survey on the compilers that influenced the design of structured and retargetable code generation. Readers interested in general IR design may refer to \cite{lattner2021mlir}.

\subsection{Lessons from ONNX}
ONNX (\url{https://onnx.ai}) is an open format for the interoperability of ML models and workloads.
As such, it is predominantly driven by the expressiveness requirements of ML, and less by the  considerations of IR design for high-performance code generation.

Similarly to ONNX, we define ``semantically charged'', ``named'' operations matching the needs of popular ML models.
But we do not stop here: we also consider \emph{transformations on these operations} as a key part of the design, and we define supporting IR abstractions \emph{favoring transformations over expressiveness where necessary}.
We also minimize the range of ``named'' operations by making them simple declarative configurations of a small set of \lstinline|generic| operations. Compared to a typical numerical library interface (the de facto standard until now), this greatly reduces maintenance burden and increases portability.
Finally, we provide multiple levels of composable abstractions, with operations on tensors, vectors, memrefs and transformations thereof, not limiting ourselves to a framework interoperability interface.

\subsection{Lessons from XLA}
XLA~\cite{XLA} is one of the first post-Theano ML compilers, introduced as a domain-specific optimizing compiler for TensorFlow. It shines on Google's Cloud TPU hardware. XLA followed a pragmatic design process where the compiler is given perfect knowledge of the semantics of each operation. The set of operations known to the XLA system is called HLOs. 
XLA code generation consists of composing C++ functions known as emitters. 
This approach has two major benefits:
(1) transformations are correct by construction
(2) strong performance on specialized hardware such as TPUs.
On the other hand, due to its approach relying on C++ function composition, extending XLA can involve significant amounts of code duplication. XLA also involves its own set of intermediate representations, optimization passes, domain-specific abstractions, resulting in the replication of similar features in contemporary frameworks.

XLA inspired how operations ``know their semantics'' and ``how to transform and lower themselves''. 
Yet the encoding of this information and its use in transformations differ:
\begin{enumerate}
\item Individual HLOs have parameterized yet special-purpose semantics, operating on a single level of abstraction (tensors), while lower level data representations come with their own set of operations. As a consequence, HLOs have evolved into a large set of operations whose semantics intersect. This echoes the operation proliferation problem also exhibited by ONNX.
\item Over-reliance on perfect op knowledge leads to situations where transformations and ops end up needing to know about each other's semantics (e.g. during fusion). Since the transformations themselves are not simple local rewrite patterns code complexity grows quickly.
\item XLA lacks a serializable IR that can be inspected, unit-tested and composed with independent compiler flows. 
This monolithic design impacts portability. 
For instance, the TPU and GPU compilers do not share much code.
\end{enumerate}

\subsection{Lessons from Halide, TVM and related languages}
Halide~\cite{ragan2013halide} is a DSL embedded in C++ that provides a way of metaprogramming an algorithmic specification in HalideIR, applying transformations declaratively. This allows expert users to transform and optimize a high level program in tailored ways. Halide, initially targeted the graphics community but is now more generally applicable.
TVM~\cite{chen2018tvm} is an evolution of Halide into the machine learning and deep-neural network space, based on HalideIR.

Halide schedules borrow from the decoupling principle of the URUK~\cite{URUK} and CHiLL~\cite{CHiLL} polyhedral transformation frameworks. To a large extent, schedules are declarative, unlike LIFT, URUK and CHiLL. It is more restricted than polyhedral schedules, capturing rectangular shapes and limited forms of fusion and fine-grain scheduling. This avoids many of the complexities of polyhedral scheduling and code generation, yet Halide schedules remain powerful enough for its application domain, and explores a large swath of possible transformations.  Halide shines at making a loop nest optimization methodology accessible to $\Omega(10-100)$ users, at a time when polyhedral tools are still only accessible to $\Omega(1-10)$ users. It makes heavy usage of canonicalization rules\,\footnote{\url{https://sunfishcode.github.io/blog/2018/10/22/Canonicalization.html}.} that are also very prevalent in LLVM and MLIR.

While these features are shared with structured and retargetable codegen, Halide diverges from our approach in a number of ways.
Counter-current to our lowering rather than raising principle, Halide's front-end uses explicit indexing and arithmetic on scalar values, which makes it difficult to target BLAS/DNN libraries, including hardware instructions such as GPU tensor cores. TVM provides a \lstinline{tensorize} transformation for that purpose, but it remains ad-hoc to use or rely on pattern matching and raising abstractions from lower level indexing and iteration. Along the same lines, reductions and scans are expressed using serial iteration, again requiring pattern matching before they can be transformed into more efficient, target-specific parallel versions (recursion tree, atomic operations, etc.). Finally, Halide only operates on its own IR and abstractions, while we provide a composable set of abstractions, and genericity over a range of data representations.

Fireiron~\cite{Fireiron}, recent work in the Halide family of languages, shares with our work the focus on decomposing an operation --- originally restricted to matrix multiplication --- into smaller kernels. It also enables the substitution of any such resulting kernel with a manually written microkernel implementation or hardware instructions of arbitrary granularity. In addition, Fireiron provides a bottom-up approach to performance modeling and extends Halide schedules with explicit data transfers; both developments will be influential in the future evolutions of our approach.

\subsection{Lessons from LIFT and Combinator Languages with Rewriting Rules}
LIFT~\cite{LIFT} is a system to write and optimized computational kernels based on functional abstractions. 
Transformations are represented by means of local rewrites to the IR, including the insertion of combinators such as map, reduce, zip... as well as decorating these with distribution, parallelization, vectorization and storage mapping information.

Similarly to LIFT, we make heavy use of local rewrite rules through the MLIR \lstinline|PatternRewrite| infrastructure.
There are important differences however:
\begin{enumerate}
    \item Our transformations are co-designed with the IR in a structural decomposition approach, while LIFT rewrites generic nested combinators.
    \item The unit of transformation and tuning is a generic $n$-D operation. Compared to LIFT, we believe this makes the optimization space more structured and tractable for search algorithms, and this will be the subject of another study.
    \item We make operations generic over the representation of tensor values. This includes vector-level primitives and side-effecting operations.
    \item Lowering to nested loops over \lstinline{vector}s and \lstinline{memref}s is more versatile than decorating combinators on tensor values, while ensuring transformation correctness (including lowering steps) by design of the structural decomposition approach.
\end{enumerate}

LIFT is expected to further influence our design in the future, leveraging experience with LIFT abstractions for sparse and position-dependent arrays.

Related work on performance portability include the Multi-Dimensional Homomorphisms (MDH) approach~\cite{Rasch2019GeneratingPH}. Emphasizing the algebraic nature of parallelization and distribution rules over tensor operations, MDH leverages a specialized code generator and facilitates autotuning.

Elevate~\cite{Elevate} addresses some of the expressiveness shortcomings of LIFT while establishing solid foundations for decoupled schedules. The proposed formalization applies to schedules driving the application of rewrite rules, which is different from Halide's code-generation-centered API.

In the same family of languages, Glenside~\cite{Glenside} introduces ``access patterns'' as representation unifying pure tensor operations and storage mapping considerations such as layouts and hardware-centered memory management. Access patterns are amenable to automatic synthesis and lowering to coarse-grained hardware instructions.

\subsection{Lessons from Tensor Comprehensions}
Tensor Comprehensions~\cite{vasilache2019next} is a high-level language to express tensor computations with a syntax generalizing the Einstein notation, coupled to an end-to-end compilation flow capable of lowering to efficient GPU code. It was integrated with 2 ML frameworks: Caffe2 (the successor to the popular Caffe~\cite{Caffe} framework), as well as 
PyTorch~\cite{PyTorch}.

The compilation flow combines Halide with a Polyhedral Compiler derived from isl~\cite{isl} and uses both HalideIR and the isl schedule-tree IR. The compiler provides a collection of polyhedral compilation algorithms to perform fusion and favor multi-level parallelism and promotion to deeper levels of the memory hierarchy. Tensor Comprehensions showed that, building on a affine framework expressive enough to cover the main compositions of tiling, fusion and parallelization transformations required to reach competitive performance, it is sufficient to expose a few knobs over predefined affine scheduling strategies to provide strong performance---in the bandwidth and latency bound regimes.
In particular, a simple genetic search combined with an autotuning framework and affine scheduling strategies reached high performance on sub-graphs or even full models in the non-compute bound regime.

However, Tensor Comprehensions lacks the IR to reason about lower level rewrites, related to vectorization, register reuse, unrolling, necessary to perform well in the \emph{compute-bound regime}. 
In particular, it lacks an SSA representation combining higher level abstractions with vector and register-level ones. 
It also relies on parameterized affine scheduling strategies, relying on integer linear programming, which remain difficult to steer towards peak performance~\cite{Vas12,Zin18}.
Most of these issues are naturally addressed in our \MLIR code generation flow, including vector abstractions and the ability to implement a wide variety of transformation and lowering strategies.

\subsection{Lessons from Polyhedral compilers}
The polyhedral model has been on the cutting edge of loop nest optimization for decades, with several incarnations in production compilers such as GRAPHITE~\cite{GRAPHITE} for GCC and Polly~\cite{Polly} for LLVM. 
Although it has proven capable of competing with the best libraries and domain-specific frameworks, in incarnations like PolyMage~\cite{PolyMage} 
and Tensor Comprehensions~\cite{vasilache2019next}, it has never been fully adopted into mainstream optimization flows.
Some reasons include:
\begin{enumerate}
    \item The IR gets more complex than affine schedules, when representing multiple levels of tiling and parallelism, data transfers, unrolling, etc., and requires complex abstractions such as isl schedule trees \cite{isl}.
    \item Scheduling and code generation from general affine representations rely on exponential algorithms \cite{Fea92a,Fea92b,Bas04,Pluto,RStream,Gro15,isl}.
    \item Affine representations are not composable with the SSA form,
    upon which most optimizing compilers are built today; this induces pass ordering conflicts with induction variable canonicalization, loop-invariant code motion, vectorization, etc.\ \cite{DeLICM}.
    \item Expressiveness limitations remain in polyhedral compilers, despite progresses towards interprocedural analysis, dynamic control flow, irregular indexing \cite{PIPS,PENCIL,Zha18}.
\end{enumerate}

The affine dialect in MLIR addresses these long-standing issues. 
In particular, it maintains the IR in the same form --- loops with additional constraints on how the bounds are expressed ---
throughout the compilation flow.
It also embeds the polyhedral representation into the SSA form through attributes and regions, facilitating the combination of
polyhedral and SSA-based transformations.
Yet some of the above-mentioned issues remain, including the algorithmic and software engineering challenges of building and tuning competitive optimization strategies by means of affine transformations. Structured operations avoid these problems by operating on a higher level of abstraction, involving tensor-operation-specific optimizations and lowering strategies instead.

\subsection{Comparison With Low-Level Code Generators}

On the back-end compiler side, an alternative approach consists in bypassing the \LLVM level completely, relying on non-portable assembly code generation instead~\cite{heinecke2016libxsmm}.
While the performance one can achieve with perfectly accurate models of the hardware is impressive, such solutions jump over multiple levels of abstraction, forbidding any reuse of low-level compiler passes and abstractions.
While such low-level bricks can be the ultimate target when decomposing structured operations, the reliance on non-portable, hand-tuned generators and proprietary tooling stands a bridge too far.
Instead, we are more interested in leveraging program synthesis and superoptimization at the lower levels of abstraction~\cite{SwizzleInventor,VGen}.

\section{Conclusion and Future Work}

We presented the composable multi-level intermediate representation and transformations that underpin tensor code generation in \MLIR.
This work heavily leverages \MLIR's progressivity design principle, which it helped to shape by designing and implementing multiple instances of progressive lowering.
The approach is characterized by a \emph{transformation-oriented IR design}: doing away with legality analyses and applicability checks on low-level IR, relying systematically on the progressive decomposition of carefully designed abstractions.
The resulting design is modular and built with optionality in mind; abstractions span data structures and control flow with both functional (SSA form) and imperative (side-effecting) semantics; they serve as generic building blocks for retargetable tensor compilers.
Transformations are systematically applied as compositions of declarative patterns.
This allows implementing advanced forms of pass fusion, known to alleviate the dreaded phase-ordering problem.
Early single threaded CPU code generation shows strong performance, even in the absence of systematic tuning.

Multiple extensions and generalizations of this work are actively being pursued:
\begin{itemize}
    \item Functionally inspired abstractions for parallel semantics on immutable tensors will extend our results to multi-threaded CPUs and beyond.
    \item Generalization of the approach to other hardware targets thanks to our retargetable vector abstractions.
    \item More powerful vector abstractions including generalized masking and warp-level GPU programming.
    \item More advanced data structures and iterators than dense or sparse to enable cross-pollination with other fields such as databases.
    \item Autotuning infrastructure to systematically find transformation compositions and parameters that produce code running at a high fraction of peak, for any hardware.
    \item Controllable framework for selecting and composing transformation patterns, making production compilers effectively extensible with domain-specific code generation strategies.
\end{itemize}

We believe these contributions will further demystify the field of high-performance compilation,
give a high-level of control to users of the system and provide a strong basis on which future research and near-peak performance production systems will be built.

\bibliographystyle{ACM-Reference-Format}
\bibliography{bibliography}


\begin{thebibliography}{60}

%%% the Web address.

\ifx \showCODEN    \undefined \def \showCODEN     #1{\unskip}     \fi
\ifx \showDOI      \undefined \def \showDOI       #1{#1}\fi
\ifx \showISBNx    \undefined \def \showISBNx     #1{\unskip}     \fi
\ifx \showISBNxiii \undefined \def \showISBNxiii  #1{\unskip}     \fi
\ifx \showISSN     \undefined \def \showISSN      #1{\unskip}     \fi
\ifx \showLCCN     \undefined \def \showLCCN      #1{\unskip}     \fi
\ifx \shownote     \undefined \def \shownote      #1{#1}          \fi
\ifx \showarticletitle \undefined \def \showarticletitle #1{#1}   \fi
\ifx \showURL      \undefined \def \showURL       {\relax}        \fi
% invisible to TeX
\providecommand\bibfield[2]{#2}
\providecommand\bibinfo[2]{#2}
\providecommand\natexlab[1]{#1}
\providecommand\showeprint[2][]{arXiv:#2}

\bibitem[\protect\citeauthoryear{Allen and Kennedy}{Allen and Kennedy}{2001}]%
        {AllenKennedy}
\bibfield{author}{\bibinfo{person}{R. Allen} {and} \bibinfo{person}{K.
  Kennedy}.} \bibinfo{year}{2001}\natexlab{}.
\newblock \bibinfo{booktitle}{\emph{Optimizing Compilers for Modern
  Architectures: A Dependence-Based Approach}}.
\newblock \bibinfo{publisher}{Morgan Kaufmann Publishers}.
\newblock
\showISBNx{9781493303540}
\urldef\tempurl%
\url{https://books.google.ch/books?id=X1QfogEACAAJ}
\showURL{%
\tempurl}


\bibitem[\protect\citeauthoryear{Baghdadi, Beaugnon, Cohen, Grosser, Kruse,
  Reddy, Verdoolaege, Betts, Donaldson, Ketema, Absar, van Haastregt, Kravets,
  Lokhmotov, David, and Hajiyev}{Baghdadi et~al\mbox{.}}{2015}]%
        {PENCIL}
\bibfield{author}{\bibinfo{person}{Riyadh Baghdadi}, \bibinfo{person}{Ulysse
  Beaugnon}, \bibinfo{person}{Albert Cohen}, \bibinfo{person}{Tobias Grosser},
  \bibinfo{person}{Michael Kruse}, \bibinfo{person}{Chandan Reddy},
  \bibinfo{person}{Sven Verdoolaege}, \bibinfo{person}{Adam Betts},
  \bibinfo{person}{Alastair~F. Donaldson}, \bibinfo{person}{Jeroen Ketema},
  \bibinfo{person}{Javed Absar}, \bibinfo{person}{Sven van Haastregt},
  \bibinfo{person}{Alexey Kravets}, \bibinfo{person}{Anton Lokhmotov},
  \bibinfo{person}{Robert David}, {and} \bibinfo{person}{Elnar Hajiyev}.}
  \bibinfo{year}{2015}\natexlab{}.
\newblock \showarticletitle{{PENCIL:} {A} Platform-Neutral Compute Intermediate
  Language for Accelerator Programming}. In \bibinfo{booktitle}{\emph{2015
  International Conference on Parallel Architectures and Compilation, {PACT}
  2015, San Francisco, CA, USA, October 18-21, 2015}}.
  \bibinfo{publisher}{{IEEE} Computer Society}, \bibinfo{pages}{138--149}.
\newblock
\urldef\tempurl%
\url{https://doi.org/10.1109/PACT.2015.17}
\showDOI{\tempurl}


\bibitem[\protect\citeauthoryear{Barham and Isard}{Barham and Isard}{2019}]%
        {ml_rut}
\bibfield{author}{\bibinfo{person}{Paul Barham} {and} \bibinfo{person}{Michael
  Isard}.} \bibinfo{year}{2019}\natexlab{}.
\newblock \showarticletitle{Machine Learning Systems Are Stuck in a Rut}. In
  \bibinfo{booktitle}{\emph{Proceedings of the Workshop on Hot Topics in
  Operating Systems}} (Bertinoro, Italy) \emph{(\bibinfo{series}{HotOS '19})}.
  \bibinfo{publisher}{Association for Computing Machinery},
  \bibinfo{pages}{177–183}.
\newblock
\showISBNx{9781450367271}
\urldef\tempurl%
\url{https://doi.org/10.1145/3317550.3321441}
\showDOI{\tempurl}


\bibitem[\protect\citeauthoryear{Bastoul}{Bastoul}{2004}]%
        {Bas04}
\bibfield{author}{\bibinfo{person}{C{\'{e}}dric Bastoul}.}
  \bibinfo{year}{2004}\natexlab{}.
\newblock \showarticletitle{Code Generation in the Polyhedral Model Is Easier
  Than You Think}. In \bibinfo{booktitle}{\emph{13th International Conference
  on Parallel Architectures and Compilation Techniques {(PACT} 2004), 29
  September - 3 October 2004, Antibes Juan-les-Pins, France}}.
  \bibinfo{publisher}{{IEEE} Computer Society}, \bibinfo{pages}{7--16}.
\newblock
\urldef\tempurl%
\url{https://doi.org/10.1109/PACT.2004.10018}
\showDOI{\tempurl}


\bibitem[\protect\citeauthoryear{Bik}{Bik}{1996}]%
        {bik96}
\bibfield{author}{\bibinfo{person}{Aart~J.C. Bik}.}
  \bibinfo{year}{1996}\natexlab{}.
\newblock \emph{\bibinfo{title}{Compiler Support for Sparse Matrix
  Computations}}.
\newblock \bibinfo{thesistype}{Ph.\,D. Dissertation}.
  \bibinfo{school}{Department of Computer Science, Leiden University}.
\newblock
\newblock
\shownote{ISBN 90-9009442-3}.


\bibitem[\protect\citeauthoryear{Bik, Brinkhaus, Knijnenburg, and Wijshoff}{Bik
  et~al\mbox{.}}{1998}]%
        {biktms}
\bibfield{author}{\bibinfo{person}{Aart~J.C. Bik}, \bibinfo{person}{Peter~J.H.
  Brinkhaus}, \bibinfo{person}{Peter~M.W. Knijnenburg}, {and}
  \bibinfo{person}{Harry~A.G. Wijshoff}.} \bibinfo{year}{1998}\natexlab{}.
\newblock \showarticletitle{The Automatic Generation of Sparse Primitives}.
\newblock \bibinfo{journal}{\emph{Transactions on Mathematical Software}}
  \bibinfo{volume}{24} (\bibinfo{year}{1998}), \bibinfo{pages}{190--225}.
\newblock


\bibitem[\protect\citeauthoryear{Bondhugula, Hartono, Ramanujam, and
  Sadayappan}{Bondhugula et~al\mbox{.}}{2008}]%
        {Pluto}
\bibfield{author}{\bibinfo{person}{Uday Bondhugula}, \bibinfo{person}{Albert
  Hartono}, \bibinfo{person}{J. Ramanujam}, {and} \bibinfo{person}{P.
  Sadayappan}.} \bibinfo{year}{2008}\natexlab{}.
\newblock \showarticletitle{A practical automatic polyhedral parallelizer and
  locality optimizer}. In \bibinfo{booktitle}{\emph{Proceedings of the {ACM}
  {SIGPLAN} 2008 Conference on Programming Language Design and Implementation,
  Tucson, AZ, USA, June 7-13, 2008}}, \bibfield{editor}{\bibinfo{person}{Rajiv
  Gupta} {and} \bibinfo{person}{Saman~P. Amarasinghe}} (Eds.).
  \bibinfo{publisher}{{ACM}}, \bibinfo{pages}{101--113}.
\newblock
\urldef\tempurl%
\url{https://doi.org/10.1145/1375581.1375595}
\showDOI{\tempurl}


\bibitem[\protect\citeauthoryear{Brockman, Cheung, Pettersson, Schneider,
  Schulman, Tang, and Zaremba}{Brockman et~al\mbox{.}}{2016}]%
        {OpenAIGym}
\bibfield{author}{\bibinfo{person}{Greg Brockman}, \bibinfo{person}{Vicki
  Cheung}, \bibinfo{person}{Ludwig Pettersson}, \bibinfo{person}{Jonas
  Schneider}, \bibinfo{person}{John Schulman}, \bibinfo{person}{Jie Tang},
  {and} \bibinfo{person}{Wojciech Zaremba}.} \bibinfo{year}{2016}\natexlab{}.
\newblock \showarticletitle{OpenAI Gym}.
\newblock \bibinfo{journal}{\emph{CoRR}}  \bibinfo{volume}{abs/1606.01540}
  (\bibinfo{year}{2016}).
\newblock
\showeprint[arXiv]{1606.01540}
\urldef\tempurl%
\url{http://arxiv.org/abs/1606.01540}
\showURL{%
\tempurl}


\bibitem[\protect\citeauthoryear{Chen, Moreau, Jiang, Zheng, Yan, Shen, Cowan,
  Wang, Hu, Ceze, et~al\mbox{.}}{Chen et~al\mbox{.}}{2018}]%
        {chen2018tvm}
\bibfield{author}{\bibinfo{person}{Tianqi Chen}, \bibinfo{person}{Thierry
  Moreau}, \bibinfo{person}{Ziheng Jiang}, \bibinfo{person}{Lianmin Zheng},
  \bibinfo{person}{Eddie Yan}, \bibinfo{person}{Haichen Shen},
  \bibinfo{person}{Meghan Cowan}, \bibinfo{person}{Leyuan Wang},
  \bibinfo{person}{Yuwei Hu}, \bibinfo{person}{Luis Ceze}, {et~al\mbox{.}}}
  \bibinfo{year}{2018}\natexlab{}.
\newblock \showarticletitle{{TVM}: An automated end-to-end optimizing compiler
  for deep learning}. In \bibinfo{booktitle}{\emph{13th {USENIX} Symposium on
  Operating Systems Design and Implementation ({OSDI} 18)}}.
  \bibinfo{publisher}{ACM}, \bibinfo{pages}{578--594}.
\newblock


\bibitem[\protect\citeauthoryear{Chen, Mendis, Carbin, and Amarasinghe}{Chen
  et~al\mbox{.}}{2021}]%
        {VGen}
\bibfield{author}{\bibinfo{person}{Yishen Chen}, \bibinfo{person}{Charith
  Mendis}, \bibinfo{person}{Michael Carbin}, {and} \bibinfo{person}{Saman~P.
  Amarasinghe}.} \bibinfo{year}{2021}\natexlab{}.
\newblock \showarticletitle{VeGen: a vectorizer generator for {SIMD} and
  beyond}. In \bibinfo{booktitle}{\emph{{ASPLOS} '21: 26th {ACM} International
  Conference on Architectural Support for Programming Languages and Operating
  Systems, Virtual Event, USA, April 19-23, 2021}},
  \bibfield{editor}{\bibinfo{person}{Tim Sherwood}, \bibinfo{person}{Emery~D.
  Berger}, {and} \bibinfo{person}{Christos Kozyrakis}} (Eds.).
  \bibinfo{publisher}{{ACM}}, \bibinfo{pages}{902--914}.
\newblock
\urldef\tempurl%
\url{https://doi.org/10.1145/3445814.3446692}
\showDOI{\tempurl}


\bibitem[\protect\citeauthoryear{Click and Cooper}{Click and Cooper}{1995}]%
        {Click95}
\bibfield{author}{\bibinfo{person}{Cliff Click} {and} \bibinfo{person}{Keith~D.
  Cooper}.} \bibinfo{year}{1995}\natexlab{}.
\newblock \showarticletitle{Combining Analyses, Combining Optimizations}.
\newblock \bibinfo{journal}{\emph{{ACM} Trans. Program. Lang. Syst.}}
  \bibinfo{volume}{17}, \bibinfo{number}{2} (\bibinfo{year}{1995}),
  \bibinfo{pages}{181--196}.
\newblock


\bibitem[\protect\citeauthoryear{Corp.}{Corp.}{2021}]%
        {IntelOptimizationManual}
\bibfield{author}{\bibinfo{person}{Intel Corp.}}
  \bibinfo{year}{2021}\natexlab{}.
\newblock \bibinfo{title}{Intel Optimization Reference Manual}.
\newblock
  \bibinfo{howpublished}{\url{https://www.intel.com/content/www/us/en/develop/download/intel-64-and-ia-32-architectures-optimization-reference-manual.html}}.
\newblock


\bibitem[\protect\citeauthoryear{Cummins, Wasti, Guo, Cui, Ansel, Gomez, Jain,
  Liu, Teytaud, Steiner, Tian, and Leather}{Cummins et~al\mbox{.}}{2021}]%
        {CompilerGym}
\bibfield{author}{\bibinfo{person}{Chris Cummins}, \bibinfo{person}{Bram
  Wasti}, \bibinfo{person}{Jiadong Guo}, \bibinfo{person}{Brandon Cui},
  \bibinfo{person}{Jason Ansel}, \bibinfo{person}{Sahir Gomez},
  \bibinfo{person}{Somya Jain}, \bibinfo{person}{Jia Liu},
  \bibinfo{person}{Olivier Teytaud}, \bibinfo{person}{Benoit Steiner},
  \bibinfo{person}{Yuandong Tian}, {and} \bibinfo{person}{Hugh Leather}.}
  \bibinfo{year}{2021}\natexlab{}.
\newblock \showarticletitle{CompilerGym: Robust, Performant Compiler
  Optimization Environments for {AI} Research}.
\newblock \bibinfo{journal}{\emph{CoRR}}  \bibinfo{volume}{abs/2109.08267}
  (\bibinfo{year}{2021}).
\newblock
\showeprint[arXiv]{2109.08267}
\urldef\tempurl%
\url{https://arxiv.org/abs/2109.08267}
\showURL{%
\tempurl}


\bibitem[\protect\citeauthoryear{Developers}{Developers}{2021}]%
        {iree}
\bibfield{author}{\bibinfo{person}{{IREE} Developers}.}
  \bibinfo{year}{2021}\natexlab{}.
\newblock \bibinfo{title}{{IREE} (Intermediate Representation Execution
  Environment}.
\newblock
\newblock
\urldef\tempurl%
\url{https://google.github.io/iree/}
\showURL{%
\tempurl}


\bibitem[\protect\citeauthoryear{Devlin, Chang, Lee, and Toutanova}{Devlin
  et~al\mbox{.}}{2019}]%
        {Bert}
\bibfield{author}{\bibinfo{person}{Jacob Devlin}, \bibinfo{person}{Ming-Wei
  Chang}, \bibinfo{person}{Kenton Lee}, {and} \bibinfo{person}{Kristina
  Toutanova}.} \bibinfo{year}{2019}\natexlab{}.
\newblock \bibinfo{title}{BERT: Pre-training of Deep Bidirectional Transformers
  for Language Understanding}.
\newblock
\newblock
\showeprint[arxiv]{1810.04805}~[cs.CL]


\bibitem[\protect\citeauthoryear{Documentation}{Documentation}{2021}]%
        {LLVM_benchmarking}
\bibfield{author}{\bibinfo{person}{LLVM Documentation}.}
  \bibinfo{year}{2021}\natexlab{}.
\newblock \bibinfo{title}{Benchmarking tips}.
\newblock
  \bibinfo{howpublished}{\url{https://llvm.org/docs/Benchmarking.html}}.
\newblock


\bibitem[\protect\citeauthoryear{Feautrier}{Feautrier}{1992a}]%
        {Fea92a}
\bibfield{author}{\bibinfo{person}{Paul Feautrier}.}
  \bibinfo{year}{1992}\natexlab{a}.
\newblock \showarticletitle{Some efficient solutions to the affine scheduling
  problem. I. One-dimensional time}.
\newblock \bibinfo{journal}{\emph{Int. J. Parallel Program.}}
  \bibinfo{volume}{21}, \bibinfo{number}{5} (\bibinfo{year}{1992}),
  \bibinfo{pages}{313--347}.
\newblock
\urldef\tempurl%
\url{https://doi.org/10.1007/BF01407835}
\showDOI{\tempurl}


\bibitem[\protect\citeauthoryear{Feautrier}{Feautrier}{1992b}]%
        {Fea92b}
\bibfield{author}{\bibinfo{person}{Paul Feautrier}.}
  \bibinfo{year}{1992}\natexlab{b}.
\newblock \showarticletitle{Some efficient solutions to the affine scheduling
  problem. Part {II.} Multidimensional time}.
\newblock \bibinfo{journal}{\emph{Int. J. Parallel Program.}}
  \bibinfo{volume}{21}, \bibinfo{number}{6} (\bibinfo{year}{1992}),
  \bibinfo{pages}{389--420}.
\newblock
\urldef\tempurl%
\url{https://doi.org/10.1007/BF01379404}
\showDOI{\tempurl}


\bibitem[\protect\citeauthoryear{Girbal, Vasilache, Bastoul, Cohen, Parello,
  Sigler, and Temam}{Girbal et~al\mbox{.}}{2006}]%
        {URUK}
\bibfield{author}{\bibinfo{person}{Sylvain Girbal}, \bibinfo{person}{Nicolas
  Vasilache}, \bibinfo{person}{C{\'{e}}dric Bastoul}, \bibinfo{person}{Albert
  Cohen}, \bibinfo{person}{David Parello}, \bibinfo{person}{Marc Sigler}, {and}
  \bibinfo{person}{Olivier Temam}.} \bibinfo{year}{2006}\natexlab{}.
\newblock \showarticletitle{Semi-Automatic Composition of Loop Transformations
  for Deep Parallelism and Memory Hierarchies}.
\newblock \bibinfo{journal}{\emph{Int. J. Parallel Program.}}
  \bibinfo{volume}{34}, \bibinfo{number}{3} (\bibinfo{year}{2006}),
  \bibinfo{pages}{261--317}.
\newblock
\urldef\tempurl%
\url{https://doi.org/10.1007/s10766-006-0012-3}
\showDOI{\tempurl}


\bibitem[\protect\citeauthoryear{Grosser, Gr{\"{o}}{\ss}linger, and
  Lengauer}{Grosser et~al\mbox{.}}{2012}]%
        {Polly}
\bibfield{author}{\bibinfo{person}{Tobias Grosser}, \bibinfo{person}{Armin
  Gr{\"{o}}{\ss}linger}, {and} \bibinfo{person}{Christian Lengauer}.}
  \bibinfo{year}{2012}\natexlab{}.
\newblock \showarticletitle{Polly - Performing Polyhedral Optimizations on a
  Low-Level Intermediate Representation}.
\newblock \bibinfo{journal}{\emph{Parallel Process. Lett.}}
  \bibinfo{volume}{22}, \bibinfo{number}{4} (\bibinfo{year}{2012}).
\newblock
\urldef\tempurl%
\url{https://doi.org/10.1142/S0129626412500107}
\showDOI{\tempurl}


\bibitem[\protect\citeauthoryear{Grosser, Verdoolaege, and Cohen}{Grosser
  et~al\mbox{.}}{2015}]%
        {Gro15}
\bibfield{author}{\bibinfo{person}{Tobias Grosser}, \bibinfo{person}{Sven
  Verdoolaege}, {and} \bibinfo{person}{Albert Cohen}.}
  \bibinfo{year}{2015}\natexlab{}.
\newblock \showarticletitle{Polyhedral {AST} Generation Is More Than Scanning
  Polyhedra}.
\newblock \bibinfo{journal}{\emph{{ACM} Trans. Program. Lang. Syst.}}
  \bibinfo{volume}{37}, \bibinfo{number}{4} (\bibinfo{year}{2015}),
  \bibinfo{pages}{12:1--12:50}.
\newblock
\urldef\tempurl%
\url{https://doi.org/10.1145/2743016}
\showDOI{\tempurl}


\bibitem[\protect\citeauthoryear{Guennebaud, Jacob, et~al\mbox{.}}{Guennebaud
  et~al\mbox{.}}{2010}]%
        {eigen}
\bibfield{author}{\bibinfo{person}{Ga\"{e}l Guennebaud},
  \bibinfo{person}{Beno\^{i}t Jacob}, {et~al\mbox{.}}}
  \bibinfo{year}{2010}\natexlab{}.
\newblock \bibinfo{title}{Eigen v3}.
\newblock \bibinfo{howpublished}{http://eigen.tuxfamily.org}.
\newblock


\bibitem[\protect\citeauthoryear{Hagedorn, Elliott, Barthels, Bod{\'{\i}}k, and
  Grover}{Hagedorn et~al\mbox{.}}{2020a}]%
        {Fireiron}
\bibfield{author}{\bibinfo{person}{Bastian Hagedorn},
  \bibinfo{person}{Archibald~Samuel Elliott}, \bibinfo{person}{Henrik
  Barthels}, \bibinfo{person}{Rastislav Bod{\'{\i}}k}, {and}
  \bibinfo{person}{Vinod Grover}.} \bibinfo{year}{2020}\natexlab{a}.
\newblock \showarticletitle{Fireiron: {A} Scheduling Language for
  High-Performance Linear Algebra on GPUs}.
\newblock \bibinfo{journal}{\emph{CoRR}}  \bibinfo{volume}{abs/2003.06324}
  (\bibinfo{year}{2020}).
\newblock
\showeprint[arXiv]{2003.06324}
\urldef\tempurl%
\url{https://arxiv.org/abs/2003.06324}
\showURL{%
\tempurl}


\bibitem[\protect\citeauthoryear{Hagedorn, Lenfers, Koehler, Gorlatch, and
  Steuwer}{Hagedorn et~al\mbox{.}}{2020b}]%
        {Elevate}
\bibfield{author}{\bibinfo{person}{Bastian Hagedorn}, \bibinfo{person}{Johannes
  Lenfers}, \bibinfo{person}{Thomas Koehler}, \bibinfo{person}{Sergei
  Gorlatch}, {and} \bibinfo{person}{Michel Steuwer}.}
  \bibinfo{year}{2020}\natexlab{b}.
\newblock \showarticletitle{A Language for Describing Optimization Strategies}.
\newblock \bibinfo{journal}{\emph{CoRR}}  \bibinfo{volume}{abs/2002.02268}
  (\bibinfo{year}{2020}).
\newblock
\showeprint[arXiv]{2002.02268}
\urldef\tempurl%
\url{https://arxiv.org/abs/2002.02268}
\showURL{%
\tempurl}


\bibitem[\protect\citeauthoryear{Hagedorn, Lenfers, Kundefinedhler, Qin,
  Gorlatch, and Steuwer}{Hagedorn et~al\mbox{.}}{2020c}]%
        {rise}
\bibfield{author}{\bibinfo{person}{Bastian Hagedorn}, \bibinfo{person}{Johannes
  Lenfers}, \bibinfo{person}{Thomas Kundefinedhler}, \bibinfo{person}{Xueying
  Qin}, \bibinfo{person}{Sergei Gorlatch}, {and} \bibinfo{person}{Michel
  Steuwer}.} \bibinfo{year}{2020}\natexlab{c}.
\newblock \showarticletitle{Achieving High-Performance the Functional Way: A
  Functional Pearl on Expressing High-Performance Optimizations as Rewrite
  Strategies}.
\newblock \bibinfo{journal}{\emph{Proceedings of ACM on Programming Languages}}
  \bibinfo{volume}{4}, \bibinfo{number}{ICFP}, Article \bibinfo{articleno}{92}
  (\bibinfo{date}{Aug.} \bibinfo{year}{2020}), \bibinfo{numpages}{29}~pages.
\newblock
\urldef\tempurl%
\url{https://doi.org/10.1145/3408974}
\showDOI{\tempurl}


\bibitem[\protect\citeauthoryear{Heinecke, Henry, Hutchinson, and
  Pabst}{Heinecke et~al\mbox{.}}{2016}]%
        {heinecke2016libxsmm}
\bibfield{author}{\bibinfo{person}{Alexander Heinecke}, \bibinfo{person}{Greg
  Henry}, \bibinfo{person}{Maxwell Hutchinson}, {and} \bibinfo{person}{Hans
  Pabst}.} \bibinfo{year}{2016}\natexlab{}.
\newblock \showarticletitle{LIBXSMM: accelerating small matrix multiplications
  by runtime code generation}. In \bibinfo{booktitle}{\emph{SC'16: Proceedings
  of the International Conference for High Performance Computing, Networking,
  Storage and Analysis}}. \bibinfo{publisher}{IEEE/ACM},
  \bibinfo{pages}{981--991}.
\newblock


\bibitem[\protect\citeauthoryear{Hoefler and Belli}{Hoefler and Belli}{2015}]%
        {benchmark2015}
\bibfield{author}{\bibinfo{person}{Torsten Hoefler} {and}
  \bibinfo{person}{Roberto Belli}.} \bibinfo{year}{2015}\natexlab{}.
\newblock \showarticletitle{Scientific Benchmarking of Parallel Computing
  Systems: Twelve Ways to Tell the Masses When Reporting Performance Results}.
  In \bibinfo{booktitle}{\emph{SC'15: Proceedings of the International
  Conference for High Performance Computing, Networking, Storage and Analysis}}
  (Austin, Texas). \bibinfo{publisher}{IEEE/ACM}, Article
  \bibinfo{articleno}{73}, \bibinfo{numpages}{12}~pages.
\newblock
\showISBNx{9781450337236}
\urldef\tempurl%
\url{https://doi.org/10.1145/2807591.2807644}
\showDOI{\tempurl}


\bibitem[\protect\citeauthoryear{Irigoin, Jouvelot, and Triolet}{Irigoin
  et~al\mbox{.}}{1991}]%
        {PIPS}
\bibfield{author}{\bibinfo{person}{Fran{\c{c}}ois Irigoin},
  \bibinfo{person}{Pierre Jouvelot}, {and} \bibinfo{person}{R{\'{e}}mi
  Triolet}.} \bibinfo{year}{1991}\natexlab{}.
\newblock \showarticletitle{Semantical interprocedural parallelization: an
  overview of the {PIPS} project}. In \bibinfo{booktitle}{\emph{Proceedings of
  the 5th international conference on Supercomputing, {ICS} 1991, Cologne,
  Germany, June 17-21, 1991}}, \bibfield{editor}{\bibinfo{person}{Edward~S.
  Davidson} {and} \bibinfo{person}{Friedel Hossfeld}} (Eds.).
  \bibinfo{publisher}{{ACM}}, \bibinfo{pages}{244--251}.
\newblock
\urldef\tempurl%
\url{https://doi.org/10.1145/109025.109086}
\showDOI{\tempurl}


\bibitem[\protect\citeauthoryear{Jia, Shelhamer, Donahue, Karayev, Long,
  Girshick, Guadarrama, and Darrell}{Jia et~al\mbox{.}}{2014}]%
        {Caffe}
\bibfield{author}{\bibinfo{person}{Yangqing Jia}, \bibinfo{person}{Evan
  Shelhamer}, \bibinfo{person}{Jeff Donahue}, \bibinfo{person}{Sergey Karayev},
  \bibinfo{person}{Jonathan Long}, \bibinfo{person}{Ross Girshick},
  \bibinfo{person}{Sergio Guadarrama}, {and} \bibinfo{person}{Trevor Darrell}.}
  \bibinfo{year}{2014}\natexlab{}.
\newblock \showarticletitle{Caffe: Convolutional Architecture for Fast Feature
  Embedding}. In \bibinfo{booktitle}{\emph{Proceedings of the 22Nd ACM
  International Conference on Multimedia}} (Orlando, Florida, USA)
  \emph{(\bibinfo{series}{MM '14})}. \bibinfo{publisher}{ACM},
  \bibinfo{pages}{675--678}.
\newblock
\showISBNx{978-1-4503-3063-3}
\urldef\tempurl%
\url{https://doi.org/10.1145/2647868.2654889}
\showDOI{\tempurl}


\bibitem[\protect\citeauthoryear{Jia, Padon, Thomas, Warszawski, Zaharia, and
  Aiken}{Jia et~al\mbox{.}}{2019}]%
        {TASO}
\bibfield{author}{\bibinfo{person}{Zhihao Jia}, \bibinfo{person}{Oded Padon},
  \bibinfo{person}{James Thomas}, \bibinfo{person}{Todd Warszawski},
  \bibinfo{person}{Matei Zaharia}, {and} \bibinfo{person}{Alex Aiken}.}
  \bibinfo{year}{2019}\natexlab{}.
\newblock \showarticletitle{TASO: Optimizing Deep Learning Computation with
  Automatic Generation of Graph Substitutions}. In
  \bibinfo{booktitle}{\emph{Proceedings of the 27th ACM Symposium on Operating
  Systems Principles}} (Huntsville, Ontario, Canada)
  \emph{(\bibinfo{series}{SOSP '19})}. \bibinfo{publisher}{Association for
  Computing Machinery}, \bibinfo{pages}{47–62}.
\newblock
\showISBNx{9781450368735}
\urldef\tempurl%
\url{https://doi.org/10.1145/3341301.3359630}
\showURL{%
\tempurl}


\bibitem[\protect\citeauthoryear{Khan, Basu, Rudy, Hall, Chen, and Chame}{Khan
  et~al\mbox{.}}{2013}]%
        {CHiLL}
\bibfield{author}{\bibinfo{person}{Malik~Murtaza Khan},
  \bibinfo{person}{Protonu Basu}, \bibinfo{person}{Gabe Rudy},
  \bibinfo{person}{Mary~W. Hall}, \bibinfo{person}{Chun Chen}, {and}
  \bibinfo{person}{Jacqueline Chame}.} \bibinfo{year}{2013}\natexlab{}.
\newblock \showarticletitle{A script-based autotuning compiler system to
  generate high-performance {CUDA} code}.
\newblock \bibinfo{journal}{\emph{{ACM} Trans. Archit. Code Optim.}}
  \bibinfo{volume}{9}, \bibinfo{number}{4} (\bibinfo{year}{2013}),
  \bibinfo{pages}{31:1--31:25}.
\newblock
\urldef\tempurl%
\url{https://doi.org/10.1145/2400682.2400690}
\showDOI{\tempurl}


\bibitem[\protect\citeauthoryear{Kjolstad, Kamil, Chou, Lugato, and
  Amarasinghe}{Kjolstad et~al\mbox{.}}{2017}]%
        {kjolstad2017taco}
\bibfield{author}{\bibinfo{person}{Fredrik Kjolstad}, \bibinfo{person}{Shoaib
  Kamil}, \bibinfo{person}{Stephen Chou}, \bibinfo{person}{David Lugato}, {and}
  \bibinfo{person}{Saman Amarasinghe}.} \bibinfo{year}{2017}\natexlab{}.
\newblock \showarticletitle{The Tensor Algebra Compiler}.
\newblock \bibinfo{journal}{\emph{Proc. ACM Program. Lang.}}
  \bibinfo{volume}{1}, \bibinfo{number}{OOPSLA}, Article
  \bibinfo{articleno}{77} (\bibinfo{date}{Oct.} \bibinfo{year}{2017}),
  \bibinfo{numpages}{29}~pages.
\newblock
\urldef\tempurl%
\url{https://doi.org/10.1145/3133901}
\showDOI{\tempurl}


\bibitem[\protect\citeauthoryear{Kruse and Grosser}{Kruse and Grosser}{2018}]%
        {DeLICM}
\bibfield{author}{\bibinfo{person}{Michael Kruse} {and} \bibinfo{person}{Tobias
  Grosser}.} \bibinfo{year}{2018}\natexlab{}.
\newblock \showarticletitle{DeLICM: scalar dependence removal at zero memory
  cost}. In \bibinfo{booktitle}{\emph{Proceedings of the 2018 International
  Symposium on Code Generation and Optimization, {CGO} 2018, V{\"{o}}sendorf /
  Vienna, Austria, February 24-28, 2018}},
  \bibfield{editor}{\bibinfo{person}{Jens Knoop}, \bibinfo{person}{Markus
  Schordan}, \bibinfo{person}{Teresa Johnson}, {and} \bibinfo{person}{Michael
  F.~P. O'Boyle}} (Eds.). \bibinfo{publisher}{{ACM}},
  \bibinfo{pages}{241--253}.
\newblock
\urldef\tempurl%
\url{https://doi.org/10.1145/3168815}
\showDOI{\tempurl}


\bibitem[\protect\citeauthoryear{Lattner, Amini, Bondhugula, Cohen, Davis,
  Pienaar, Riddle, Shpeisman, Vasilache, and Zinenko}{Lattner
  et~al\mbox{.}}{2021}]%
        {lattner2021mlir}
\bibfield{author}{\bibinfo{person}{Chris Lattner}, \bibinfo{person}{Mehdi
  Amini}, \bibinfo{person}{Uday Bondhugula}, \bibinfo{person}{Albert Cohen},
  \bibinfo{person}{Andy Davis}, \bibinfo{person}{Jacques Pienaar},
  \bibinfo{person}{River Riddle}, \bibinfo{person}{Tatiana Shpeisman},
  \bibinfo{person}{Nicolas Vasilache}, {and} \bibinfo{person}{Oleksandr
  Zinenko}.} \bibinfo{year}{2021}\natexlab{}.
\newblock \showarticletitle{Mlir: Scaling compiler infrastructure for domain
  specific computation}. In \bibinfo{booktitle}{\emph{2021 IEEE/ACM
  International Symposium on Code Generation and Optimization (CGO)}}.
  IEEE/ACM, \bibinfo{publisher}{IEEE/ACM}, \bibinfo{pages}{2--14}.
\newblock


\bibitem[\protect\citeauthoryear{Meister, Vasilache, Wohlford, Baskaran, Leung,
  and Lethin}{Meister et~al\mbox{.}}{2011}]%
        {RStream}
\bibfield{author}{\bibinfo{person}{Beno{\^{\i}}t Meister},
  \bibinfo{person}{Nicolas Vasilache}, \bibinfo{person}{David Wohlford},
  \bibinfo{person}{Muthu~Manikandan Baskaran}, \bibinfo{person}{Allen Leung},
  {and} \bibinfo{person}{Richard Lethin}.} \bibinfo{year}{2011}\natexlab{}.
\newblock \showarticletitle{R-Stream Compiler}.
\newblock In \bibinfo{booktitle}{\emph{Encyclopedia of Parallel Computing}},
  \bibfield{editor}{\bibinfo{person}{David~A. Padua}} (Ed.).
  \bibinfo{publisher}{Springer}, \bibinfo{pages}{1756--1765}.
\newblock
\urldef\tempurl%
\url{https://doi.org/10.1007/978-0-387-09766-4\_515}
\showDOI{\tempurl}


\bibitem[\protect\citeauthoryear{Mullapudi, Vasista, and Bondhugula}{Mullapudi
  et~al\mbox{.}}{2015}]%
        {PolyMage}
\bibfield{author}{\bibinfo{person}{Ravi~Teja Mullapudi}, \bibinfo{person}{Vinay
  Vasista}, {and} \bibinfo{person}{Uday Bondhugula}.}
  \bibinfo{year}{2015}\natexlab{}.
\newblock \showarticletitle{PolyMage: Automatic Optimization for Image
  Processing Pipelines}. In \bibinfo{booktitle}{\emph{Proceedings of the
  Twentieth International Conference on Architectural Support for Programming
  Languages and Operating Systems, {ASPLOS} 2015, Istanbul, Turkey, March
  14-18, 2015}}, \bibfield{editor}{\bibinfo{person}{{\"{O}}zcan {\"{O}}zturk},
  \bibinfo{person}{Kemal Ebcioglu}, {and} \bibinfo{person}{Sandhya Dwarkadas}}
  (Eds.). \bibinfo{publisher}{{ACM}}, \bibinfo{pages}{429--443}.
\newblock
\urldef\tempurl%
\url{https://doi.org/10.1145/2694344.2694364}
\showDOI{\tempurl}


\bibitem[\protect\citeauthoryear{Nod.AI}{Nod.AI}{2022a}]%
        {NodeAi-tuner}
\bibfield{author}{\bibinfo{person}{Nod.AI}.} \bibinfo{year}{2022}\natexlab{a}.
\newblock \bibinfo{title}{Outperforming Intel’s MKL and OctoML/Apache TVM
  with MLIR and Nod.ai’s Codegen Search}.
\newblock
  \bibinfo{howpublished}{\url{https://nod.ai/outperforming-octoml-apache-tvm-and-intel-mkl/}}.
\newblock


\bibitem[\protect\citeauthoryear{Nod.AI}{Nod.AI}{2022b}]%
        {NodeAi-runtime}
\bibfield{author}{\bibinfo{person}{Nod.AI}.} \bibinfo{year}{2022}\natexlab{b}.
\newblock \bibinfo{title}{SHARK: The fastest PyTorch runtime – 3x over
  Torchscript, 1.6x over TF/XLA, 23\% faster than ONNXRuntime}.
\newblock
  \bibinfo{howpublished}{\url{https://nod.ai/shark-the-fastest-runtime/}}.
\newblock


\bibitem[\protect\citeauthoryear{Paszke, Gross, Massa, Lerer, Bradbury, Chanan,
  Killeen, Lin, Gimelshein, Antiga, Desmaison, Kopf, Yang, DeVito, Raison,
  Tejani, Chilamkurthy, Steiner, Fang, Bai, and Chintala}{Paszke
  et~al\mbox{.}}{2019}]%
        {PyTorch}
\bibfield{author}{\bibinfo{person}{Adam Paszke}, \bibinfo{person}{Sam Gross},
  \bibinfo{person}{Francisco Massa}, \bibinfo{person}{Adam Lerer},
  \bibinfo{person}{James Bradbury}, \bibinfo{person}{Gregory Chanan},
  \bibinfo{person}{Trevor Killeen}, \bibinfo{person}{Zeming Lin},
  \bibinfo{person}{Natalia Gimelshein}, \bibinfo{person}{Luca Antiga},
  \bibinfo{person}{Alban Desmaison}, \bibinfo{person}{Andreas Kopf},
  \bibinfo{person}{Edward Yang}, \bibinfo{person}{Zachary DeVito},
  \bibinfo{person}{Martin Raison}, \bibinfo{person}{Alykhan Tejani},
  \bibinfo{person}{Sasank Chilamkurthy}, \bibinfo{person}{Benoit Steiner},
  \bibinfo{person}{Lu Fang}, \bibinfo{person}{Junjie Bai}, {and}
  \bibinfo{person}{Soumith Chintala}.} \bibinfo{year}{2019}\natexlab{}.
\newblock \showarticletitle{PyTorch: An Imperative Style, High-Performance Deep
  Learning Library}.
\newblock In \bibinfo{booktitle}{\emph{Advances in Neural Information
  Processing Systems 32}}, \bibfield{editor}{\bibinfo{person}{H.~Wallach},
  \bibinfo{person}{H.~Larochelle}, \bibinfo{person}{A.~Beygelzimer},
  \bibinfo{person}{F.~d\textquotesingle Alch\'{e}-Buc},
  \bibinfo{person}{E.~Fox}, {and} \bibinfo{person}{R.~Garnett}} (Eds.).
  \bibinfo{publisher}{Curran Associates, Inc.}, \bibinfo{pages}{8024--8035}.
\newblock
\urldef\tempurl%
\url{http://papers.neurips.cc/paper/9015-pytorch-an-imperative-style-high-performance-deep-learning-library.pdf}
\showURL{%
\tempurl}


\bibitem[\protect\citeauthoryear{Phothilimthana, Elliott, Wang, Jangda,
  Hagedorn, Barthels, Kaufman, Grover, Torlak, and Bod{\'{\i}}k}{Phothilimthana
  et~al\mbox{.}}{2019}]%
        {SwizzleInventor}
\bibfield{author}{\bibinfo{person}{Phitchaya~Mangpo Phothilimthana},
  \bibinfo{person}{Archibald~Samuel Elliott}, \bibinfo{person}{An Wang},
  \bibinfo{person}{Abhinav Jangda}, \bibinfo{person}{Bastian Hagedorn},
  \bibinfo{person}{Henrik Barthels}, \bibinfo{person}{Samuel~J. Kaufman},
  \bibinfo{person}{Vinod Grover}, \bibinfo{person}{Emina Torlak}, {and}
  \bibinfo{person}{Rastislav Bod{\'{\i}}k}.} \bibinfo{year}{2019}\natexlab{}.
\newblock \showarticletitle{Swizzle Inventor: Data Movement Synthesis for {GPU}
  Kernels}. In \bibinfo{booktitle}{\emph{Proceedings of the Twenty-Fourth
  International Conference on Architectural Support for Programming Languages
  and Operating Systems, {ASPLOS} 2019, Providence, RI, USA, April 13-17,
  2019}}, \bibfield{editor}{\bibinfo{person}{Iris Bahar},
  \bibinfo{person}{Maurice Herlihy}, \bibinfo{person}{Emmett Witchel}, {and}
  \bibinfo{person}{Alvin~R. Lebeck}} (Eds.). \bibinfo{publisher}{{ACM}},
  \bibinfo{pages}{65--78}.
\newblock
\urldef\tempurl%
\url{https://doi.org/10.1145/3297858.3304059}
\showDOI{\tempurl}


\bibitem[\protect\citeauthoryear{Ragan-Kelley, Barnes, Adams, Paris, Durand,
  and Amarasinghe}{Ragan-Kelley et~al\mbox{.}}{2013}]%
        {ragan2013halide}
\bibfield{author}{\bibinfo{person}{Jonathan Ragan-Kelley},
  \bibinfo{person}{Connelly Barnes}, \bibinfo{person}{Andrew Adams},
  \bibinfo{person}{Sylvain Paris}, \bibinfo{person}{Fr{\'e}do Durand}, {and}
  \bibinfo{person}{Saman Amarasinghe}.} \bibinfo{year}{2013}\natexlab{}.
\newblock \showarticletitle{Halide: a language and compiler for optimizing
  parallelism, locality, and recomputation in image processing pipelines}.
\newblock \bibinfo{journal}{\emph{Acm Sigplan Notices}} \bibinfo{volume}{48},
  \bibinfo{number}{6} (\bibinfo{year}{2013}), \bibinfo{pages}{519--530}.
\newblock


\bibitem[\protect\citeauthoryear{Rasch, Schulze, and Gorlatch}{Rasch
  et~al\mbox{.}}{2019}]%
        {Rasch2019GeneratingPH}
\bibfield{author}{\bibinfo{person}{Ari Rasch}, \bibinfo{person}{Richard
  Schulze}, {and} \bibinfo{person}{Sergei Gorlatch}.}
  \bibinfo{year}{2019}\natexlab{}.
\newblock \showarticletitle{Generating Portable High-Performance Code via
  Multi-Dimensional Homomorphisms}. In \bibinfo{booktitle}{\emph{2019 28th
  International Conference on Parallel Architectures and Compilation Techniques
  (PACT)}}. \bibinfo{publisher}{IEEE}, \bibinfo{pages}{354--369}.
\newblock


\bibitem[\protect\citeauthoryear{Riddle}{Riddle}{2021}]%
        {PDLL}
\bibfield{author}{\bibinfo{person}{River Riddle}.}
  \bibinfo{year}{2021}\natexlab{}.
\newblock \bibinfo{title}{PDLL: a new declarative rewrite frontend for MLIR}.
\newblock
  \bibinfo{howpublished}{\url{https://llvm.discourse.group/t/rfc-pdll-a-new-declarative-rewrite-frontend-for-mlir/4798}}.
\newblock


\bibitem[\protect\citeauthoryear{Rossini}{Rossini}{2021}]%
        {OuterproductSpilling}
\bibfield{author}{\bibinfo{person}{Giuseppe Rossini}.}
  \bibinfo{year}{2021}\natexlab{}.
\newblock \bibinfo{title}{Outerproduct spilling}.
\newblock
  \bibinfo{howpublished}{\url{https://llvm.discourse.group/t/outer-product-spilling/4405}}.
\newblock


\bibitem[\protect\citeauthoryear{Rotem, Fix, Abdulrasool, Deng, Dzhabarov,
  Hegeman, Levenstein, Maher, Satish, Olesen, Park, Rakhov, and
  Smelyanskiy}{Rotem et~al\mbox{.}}{2018}]%
        {rotem2018glow}
\bibfield{author}{\bibinfo{person}{Nadav Rotem}, \bibinfo{person}{Jordan Fix},
  \bibinfo{person}{Saleem Abdulrasool}, \bibinfo{person}{Summer Deng},
  \bibinfo{person}{Roman Dzhabarov}, \bibinfo{person}{James Hegeman},
  \bibinfo{person}{Roman Levenstein}, \bibinfo{person}{Bert Maher},
  \bibinfo{person}{Nadathur Satish}, \bibinfo{person}{Jakob Olesen},
  \bibinfo{person}{Jongsoo Park}, \bibinfo{person}{Artem Rakhov}, {and}
  \bibinfo{person}{Misha Smelyanskiy}.} \bibinfo{year}{2018}\natexlab{}.
\newblock \showarticletitle{Glow: Graph Lowering Compiler Techniques for Neural
  Networks}.
\newblock \bibinfo{journal}{\emph{CoRR}}  \bibinfo{volume}{abs/1805.00907}
  (\bibinfo{year}{2018}).
\newblock
\showeprint[arXiv]{1805.00907}
\urldef\tempurl%
\url{http://arxiv.org/abs/1805.00907}
\showURL{%
\tempurl}


\bibitem[\protect\citeauthoryear{Sandler, Howard, Zhu, Zhmoginov, and
  Chen}{Sandler et~al\mbox{.}}{2018}]%
        {MobileNetV2}
\bibfield{author}{\bibinfo{person}{Mark Sandler}, \bibinfo{person}{Andrew~G.
  Howard}, \bibinfo{person}{Menglong Zhu}, \bibinfo{person}{Andrey Zhmoginov},
  {and} \bibinfo{person}{Liang{-}Chieh Chen}.} \bibinfo{year}{2018}\natexlab{}.
\newblock \showarticletitle{Inverted Residuals and Linear Bottlenecks: Mobile
  Networks for Classification, Detection and Segmentation}.
\newblock \bibinfo{journal}{\emph{CoRR}}  \bibinfo{volume}{abs/1801.04381}
  (\bibinfo{year}{2018}).
\newblock
\showeprint[arXiv]{1801.04381}
\urldef\tempurl%
\url{http://arxiv.org/abs/1801.04381}
\showURL{%
\tempurl}


\bibitem[\protect\citeauthoryear{Schrijver}{Schrijver}{1986}]%
        {schrijver86}
\bibfield{author}{\bibinfo{person}{Alexander Schrijver}.}
  \bibinfo{year}{1986}\natexlab{}.
\newblock \bibinfo{booktitle}{\emph{Theory of Linear and Integer Programming}}.
\newblock \bibinfo{publisher}{John Wiley \& Sons}.
\newblock


\bibitem[\protect\citeauthoryear{Smith, Liu, Lyubomirsky, Davidson, McMahan,
  Taylor, Ceze, and Tatlock}{Smith et~al\mbox{.}}{2021}]%
        {Glenside}
\bibfield{author}{\bibinfo{person}{Gus~Henry Smith}, \bibinfo{person}{Andrew
  Liu}, \bibinfo{person}{Steven Lyubomirsky}, \bibinfo{person}{Scott Davidson},
  \bibinfo{person}{Joseph McMahan}, \bibinfo{person}{Michael~B. Taylor},
  \bibinfo{person}{Luis Ceze}, {and} \bibinfo{person}{Zachary Tatlock}.}
  \bibinfo{year}{2021}\natexlab{}.
\newblock \showarticletitle{Pure Tensor Program Rewriting via Access Patterns
  (Representation Pearl)}.
\newblock \bibinfo{journal}{\emph{CoRR}}  \bibinfo{volume}{abs/2105.09377}
  (\bibinfo{year}{2021}).
\newblock
\showeprint[arXiv]{2105.09377}
\urldef\tempurl%
\url{https://arxiv.org/abs/2105.09377}
\showURL{%
\tempurl}


\bibitem[\protect\citeauthoryear{Steuwer, Remmelg, and Dubach}{Steuwer
  et~al\mbox{.}}{2017}]%
        {LIFT}
\bibfield{author}{\bibinfo{person}{Michel Steuwer}, \bibinfo{person}{Toomas
  Remmelg}, {and} \bibinfo{person}{Christophe Dubach}.}
  \bibinfo{year}{2017}\natexlab{}.
\newblock \showarticletitle{LIFT: A functional data-parallel IR for
  high-performance GPU code generation}. In \bibinfo{booktitle}{\emph{2017
  IEEE/ACM International Symposium on Code Generation and Optimization (CGO)}}.
  \bibinfo{publisher}{IEEE/ACM}, \bibinfo{pages}{74--85}.
\newblock
\urldef\tempurl%
\url{https://doi.org/10.1109/CGO.2017.7863730}
\showDOI{\tempurl}


\bibitem[\protect\citeauthoryear{team~within Google}{team~within
  Google}{2017}]%
        {XLA}
\bibfield{author}{\bibinfo{person}{XLA team~within Google}.}
  \bibinfo{year}{2017}\natexlab{}.
\newblock \bibinfo{title}{XLA: TensorFlow, Compiled.}
\newblock \bibinfo{howpublished}{Google Developers Blog}.
\newblock
\urldef\tempurl%
\url{https://developers.googleblog.com/2017/03/xla-tensorflow-compiled.html}
\showURL{%
\tempurl}


\bibitem[\protect\citeauthoryear{Tian, Saito, Preis, Garcia, Kozhukhov, Masten,
  Cherkasov, and Panchenko}{Tian et~al\mbox{.}}{2013}]%
        {IPDPS2013}
\bibfield{author}{\bibinfo{person}{Xinmin Tian}, \bibinfo{person}{Hideki
  Saito}, \bibinfo{person}{Serguei~V. Preis}, \bibinfo{person}{Eric~N. Garcia},
  \bibinfo{person}{Sergey~S. Kozhukhov}, \bibinfo{person}{Matt Masten},
  \bibinfo{person}{Aleksei~G. Cherkasov}, {and} \bibinfo{person}{Nikolay
  Panchenko}.} \bibinfo{year}{2013}\natexlab{}.
\newblock \showarticletitle{Practical SIMD Vectorization Techniques for Intel®
  Xeon Phi Coprocessors}. In \bibinfo{booktitle}{\emph{2013 IEEE International
  Symposium on Parallel Distributed Processing, Workshops and Phd Forum}}.
  \bibinfo{publisher}{IEEE}, \bibinfo{pages}{1149--1158}.
\newblock
\urldef\tempurl%
\url{https://doi.org/10.1109/IPDPSW.2013.245}
\showDOI{\tempurl}


\bibitem[\protect\citeauthoryear{Trifunovic, Cohen, Edelsohn, Li, Grosser,
  Jagasia, Ladelsky, Pop, Sj{\"o}din, and Upadrasta}{Trifunovic
  et~al\mbox{.}}{2010}]%
        {GRAPHITE}
\bibfield{author}{\bibinfo{person}{Konrad Trifunovic}, \bibinfo{person}{Albert
  Cohen}, \bibinfo{person}{David Edelsohn}, \bibinfo{person}{Feng Li},
  \bibinfo{person}{Tobias Grosser}, \bibinfo{person}{Harsha Jagasia},
  \bibinfo{person}{Razya Ladelsky}, \bibinfo{person}{Sebastian Pop},
  \bibinfo{person}{Jan Sj{\"o}din}, {and} \bibinfo{person}{Ramakrishna
  Upadrasta}.} \bibinfo{year}{2010}\natexlab{}.
\newblock \showarticletitle{{GRAPHITE Two Years After: First Lessons Learned
  From Real-World Polyhedral Compilation}}. In \bibinfo{booktitle}{\emph{{GCC
  Research Opportunities Workshop (GROW'10)}}}. \bibinfo{publisher}{HAL}.
\newblock
\urldef\tempurl%
\url{https://hal.inria.fr/inria-00551516}
\showURL{%
\tempurl}


\bibitem[\protect\citeauthoryear{Vasilache}{Vasilache}{2021}]%
        {VBlendPsLimitation}
\bibfield{author}{\bibinfo{person}{Nicolas Vasilache}.}
  \bibinfo{year}{2021}\natexlab{}.
\newblock \bibinfo{title}{Understanding and controlling some of the AVX shuffle
  emission paths}.
\newblock
  \bibinfo{howpublished}{\url{https://groups.google.com/g/llvm-dev/c/P_CXf5hPro8/m/RI6Fm3bzAQAJ}}.
\newblock


\bibitem[\protect\citeauthoryear{Vasilache, Meister, Baskaran, and
  Lethin}{Vasilache et~al\mbox{.}}{2012}]%
        {Vas12}
\bibfield{author}{\bibinfo{person}{Nicolas Vasilache}, \bibinfo{person}{Benoit
  Meister}, \bibinfo{person}{Muthu Baskaran}, {and} \bibinfo{person}{Richard
  Lethin}.} \bibinfo{year}{2012}\natexlab{}.
\newblock \showarticletitle{Joint scheduling and layout optimization to enable
  multi-level vectorization}. In \bibinfo{booktitle}{\emph{In Second Intl.\
  Workshop on Polyhedral Compilation Techniques (IMPACT)}}.
  \bibinfo{publisher}{Informal proceedings}.
\newblock


\bibitem[\protect\citeauthoryear{Vasilache, Zinenko, Theodoridis, Goyal,
  Devito, Moses, Verdoolaege, Adams, and Cohen}{Vasilache
  et~al\mbox{.}}{2019}]%
        {vasilache2019next}
\bibfield{author}{\bibinfo{person}{Nicolas Vasilache},
  \bibinfo{person}{Oleksandr Zinenko}, \bibinfo{person}{Theodoros Theodoridis},
  \bibinfo{person}{Priya Goyal}, \bibinfo{person}{Zachary Devito},
  \bibinfo{person}{William~S Moses}, \bibinfo{person}{Sven Verdoolaege},
  \bibinfo{person}{Andrew Adams}, {and} \bibinfo{person}{Albert Cohen}.}
  \bibinfo{year}{2019}\natexlab{}.
\newblock \showarticletitle{The next 700 accelerated layers: From mathematical
  expressions of network computation graphs to accelerated GPU kernels,
  automatically}.
\newblock \bibinfo{journal}{\emph{ACM Transactions on Architecture and Code
  Optimization (TACO)}} \bibinfo{volume}{16}, \bibinfo{number}{4}
  (\bibinfo{year}{2019}), \bibinfo{pages}{1--26}.
\newblock


\bibitem[\protect\citeauthoryear{Verdoolaege}{Verdoolaege}{2010}]%
        {isl}
\bibfield{author}{\bibinfo{person}{Sven Verdoolaege}.}
  \bibinfo{year}{2010}\natexlab{}.
\newblock \showarticletitle{\emph{isl}: An Integer Set Library for the
  Polyhedral Model}. In \bibinfo{booktitle}{\emph{Mathematical Software --
  {ICMS} 2010, Third International Congress on Mathematical Software}}
  \emph{(\bibinfo{series}{Lecture Notes in Computer Science},
  Vol.~\bibinfo{volume}{6327})}, \bibfield{editor}{\bibinfo{person}{Komei
  Fukuda}, \bibinfo{person}{Joris van~der Hoeven}, \bibinfo{person}{Michael
  Joswig}, {and} \bibinfo{person}{Nobuki Takayama}} (Eds.).
  \bibinfo{publisher}{Springer}, \bibinfo{pages}{299--302}.
\newblock
\urldef\tempurl%
\url{https://doi.org/10.1007/978-3-642-15582-6\_49}
\showDOI{\tempurl}


\bibitem[\protect\citeauthoryear{Williams, Waterman, and Patterson}{Williams
  et~al\mbox{.}}{2009}]%
        {roofline}
\bibfield{author}{\bibinfo{person}{Samuel Williams}, \bibinfo{person}{Andrew
  Waterman}, {and} \bibinfo{person}{David~A. Patterson}.}
  \bibinfo{year}{2009}\natexlab{}.
\newblock \showarticletitle{Roofline: an insightful visual performance model
  for multicore architectures}.
\newblock \bibinfo{journal}{\emph{Commun. {ACM}}} \bibinfo{volume}{52},
  \bibinfo{number}{4} (\bibinfo{year}{2009}), \bibinfo{pages}{65--76}.
\newblock
\urldef\tempurl%
\url{https://doi.org/10.1145/1498765.1498785}
\showDOI{\tempurl}


\bibitem[\protect\citeauthoryear{Zhao, Kruse, and Cohen}{Zhao
  et~al\mbox{.}}{2018}]%
        {Zha18}
\bibfield{author}{\bibinfo{person}{Jie Zhao}, \bibinfo{person}{Michael Kruse},
  {and} \bibinfo{person}{Albert Cohen}.} \bibinfo{year}{2018}\natexlab{}.
\newblock \showarticletitle{A polyhedral compilation framework for loops with
  dynamic data-dependent bounds}. In \bibinfo{booktitle}{\emph{Proceedings of
  the 27th International Conference on Compiler Construction, {CC} 2018,
  February 24-25, 2018, Vienna, Austria}},
  \bibfield{editor}{\bibinfo{person}{Christophe Dubach} {and}
  \bibinfo{person}{Jingling Xue}} (Eds.). \bibinfo{publisher}{{ACM}},
  \bibinfo{pages}{14--24}.
\newblock
\urldef\tempurl%
\url{https://doi.org/10.1145/3178372.3179509}
\showDOI{\tempurl}


\bibitem[\protect\citeauthoryear{Zheng, Jia, Sun, Wu, Yu, Haj{-}Ali, Wang,
  Yang, Zhuo, Sen, Gonzalez, and Stoica}{Zheng et~al\mbox{.}}{2020}]%
        {zheng2020ansor}
\bibfield{author}{\bibinfo{person}{Lianmin Zheng}, \bibinfo{person}{Chengfan
  Jia}, \bibinfo{person}{Minmin Sun}, \bibinfo{person}{Zhao Wu},
  \bibinfo{person}{Cody~Hao Yu}, \bibinfo{person}{Ameer Haj{-}Ali},
  \bibinfo{person}{Yida Wang}, \bibinfo{person}{Jun Yang},
  \bibinfo{person}{Danyang Zhuo}, \bibinfo{person}{Koushik Sen},
  \bibinfo{person}{Joseph~E. Gonzalez}, {and} \bibinfo{person}{Ion Stoica}.}
  \bibinfo{year}{2020}\natexlab{}.
\newblock \showarticletitle{Ansor: Generating High-Performance Tensor Programs
  for Deep Learning}. In \bibinfo{booktitle}{\emph{14th {USENIX} Symposium on
  Operating Systems Design and Implementation, {OSDI} 2020, Virtual Event,
  November 4-6, 2020}}. \bibinfo{publisher}{{USENIX} Association},
  \bibinfo{pages}{863--879}.
\newblock
\urldef\tempurl%
\url{https://www.usenix.org/conference/osdi20/presentation/zheng}
\showURL{%
\tempurl}


\bibitem[\protect\citeauthoryear{Zinenko, Verdoolaege, Reddy, Shirako, Grosser,
  Sarkar, and Cohen}{Zinenko et~al\mbox{.}}{2018}]%
        {Zin18}
\bibfield{author}{\bibinfo{person}{Oleksandr Zinenko}, \bibinfo{person}{Sven
  Verdoolaege}, \bibinfo{person}{Chandan Reddy}, \bibinfo{person}{Jun Shirako},
  \bibinfo{person}{Tobias Grosser}, \bibinfo{person}{Vivek Sarkar}, {and}
  \bibinfo{person}{Albert Cohen}.} \bibinfo{year}{2018}\natexlab{}.
\newblock \showarticletitle{Modeling the conflicting demands of parallelism and
  Temporal/Spatial locality in affine scheduling}. In
  \bibinfo{booktitle}{\emph{Proceedings of the 27th International Conference on
  Compiler Construction, {CC} 2018, February 24-25, 2018, Vienna, Austria}},
  \bibfield{editor}{\bibinfo{person}{Christophe Dubach} {and}
  \bibinfo{person}{Jingling Xue}} (Eds.). \bibinfo{publisher}{{ACM}},
  \bibinfo{pages}{3--13}.
\newblock
\urldef\tempurl%
\url{https://doi.org/10.1145/3178372.3179507}
\showDOI{\tempurl}


\end{thebibliography}



\begin{thebibliography}{60}

%%% the Web address.

\ifx \showCODEN    \undefined \def \showCODEN     #1{\unskip}     \fi
\ifx \showDOI      \undefined \def \showDOI       #1{#1}\fi
\ifx \showISBNx    \undefined \def \showISBNx     #1{\unskip}     \fi
\ifx \showISBNxiii \undefined \def \showISBNxiii  #1{\unskip}     \fi
\ifx \showISSN     \undefined \def \showISSN      #1{\unskip}     \fi
\ifx \showLCCN     \undefined \def \showLCCN      #1{\unskip}     \fi
\ifx \shownote     \undefined \def \shownote      #1{#1}          \fi
\ifx \showarticletitle \undefined \def \showarticletitle #1{#1}   \fi
\ifx \showURL      \undefined \def \showURL       {\relax}        \fi
% invisible to TeX
\providecommand\bibfield[2]{#2}
\providecommand\bibinfo[2]{#2}
\providecommand\natexlab[1]{#1}
\providecommand\showeprint[2][]{arXiv:#2}

\bibitem[\protect\citeauthoryear{Allen and Kennedy}{Allen and Kennedy}{2001}]%
        {AllenKennedy}
\bibfield{author}{\bibinfo{person}{Randy Allen} {and} \bibinfo{person}{Ken
  Kennedy}.} \bibinfo{year}{2001}\natexlab{}.
\newblock \bibinfo{booktitle}{\emph{Optimizing Compilers for Modern
  Architectures: A Dependence-Based Approach}}.
\newblock \bibinfo{publisher}{Morgan Kaufmann Publishers}.
\newblock
\showISBNx{9781493303540}


\bibitem[\protect\citeauthoryear{Baghdadi, Beaugnon, Cohen, Grosser, Kruse,
  Reddy, Verdoolaege, Betts, Donaldson, Ketema, Absar, van Haastregt, Kravets,
  Lokhmotov, David, and Hajiyev}{Baghdadi et~al\mbox{.}}{2015}]%
        {PENCIL}
\bibfield{author}{\bibinfo{person}{Riyadh Baghdadi}, \bibinfo{person}{Ulysse
  Beaugnon}, \bibinfo{person}{Albert Cohen}, \bibinfo{person}{Tobias Grosser},
  \bibinfo{person}{Michael Kruse}, \bibinfo{person}{Chandan Reddy},
  \bibinfo{person}{Sven Verdoolaege}, \bibinfo{person}{Adam Betts},
  \bibinfo{person}{Alastair~F. Donaldson}, \bibinfo{person}{Jeroen Ketema},
  \bibinfo{person}{Javed Absar}, \bibinfo{person}{Sven van Haastregt},
  \bibinfo{person}{Alexey Kravets}, \bibinfo{person}{Anton Lokhmotov},
  \bibinfo{person}{Robert David}, {and} \bibinfo{person}{Elnar Hajiyev}.}
  \bibinfo{year}{2015}\natexlab{}.
\newblock \showarticletitle{{PENCIL:} {A} Platform-Neutral Compute Intermediate
  Language for Accelerator Programming}. In \bibinfo{booktitle}{\emph{2015
  International Conference on Parallel Architectures and Compilation, {PACT}
  2015, San Francisco, CA, USA, October 18-21, 2015}}.
  \bibinfo{publisher}{{IEEE} Computer Society}, \bibinfo{pages}{138--149}.
\newblock
\urldef\tempurl%
\url{https://doi.org/10.1109/PACT.2015.17}
\showDOI{\tempurl}


\bibitem[\protect\citeauthoryear{Barham and Isard}{Barham and Isard}{2019}]%
        {ml_rut}
\bibfield{author}{\bibinfo{person}{Paul Barham} {and} \bibinfo{person}{Michael
  Isard}.} \bibinfo{year}{2019}\natexlab{}.
\newblock \showarticletitle{Machine Learning Systems Are Stuck in a Rut}. In
  \bibinfo{booktitle}{\emph{Proceedings of the Workshop on Hot Topics in
  Operating Systems}} (Bertinoro, Italy) \emph{(\bibinfo{series}{HotOS '19})}.
  \bibinfo{publisher}{Association for Computing Machinery},
  \bibinfo{pages}{177–183}.
\newblock
\showISBNx{9781450367271}
\urldef\tempurl%
\url{https://doi.org/10.1145/3317550.3321441}
\showDOI{\tempurl}


\bibitem[\protect\citeauthoryear{Bastoul}{Bastoul}{2004}]%
        {Bas04}
\bibfield{author}{\bibinfo{person}{C{\'{e}}dric Bastoul}.}
  \bibinfo{year}{2004}\natexlab{}.
\newblock \showarticletitle{Code Generation in the Polyhedral Model Is Easier
  Than You Think}. In \bibinfo{booktitle}{\emph{13th International Conference
  on Parallel Architectures and Compilation Techniques {(PACT} 2004), 29
  September - 3 October 2004, Antibes Juan-les-Pins, France}}.
  \bibinfo{publisher}{{IEEE} Computer Society}, \bibinfo{pages}{7--16}.
\newblock
\urldef\tempurl%
\url{https://doi.org/10.1109/PACT.2004.10018}
\showDOI{\tempurl}


\bibitem[\protect\citeauthoryear{Bik}{Bik}{1996}]%
        {bik96}
\bibfield{author}{\bibinfo{person}{Aart~J.C. Bik}.}
  \bibinfo{year}{1996}\natexlab{}.
\newblock \emph{\bibinfo{title}{Compiler Support for Sparse Matrix
  Computations}}.
\newblock \bibinfo{thesistype}{Ph.\,D. Dissertation}.
  \bibinfo{school}{Department of Computer Science, Leiden University}.
\newblock
\newblock
\shownote{ISBN 90-9009442-3}.


\bibitem[\protect\citeauthoryear{Bik, Brinkhaus, Knijnenburg, and Wijshoff}{Bik
  et~al\mbox{.}}{1998}]%
        {biktms}
\bibfield{author}{\bibinfo{person}{Aart~J.C. Bik}, \bibinfo{person}{Peter~J.H.
  Brinkhaus}, \bibinfo{person}{Peter~M.W. Knijnenburg}, {and}
  \bibinfo{person}{Harry~A.G. Wijshoff}.} \bibinfo{year}{1998}\natexlab{}.
\newblock \showarticletitle{The Automatic Generation of Sparse Primitives}.
\newblock \bibinfo{journal}{\emph{Transactions on Mathematical Software}}
  \bibinfo{volume}{24} (\bibinfo{year}{1998}), \bibinfo{pages}{190--225}.
\newblock
\urldef\tempurl%
\url{https://doi.org/10.1145/290200.287636}
\showURL{%
\tempurl}


\bibitem[\protect\citeauthoryear{Bondhugula, Hartono, Ramanujam, and
  Sadayappan}{Bondhugula et~al\mbox{.}}{2008}]%
        {Pluto}
\bibfield{author}{\bibinfo{person}{Uday Bondhugula}, \bibinfo{person}{Albert
  Hartono}, \bibinfo{person}{J. Ramanujam}, {and} \bibinfo{person}{P.
  Sadayappan}.} \bibinfo{year}{2008}\natexlab{}.
\newblock \showarticletitle{A practical automatic polyhedral parallelizer and
  locality optimizer}. In \bibinfo{booktitle}{\emph{Proceedings of the {ACM}
  {SIGPLAN} 2008 Conference on Programming Language Design and Implementation,
  Tucson, AZ, USA, June 7-13, 2008}}, \bibfield{editor}{\bibinfo{person}{Rajiv
  Gupta} {and} \bibinfo{person}{Saman~P. Amarasinghe}} (Eds.).
  \bibinfo{publisher}{{ACM}}, \bibinfo{pages}{101--113}.
\newblock
\urldef\tempurl%
\url{https://doi.org/10.1145/1375581.1375595}
\showDOI{\tempurl}


\bibitem[\protect\citeauthoryear{Brockman, Cheung, Pettersson, Schneider,
  Schulman, Tang, and Zaremba}{Brockman et~al\mbox{.}}{2016}]%
        {OpenAIGym}
\bibfield{author}{\bibinfo{person}{Greg Brockman}, \bibinfo{person}{Vicki
  Cheung}, \bibinfo{person}{Ludwig Pettersson}, \bibinfo{person}{Jonas
  Schneider}, \bibinfo{person}{John Schulman}, \bibinfo{person}{Jie Tang},
  {and} \bibinfo{person}{Wojciech Zaremba}.} \bibinfo{year}{2016}\natexlab{}.
\newblock \showarticletitle{OpenAI Gym}.
\newblock \bibinfo{journal}{\emph{CoRR}}  \bibinfo{volume}{abs/1606.01540}
  (\bibinfo{year}{2016}).
\newblock
\showeprint[arXiv]{1606.01540}
\urldef\tempurl%
\url{http://arxiv.org/abs/1606.01540}
\showURL{%
\tempurl}


\bibitem[\protect\citeauthoryear{Chen, Moreau, Jiang, Zheng, Yan, Shen, Cowan,
  Wang, Hu, Ceze, et~al\mbox{.}}{Chen et~al\mbox{.}}{2018}]%
        {chen2018tvm}
\bibfield{author}{\bibinfo{person}{Tianqi Chen}, \bibinfo{person}{Thierry
  Moreau}, \bibinfo{person}{Ziheng Jiang}, \bibinfo{person}{Lianmin Zheng},
  \bibinfo{person}{Eddie Yan}, \bibinfo{person}{Haichen Shen},
  \bibinfo{person}{Meghan Cowan}, \bibinfo{person}{Leyuan Wang},
  \bibinfo{person}{Yuwei Hu}, \bibinfo{person}{Luis Ceze}, {et~al\mbox{.}}}
  \bibinfo{year}{2018}\natexlab{}.
\newblock \showarticletitle{{TVM}: An automated end-to-end optimizing compiler
  for deep learning}. In \bibinfo{booktitle}{\emph{13th {USENIX} Symposium on
  Operating Systems Design and Implementation ({OSDI} 18)}}.
  \bibinfo{publisher}{{USENIX} Association}, \bibinfo{pages}{578--594}.
\newblock
\urldef\tempurl%
\url{https://www.usenix.org/conference/osdi18/presentation/chen}
\showURL{%
\tempurl}


\bibitem[\protect\citeauthoryear{Chen, Mendis, Carbin, and Amarasinghe}{Chen
  et~al\mbox{.}}{2021}]%
        {VGen}
\bibfield{author}{\bibinfo{person}{Yishen Chen}, \bibinfo{person}{Charith
  Mendis}, \bibinfo{person}{Michael Carbin}, {and} \bibinfo{person}{Saman~P.
  Amarasinghe}.} \bibinfo{year}{2021}\natexlab{}.
\newblock \showarticletitle{VeGen: a vectorizer generator for {SIMD} and
  beyond}. In \bibinfo{booktitle}{\emph{{ASPLOS} '21: 26th {ACM} International
  Conference on Architectural Support for Programming Languages and Operating
  Systems, Virtual Event, USA, April 19-23, 2021}},
  \bibfield{editor}{\bibinfo{person}{Tim Sherwood}, \bibinfo{person}{Emery~D.
  Berger}, {and} \bibinfo{person}{Christos Kozyrakis}} (Eds.).
  \bibinfo{publisher}{{ACM}}, \bibinfo{pages}{902--914}.
\newblock
\urldef\tempurl%
\url{https://doi.org/10.1145/3445814.3446692}
\showDOI{\tempurl}


\bibitem[\protect\citeauthoryear{Click and Cooper}{Click and Cooper}{1995}]%
        {Click95}
\bibfield{author}{\bibinfo{person}{Cliff Click} {and} \bibinfo{person}{Keith~D.
  Cooper}.} \bibinfo{year}{1995}\natexlab{}.
\newblock \showarticletitle{Combining Analyses, Combining Optimizations}.
\newblock \bibinfo{journal}{\emph{{ACM} Transactions on Programming Languages
  and Systems}} \bibinfo{volume}{17}, \bibinfo{number}{2}
  (\bibinfo{year}{1995}), \bibinfo{pages}{181--196}.
\newblock


\bibitem[\protect\citeauthoryear{Corp.}{Corp.}{2021}]%
        {IntelOptimizationManual}
\bibfield{author}{\bibinfo{person}{Intel Corp.}}
  \bibinfo{year}{2021}\natexlab{}.
\newblock \bibinfo{title}{Intel Optimization Reference Manual}.
\newblock
  \bibinfo{howpublished}{\url{https://www.intel.com/content/www/us/en/develop/download/intel-64-and-ia-32-architectures-optimization-reference-manual.html}}.
\newblock


\bibitem[\protect\citeauthoryear{Cummins, Wasti, Guo, Cui, Ansel, Gomez, Jain,
  Liu, Teytaud, Steiner, Tian, and Leather}{Cummins et~al\mbox{.}}{2021}]%
        {CompilerGym}
\bibfield{author}{\bibinfo{person}{Chris Cummins}, \bibinfo{person}{Bram
  Wasti}, \bibinfo{person}{Jiadong Guo}, \bibinfo{person}{Brandon Cui},
  \bibinfo{person}{Jason Ansel}, \bibinfo{person}{Sahir Gomez},
  \bibinfo{person}{Somya Jain}, \bibinfo{person}{Jia Liu},
  \bibinfo{person}{Olivier Teytaud}, \bibinfo{person}{Benoit Steiner},
  \bibinfo{person}{Yuandong Tian}, {and} \bibinfo{person}{Hugh Leather}.}
  \bibinfo{year}{2021}\natexlab{}.
\newblock \showarticletitle{CompilerGym: Robust, Performant Compiler
  Optimization Environments for {AI} Research}.
\newblock \bibinfo{journal}{\emph{CoRR}}  \bibinfo{volume}{abs/2109.08267}
  (\bibinfo{year}{2021}).
\newblock
\showeprint[arXiv]{2109.08267}
\urldef\tempurl%
\url{https://arxiv.org/abs/2109.08267}
\showURL{%
\tempurl}


\bibitem[\protect\citeauthoryear{Developers}{Developers}{2021}]%
        {iree}
\bibfield{author}{\bibinfo{person}{{IREE} Developers}.}
  \bibinfo{year}{2021}\natexlab{}.
\newblock \bibinfo{title}{{IREE} (Intermediate Representation Execution
  Environment}.
\newblock
\newblock
\urldef\tempurl%
\url{https://google.github.io/iree/}
\showURL{%
\tempurl}


\bibitem[\protect\citeauthoryear{Devlin, Chang, Lee, and Toutanova}{Devlin
  et~al\mbox{.}}{2019}]%
        {Bert}
\bibfield{author}{\bibinfo{person}{Jacob Devlin}, \bibinfo{person}{Ming-Wei
  Chang}, \bibinfo{person}{Kenton Lee}, {and} \bibinfo{person}{Kristina
  Toutanova}.} \bibinfo{year}{2019}\natexlab{}.
\newblock \bibinfo{title}{BERT: Pre-training of Deep Bidirectional Transformers
  for Language Understanding}.
\newblock
\newblock
\showeprint[arxiv]{1810.04805}~[cs.CL]


\bibitem[\protect\citeauthoryear{Documentation}{Documentation}{2021}]%
        {LLVM_benchmarking}
\bibfield{author}{\bibinfo{person}{LLVM Documentation}.}
  \bibinfo{year}{2021}\natexlab{}.
\newblock \bibinfo{title}{Benchmarking tips}.
\newblock
  \bibinfo{howpublished}{\url{https://llvm.org/docs/Benchmarking.html}}.
\newblock


\bibitem[\protect\citeauthoryear{Feautrier}{Feautrier}{1992a}]%
        {Fea92a}
\bibfield{author}{\bibinfo{person}{Paul Feautrier}.}
  \bibinfo{year}{1992}\natexlab{a}.
\newblock \showarticletitle{Some efficient solutions to the affine scheduling
  problem. Part {I.} One-dimensional time}.
\newblock \bibinfo{journal}{\emph{Int. J. Parallel Program.}}
  \bibinfo{volume}{21}, \bibinfo{number}{5} (\bibinfo{year}{1992}),
  \bibinfo{pages}{313--347}.
\newblock
\urldef\tempurl%
\url{https://doi.org/10.1007/BF01407835}
\showDOI{\tempurl}


\bibitem[\protect\citeauthoryear{Feautrier}{Feautrier}{1992b}]%
        {Fea92b}
\bibfield{author}{\bibinfo{person}{Paul Feautrier}.}
  \bibinfo{year}{1992}\natexlab{b}.
\newblock \showarticletitle{Some efficient solutions to the affine scheduling
  problem. Part {II.} Multidimensional time}.
\newblock \bibinfo{journal}{\emph{Int. J. Parallel Program.}}
  \bibinfo{volume}{21}, \bibinfo{number}{6} (\bibinfo{year}{1992}),
  \bibinfo{pages}{389--420}.
\newblock
\urldef\tempurl%
\url{https://doi.org/10.1007/BF01379404}
\showDOI{\tempurl}


\bibitem[\protect\citeauthoryear{Girbal, Vasilache, Bastoul, Cohen, Parello,
  Sigler, and Temam}{Girbal et~al\mbox{.}}{2006}]%
        {URUK}
\bibfield{author}{\bibinfo{person}{Sylvain Girbal}, \bibinfo{person}{Nicolas
  Vasilache}, \bibinfo{person}{C{\'{e}}dric Bastoul}, \bibinfo{person}{Albert
  Cohen}, \bibinfo{person}{David Parello}, \bibinfo{person}{Marc Sigler}, {and}
  \bibinfo{person}{Olivier Temam}.} \bibinfo{year}{2006}\natexlab{}.
\newblock \showarticletitle{Semi-Automatic Composition of Loop Transformations
  for Deep Parallelism and Memory Hierarchies}.
\newblock \bibinfo{journal}{\emph{Int. J. Parallel Program.}}
  \bibinfo{volume}{34}, \bibinfo{number}{3} (\bibinfo{year}{2006}),
  \bibinfo{pages}{261--317}.
\newblock
\urldef\tempurl%
\url{https://doi.org/10.1007/s10766-006-0012-3}
\showDOI{\tempurl}


\bibitem[\protect\citeauthoryear{Grosser, Gr{\"{o}}{\ss}linger, and
  Lengauer}{Grosser et~al\mbox{.}}{2012}]%
        {Polly}
\bibfield{author}{\bibinfo{person}{Tobias Grosser}, \bibinfo{person}{Armin
  Gr{\"{o}}{\ss}linger}, {and} \bibinfo{person}{Christian Lengauer}.}
  \bibinfo{year}{2012}\natexlab{}.
\newblock \showarticletitle{Polly - Performing Polyhedral Optimizations on a
  Low-Level Intermediate Representation}.
\newblock \bibinfo{journal}{\emph{Parallel Process. Lett.}}
  \bibinfo{volume}{22}, \bibinfo{number}{4} (\bibinfo{year}{2012}).
\newblock
\urldef\tempurl%
\url{https://doi.org/10.1142/S0129626412500107}
\showDOI{\tempurl}


\bibitem[\protect\citeauthoryear{Grosser, Verdoolaege, and Cohen}{Grosser
  et~al\mbox{.}}{2015}]%
        {Gro15}
\bibfield{author}{\bibinfo{person}{Tobias Grosser}, \bibinfo{person}{Sven
  Verdoolaege}, {and} \bibinfo{person}{Albert Cohen}.}
  \bibinfo{year}{2015}\natexlab{}.
\newblock \showarticletitle{Polyhedral {AST} Generation Is More Than Scanning
  Polyhedra}.
\newblock \bibinfo{journal}{\emph{{ACM} Transactions on Programming Languages
  and Systems}} \bibinfo{volume}{37}, \bibinfo{number}{4}
  (\bibinfo{year}{2015}), \bibinfo{pages}{12:1--12:50}.
\newblock
\urldef\tempurl%
\url{https://doi.org/10.1145/2743016}
\showDOI{\tempurl}


\bibitem[\protect\citeauthoryear{Guennebaud, Jacob, et~al\mbox{.}}{Guennebaud
  et~al\mbox{.}}{2010}]%
        {eigen}
\bibfield{author}{\bibinfo{person}{Ga\"{e}l Guennebaud},
  \bibinfo{person}{Beno\^{i}t Jacob}, {et~al\mbox{.}}}
  \bibinfo{year}{2010}\natexlab{}.
\newblock \bibinfo{title}{Eigen v3}.
\newblock
\newblock
\urldef\tempurl%
\url{http://eigen.tuxfamily.org}
\showURL{%
\tempurl}


\bibitem[\protect\citeauthoryear{Hagedorn, Elliott, Barthels, Bod{\'{\i}}k, and
  Grover}{Hagedorn et~al\mbox{.}}{2020a}]%
        {Fireiron}
\bibfield{author}{\bibinfo{person}{Bastian Hagedorn},
  \bibinfo{person}{Archibald~Samuel Elliott}, \bibinfo{person}{Henrik
  Barthels}, \bibinfo{person}{Rastislav Bod{\'{\i}}k}, {and}
  \bibinfo{person}{Vinod Grover}.} \bibinfo{year}{2020}\natexlab{a}.
\newblock \showarticletitle{Fireiron: {A} Scheduling Language for
  High-Performance Linear Algebra on GPUs}.
\newblock \bibinfo{journal}{\emph{CoRR}}  \bibinfo{volume}{abs/2003.06324}
  (\bibinfo{year}{2020}).
\newblock
\showeprint[arXiv]{2003.06324}
\urldef\tempurl%
\url{https://arxiv.org/abs/2003.06324}
\showURL{%
\tempurl}


\bibitem[\protect\citeauthoryear{Hagedorn, Lenfers, Koehler, Gorlatch, and
  Steuwer}{Hagedorn et~al\mbox{.}}{2020b}]%
        {Elevate}
\bibfield{author}{\bibinfo{person}{Bastian Hagedorn}, \bibinfo{person}{Johannes
  Lenfers}, \bibinfo{person}{Thomas Koehler}, \bibinfo{person}{Sergei
  Gorlatch}, {and} \bibinfo{person}{Michel Steuwer}.}
  \bibinfo{year}{2020}\natexlab{b}.
\newblock \showarticletitle{A Language for Describing Optimization Strategies}.
\newblock \bibinfo{journal}{\emph{CoRR}}  \bibinfo{volume}{abs/2002.02268}
  (\bibinfo{year}{2020}).
\newblock
\showeprint[arXiv]{2002.02268}
\urldef\tempurl%
\url{https://arxiv.org/abs/2002.02268}
\showURL{%
\tempurl}


\bibitem[\protect\citeauthoryear{Hagedorn, Lenfers, Kundefinedhler, Qin,
  Gorlatch, and Steuwer}{Hagedorn et~al\mbox{.}}{2020c}]%
        {rise}
\bibfield{author}{\bibinfo{person}{Bastian Hagedorn}, \bibinfo{person}{Johannes
  Lenfers}, \bibinfo{person}{Thomas Kundefinedhler}, \bibinfo{person}{Xueying
  Qin}, \bibinfo{person}{Sergei Gorlatch}, {and} \bibinfo{person}{Michel
  Steuwer}.} \bibinfo{year}{2020}\natexlab{c}.
\newblock \showarticletitle{Achieving High-Performance the Functional Way: A
  Functional Pearl on Expressing High-Performance Optimizations as Rewrite
  Strategies}.
\newblock \bibinfo{journal}{\emph{Proceedings of ACM on Programming Languages}}
  \bibinfo{volume}{4}, \bibinfo{number}{ICFP}, Article \bibinfo{articleno}{92}
  (\bibinfo{date}{Aug.} \bibinfo{year}{2020}), \bibinfo{numpages}{29}~pages.
\newblock
\urldef\tempurl%
\url{https://doi.org/10.1145/3408974}
\showDOI{\tempurl}


\bibitem[\protect\citeauthoryear{Heinecke, Henry, Hutchinson, and
  Pabst}{Heinecke et~al\mbox{.}}{2016}]%
        {heinecke2016libxsmm}
\bibfield{author}{\bibinfo{person}{Alexander Heinecke}, \bibinfo{person}{Greg
  Henry}, \bibinfo{person}{Maxwell Hutchinson}, {and} \bibinfo{person}{Hans
  Pabst}.} \bibinfo{year}{2016}\natexlab{}.
\newblock \showarticletitle{{LIBXSMM}: accelerating small matrix
  multiplications by runtime code generation}. In
  \bibinfo{booktitle}{\emph{SC'16: Proceedings of the International Conference
  for High Performance Computing, Networking, Storage and Analysis}}.
  \bibinfo{publisher}{IEEE/ACM}, \bibinfo{pages}{981--991}.
\newblock
\urldef\tempurl%
\url{https://doi.org/10.1109/SC.2016.83}
\showURL{%
\tempurl}


\bibitem[\protect\citeauthoryear{Hoefler and Belli}{Hoefler and Belli}{2015}]%
        {benchmark2015}
\bibfield{author}{\bibinfo{person}{Torsten Hoefler} {and}
  \bibinfo{person}{Roberto Belli}.} \bibinfo{year}{2015}\natexlab{}.
\newblock \showarticletitle{Scientific Benchmarking of Parallel Computing
  Systems: Twelve Ways to Tell the Masses When Reporting Performance Results}.
  In \bibinfo{booktitle}{\emph{SC'15: Proceedings of the International
  Conference for High Performance Computing, Networking, Storage and Analysis}}
  (Austin, Texas). \bibinfo{publisher}{IEEE/ACM}, Article
  \bibinfo{articleno}{73}, \bibinfo{numpages}{12}~pages.
\newblock
\showISBNx{9781450337236}
\urldef\tempurl%
\url{https://doi.org/10.1145/2807591.2807644}
\showDOI{\tempurl}


\bibitem[\protect\citeauthoryear{Irigoin, Jouvelot, and Triolet}{Irigoin
  et~al\mbox{.}}{1991}]%
        {PIPS}
\bibfield{author}{\bibinfo{person}{Fran{\c{c}}ois Irigoin},
  \bibinfo{person}{Pierre Jouvelot}, {and} \bibinfo{person}{R{\'{e}}mi
  Triolet}.} \bibinfo{year}{1991}\natexlab{}.
\newblock \showarticletitle{Semantical interprocedural parallelization: an
  overview of the {PIPS} project}. In \bibinfo{booktitle}{\emph{Proceedings of
  the 5th international conference on Supercomputing, {ICS} 1991, Cologne,
  Germany, June 17-21, 1991}}, \bibfield{editor}{\bibinfo{person}{Edward~S.
  Davidson} {and} \bibinfo{person}{Friedel Hossfeld}} (Eds.).
  \bibinfo{publisher}{{ACM}}, \bibinfo{pages}{244--251}.
\newblock
\urldef\tempurl%
\url{https://doi.org/10.1145/109025.109086}
\showDOI{\tempurl}


\bibitem[\protect\citeauthoryear{Jia, Shelhamer, Donahue, Karayev, Long,
  Girshick, Guadarrama, and Darrell}{Jia et~al\mbox{.}}{2014}]%
        {Caffe}
\bibfield{author}{\bibinfo{person}{Yangqing Jia}, \bibinfo{person}{Evan
  Shelhamer}, \bibinfo{person}{Jeff Donahue}, \bibinfo{person}{Sergey Karayev},
  \bibinfo{person}{Jonathan Long}, \bibinfo{person}{Ross Girshick},
  \bibinfo{person}{Sergio Guadarrama}, {and} \bibinfo{person}{Trevor Darrell}.}
  \bibinfo{year}{2014}\natexlab{}.
\newblock \showarticletitle{Caffe: Convolutional Architecture for Fast Feature
  Embedding}. In \bibinfo{booktitle}{\emph{Proceedings of the 22Nd ACM
  International Conference on Multimedia}} (Orlando, Florida, USA)
  \emph{(\bibinfo{series}{MM '14})}. \bibinfo{publisher}{ACM},
  \bibinfo{pages}{675--678}.
\newblock
\showISBNx{978-1-4503-3063-3}
\urldef\tempurl%
\url{https://doi.org/10.1145/2647868.2654889}
\showDOI{\tempurl}


\bibitem[\protect\citeauthoryear{Jia, Padon, Thomas, Warszawski, Zaharia, and
  Aiken}{Jia et~al\mbox{.}}{2019}]%
        {TASO}
\bibfield{author}{\bibinfo{person}{Zhihao Jia}, \bibinfo{person}{Oded Padon},
  \bibinfo{person}{James Thomas}, \bibinfo{person}{Todd Warszawski},
  \bibinfo{person}{Matei Zaharia}, {and} \bibinfo{person}{Alex Aiken}.}
  \bibinfo{year}{2019}\natexlab{}.
\newblock \showarticletitle{TASO: Optimizing Deep Learning Computation with
  Automatic Generation of Graph Substitutions}. In
  \bibinfo{booktitle}{\emph{Proceedings of the 27th ACM Symposium on Operating
  Systems Principles}} (Huntsville, Ontario, Canada)
  \emph{(\bibinfo{series}{SOSP '19})}. \bibinfo{publisher}{Association for
  Computing Machinery}, \bibinfo{pages}{47–62}.
\newblock
\showISBNx{9781450368735}
\urldef\tempurl%
\url{https://doi.org/10.1145/3341301.3359630}
\showURL{%
\tempurl}


\bibitem[\protect\citeauthoryear{Khan, Basu, Rudy, Hall, Chen, and Chame}{Khan
  et~al\mbox{.}}{2013}]%
        {CHiLL}
\bibfield{author}{\bibinfo{person}{Malik~Murtaza Khan},
  \bibinfo{person}{Protonu Basu}, \bibinfo{person}{Gabe Rudy},
  \bibinfo{person}{Mary~W. Hall}, \bibinfo{person}{Chun Chen}, {and}
  \bibinfo{person}{Jacqueline Chame}.} \bibinfo{year}{2013}\natexlab{}.
\newblock \showarticletitle{A script-based autotuning compiler system to
  generate high-performance {CUDA} code}.
\newblock \bibinfo{journal}{\emph{{ACM} Transactions on Architecture and Code
  Optimization}} \bibinfo{volume}{9}, \bibinfo{number}{4}
  (\bibinfo{year}{2013}), \bibinfo{pages}{31:1--31:25}.
\newblock
\urldef\tempurl%
\url{https://doi.org/10.1145/2400682.2400690}
\showDOI{\tempurl}


\bibitem[\protect\citeauthoryear{Kjolstad, Kamil, Chou, Lugato, and
  Amarasinghe}{Kjolstad et~al\mbox{.}}{2017}]%
        {kjolstad2017taco}
\bibfield{author}{\bibinfo{person}{Fredrik Kjolstad}, \bibinfo{person}{Shoaib
  Kamil}, \bibinfo{person}{Stephen Chou}, \bibinfo{person}{David Lugato}, {and}
  \bibinfo{person}{Saman Amarasinghe}.} \bibinfo{year}{2017}\natexlab{}.
\newblock \showarticletitle{The Tensor Algebra Compiler}.
\newblock \bibinfo{journal}{\emph{Proc. ACM Program. Lang.}}
  \bibinfo{volume}{1}, \bibinfo{number}{OOPSLA}, Article
  \bibinfo{articleno}{77} (\bibinfo{date}{Oct.} \bibinfo{year}{2017}),
  \bibinfo{numpages}{29}~pages.
\newblock
\urldef\tempurl%
\url{https://doi.org/10.1145/3133901}
\showDOI{\tempurl}


\bibitem[\protect\citeauthoryear{Kruse and Grosser}{Kruse and Grosser}{2018}]%
        {DeLICM}
\bibfield{author}{\bibinfo{person}{Michael Kruse} {and} \bibinfo{person}{Tobias
  Grosser}.} \bibinfo{year}{2018}\natexlab{}.
\newblock \showarticletitle{DeLICM: scalar dependence removal at zero memory
  cost}. In \bibinfo{booktitle}{\emph{Proceedings of the 2018 International
  Symposium on Code Generation and Optimization, {CGO} 2018, V{\"{o}}sendorf /
  Vienna, Austria, February 24-28, 2018}},
  \bibfield{editor}{\bibinfo{person}{Jens Knoop}, \bibinfo{person}{Markus
  Schordan}, \bibinfo{person}{Teresa Johnson}, {and} \bibinfo{person}{Michael
  F.~P. O'Boyle}} (Eds.). \bibinfo{publisher}{{ACM}},
  \bibinfo{pages}{241--253}.
\newblock
\urldef\tempurl%
\url{https://doi.org/10.1145/3168815}
\showDOI{\tempurl}


\bibitem[\protect\citeauthoryear{Lattner, Amini, Bondhugula, Cohen, Davis,
  Pienaar, Riddle, Shpeisman, Vasilache, and Zinenko}{Lattner
  et~al\mbox{.}}{2021}]%
        {lattner2021mlir}
\bibfield{author}{\bibinfo{person}{Chris Lattner}, \bibinfo{person}{Mehdi
  Amini}, \bibinfo{person}{Uday Bondhugula}, \bibinfo{person}{Albert Cohen},
  \bibinfo{person}{Andy Davis}, \bibinfo{person}{Jacques Pienaar},
  \bibinfo{person}{River Riddle}, \bibinfo{person}{Tatiana Shpeisman},
  \bibinfo{person}{Nicolas Vasilache}, {and} \bibinfo{person}{Oleksandr
  Zinenko}.} \bibinfo{year}{2021}\natexlab{}.
\newblock \showarticletitle{Mlir: Scaling compiler infrastructure for domain
  specific computation}. In \bibinfo{booktitle}{\emph{2021 IEEE/ACM
  International Symposium on Code Generation and Optimization (CGO)}}.
  IEEE/ACM, \bibinfo{publisher}{IEEE/ACM}, \bibinfo{pages}{2--14}.
\newblock
\urldef\tempurl%
\url{https://doi.org/10.1109/CGO51591.2021.9370308}
\showURL{%
\tempurl}


\bibitem[\protect\citeauthoryear{Meister, Vasilache, Wohlford, Baskaran, Leung,
  and Lethin}{Meister et~al\mbox{.}}{2011}]%
        {RStream}
\bibfield{author}{\bibinfo{person}{Beno{\^{\i}}t Meister},
  \bibinfo{person}{Nicolas Vasilache}, \bibinfo{person}{David Wohlford},
  \bibinfo{person}{Muthu~Manikandan Baskaran}, \bibinfo{person}{Allen Leung},
  {and} \bibinfo{person}{Richard Lethin}.} \bibinfo{year}{2011}\natexlab{}.
\newblock \showarticletitle{R-Stream Compiler}.
\newblock In \bibinfo{booktitle}{\emph{Encyclopedia of Parallel Computing}},
  \bibfield{editor}{\bibinfo{person}{David~A. Padua}} (Ed.).
  \bibinfo{publisher}{Springer}, \bibinfo{pages}{1756--1765}.
\newblock
\urldef\tempurl%
\url{https://doi.org/10.1007/978-0-387-09766-4\_515}
\showDOI{\tempurl}


\bibitem[\protect\citeauthoryear{Mullapudi, Vasista, and Bondhugula}{Mullapudi
  et~al\mbox{.}}{2015}]%
        {PolyMage}
\bibfield{author}{\bibinfo{person}{Ravi~Teja Mullapudi}, \bibinfo{person}{Vinay
  Vasista}, {and} \bibinfo{person}{Uday Bondhugula}.}
  \bibinfo{year}{2015}\natexlab{}.
\newblock \showarticletitle{PolyMage: Automatic Optimization for Image
  Processing Pipelines}. In \bibinfo{booktitle}{\emph{Proceedings of the
  Twentieth International Conference on Architectural Support for Programming
  Languages and Operating Systems, {ASPLOS} 2015, Istanbul, Turkey, March
  14-18, 2015}}, \bibfield{editor}{\bibinfo{person}{{\"{O}}zcan {\"{O}}zturk},
  \bibinfo{person}{Kemal Ebcioglu}, {and} \bibinfo{person}{Sandhya Dwarkadas}}
  (Eds.). \bibinfo{publisher}{{ACM}}, \bibinfo{pages}{429--443}.
\newblock
\urldef\tempurl%
\url{https://doi.org/10.1145/2694344.2694364}
\showDOI{\tempurl}


\bibitem[\protect\citeauthoryear{Nod.AI}{Nod.AI}{2022a}]%
        {NodeAi-tuner}
\bibfield{author}{\bibinfo{person}{Nod.AI}.} \bibinfo{year}{2022}\natexlab{a}.
\newblock \bibinfo{title}{Outperforming Intel’s MKL and OctoML/Apache TVM
  with MLIR and Nod.ai’s Codegen Search}.
\newblock
  \bibinfo{howpublished}{\url{https://nod.ai/outperforming-octoml-apache-tvm-and-intel-mkl/}}.
\newblock


\bibitem[\protect\citeauthoryear{Nod.AI}{Nod.AI}{2022b}]%
        {NodeAi-runtime}
\bibfield{author}{\bibinfo{person}{Nod.AI}.} \bibinfo{year}{2022}\natexlab{b}.
\newblock \bibinfo{title}{SHARK: The fastest PyTorch runtime – 3x over
  Torchscript, 1.6x over TF/XLA, 23\% faster than ONNXRuntime}.
\newblock
  \bibinfo{howpublished}{\url{https://nod.ai/shark-the-fastest-runtime/}}.
\newblock


\bibitem[\protect\citeauthoryear{Paszke, Gross, Massa, Lerer, Bradbury, Chanan,
  Killeen, Lin, Gimelshein, Antiga, Desmaison, Kopf, Yang, DeVito, Raison,
  Tejani, Chilamkurthy, Steiner, Fang, Bai, and Chintala}{Paszke
  et~al\mbox{.}}{2019}]%
        {PyTorch}
\bibfield{author}{\bibinfo{person}{Adam Paszke}, \bibinfo{person}{Sam Gross},
  \bibinfo{person}{Francisco Massa}, \bibinfo{person}{Adam Lerer},
  \bibinfo{person}{James Bradbury}, \bibinfo{person}{Gregory Chanan},
  \bibinfo{person}{Trevor Killeen}, \bibinfo{person}{Zeming Lin},
  \bibinfo{person}{Natalia Gimelshein}, \bibinfo{person}{Luca Antiga},
  \bibinfo{person}{Alban Desmaison}, \bibinfo{person}{Andreas Kopf},
  \bibinfo{person}{Edward Yang}, \bibinfo{person}{Zachary DeVito},
  \bibinfo{person}{Martin Raison}, \bibinfo{person}{Alykhan Tejani},
  \bibinfo{person}{Sasank Chilamkurthy}, \bibinfo{person}{Benoit Steiner},
  \bibinfo{person}{Lu Fang}, \bibinfo{person}{Junjie Bai}, {and}
  \bibinfo{person}{Soumith Chintala}.} \bibinfo{year}{2019}\natexlab{}.
\newblock \showarticletitle{PyTorch: An Imperative Style, High-Performance Deep
  Learning Library}.
\newblock In \bibinfo{booktitle}{\emph{Advances in Neural Information
  Processing Systems 32}}, \bibfield{editor}{\bibinfo{person}{H.~Wallach},
  \bibinfo{person}{H.~Larochelle}, \bibinfo{person}{A.~Beygelzimer},
  \bibinfo{person}{F.~d\textquotesingle Alch\'{e}-Buc},
  \bibinfo{person}{E.~Fox}, {and} \bibinfo{person}{R.~Garnett}} (Eds.).
  \bibinfo{publisher}{Curran Associates, Inc.}, \bibinfo{pages}{8024--8035}.
\newblock
\urldef\tempurl%
\url{http://papers.neurips.cc/paper/9015-pytorch-an-imperative-style-high-performance-deep-learning-library.pdf}
\showURL{%
\tempurl}


\bibitem[\protect\citeauthoryear{Phothilimthana, Elliott, Wang, Jangda,
  Hagedorn, Barthels, Kaufman, Grover, Torlak, and Bod{\'{\i}}k}{Phothilimthana
  et~al\mbox{.}}{2019}]%
        {SwizzleInventor}
\bibfield{author}{\bibinfo{person}{Phitchaya~Mangpo Phothilimthana},
  \bibinfo{person}{Archibald~Samuel Elliott}, \bibinfo{person}{An Wang},
  \bibinfo{person}{Abhinav Jangda}, \bibinfo{person}{Bastian Hagedorn},
  \bibinfo{person}{Henrik Barthels}, \bibinfo{person}{Samuel~J. Kaufman},
  \bibinfo{person}{Vinod Grover}, \bibinfo{person}{Emina Torlak}, {and}
  \bibinfo{person}{Rastislav Bod{\'{\i}}k}.} \bibinfo{year}{2019}\natexlab{}.
\newblock \showarticletitle{Swizzle Inventor: Data Movement Synthesis for {GPU}
  Kernels}. In \bibinfo{booktitle}{\emph{Proceedings of the Twenty-Fourth
  International Conference on Architectural Support for Programming Languages
  and Operating Systems, {ASPLOS} 2019, Providence, RI, USA, April 13-17,
  2019}}, \bibfield{editor}{\bibinfo{person}{Iris Bahar},
  \bibinfo{person}{Maurice Herlihy}, \bibinfo{person}{Emmett Witchel}, {and}
  \bibinfo{person}{Alvin~R. Lebeck}} (Eds.). \bibinfo{publisher}{{ACM}},
  \bibinfo{pages}{65--78}.
\newblock
\urldef\tempurl%
\url{https://doi.org/10.1145/3297858.3304059}
\showDOI{\tempurl}


\bibitem[\protect\citeauthoryear{Ragan-Kelley, Barnes, Adams, Paris, Durand,
  and Amarasinghe}{Ragan-Kelley et~al\mbox{.}}{2013}]%
        {ragan2013halide}
\bibfield{author}{\bibinfo{person}{Jonathan Ragan-Kelley},
  \bibinfo{person}{Connelly Barnes}, \bibinfo{person}{Andrew Adams},
  \bibinfo{person}{Sylvain Paris}, \bibinfo{person}{Fr{\'e}do Durand}, {and}
  \bibinfo{person}{Saman Amarasinghe}.} \bibinfo{year}{2013}\natexlab{}.
\newblock \showarticletitle{Halide: a language and compiler for optimizing
  parallelism, locality, and recomputation in image processing pipelines}.
\newblock \bibinfo{journal}{\emph{{ACM} {SigPLAN} Notices}}
  \bibinfo{volume}{48}, \bibinfo{number}{6} (\bibinfo{year}{2013}),
  \bibinfo{pages}{519--530}.
\newblock
\urldef\tempurl%
\url{https://doi.org/10.1145/2499370.2462176}
\showURL{%
\tempurl}


\bibitem[\protect\citeauthoryear{Rasch, Schulze, and Gorlatch}{Rasch
  et~al\mbox{.}}{2019}]%
        {Rasch2019GeneratingPH}
\bibfield{author}{\bibinfo{person}{Ari Rasch}, \bibinfo{person}{Richard
  Schulze}, {and} \bibinfo{person}{Sergei Gorlatch}.}
  \bibinfo{year}{2019}\natexlab{}.
\newblock \showarticletitle{Generating Portable High-Performance Code via
  Multi-Dimensional Homomorphisms}. In \bibinfo{booktitle}{\emph{2019 28th
  International Conference on Parallel Architectures and Compilation Techniques
  (PACT)}}. \bibinfo{publisher}{IEEE}, \bibinfo{pages}{354--369}.
\newblock
\urldef\tempurl%
\url{https://doi.org/10.1109/PACT.2019.00035}
\showURL{%
\tempurl}


\bibitem[\protect\citeauthoryear{Riddle}{Riddle}{2021}]%
        {PDLL}
\bibfield{author}{\bibinfo{person}{River Riddle}.}
  \bibinfo{year}{2021}\natexlab{}.
\newblock \bibinfo{title}{PDLL: a new declarative rewrite frontend for MLIR}.
\newblock
  \bibinfo{howpublished}{\url{https://llvm.discourse.group/t/rfc-pdll-a-new-declarative-rewrite-frontend-for-mlir/4798}}.
\newblock


\bibitem[\protect\citeauthoryear{Rossini}{Rossini}{2021}]%
        {OuterproductSpilling}
\bibfield{author}{\bibinfo{person}{Giuseppe Rossini}.}
  \bibinfo{year}{2021}\natexlab{}.
\newblock \bibinfo{title}{Outerproduct spilling}.
\newblock
  \bibinfo{howpublished}{\url{https://llvm.discourse.group/t/outer-product-spilling/4405}}.
\newblock


\bibitem[\protect\citeauthoryear{Rotem, Fix, Abdulrasool, Deng, Dzhabarov,
  Hegeman, Levenstein, Maher, Satish, Olesen, Park, Rakhov, and
  Smelyanskiy}{Rotem et~al\mbox{.}}{2018}]%
        {rotem2018glow}
\bibfield{author}{\bibinfo{person}{Nadav Rotem}, \bibinfo{person}{Jordan Fix},
  \bibinfo{person}{Saleem Abdulrasool}, \bibinfo{person}{Summer Deng},
  \bibinfo{person}{Roman Dzhabarov}, \bibinfo{person}{James Hegeman},
  \bibinfo{person}{Roman Levenstein}, \bibinfo{person}{Bert Maher},
  \bibinfo{person}{Nadathur Satish}, \bibinfo{person}{Jakob Olesen},
  \bibinfo{person}{Jongsoo Park}, \bibinfo{person}{Artem Rakhov}, {and}
  \bibinfo{person}{Misha Smelyanskiy}.} \bibinfo{year}{2018}\natexlab{}.
\newblock \showarticletitle{Glow: Graph Lowering Compiler Techniques for Neural
  Networks}.
\newblock \bibinfo{journal}{\emph{CoRR}}  \bibinfo{volume}{abs/1805.00907}
  (\bibinfo{year}{2018}).
\newblock
\showeprint[arXiv]{1805.00907}
\urldef\tempurl%
\url{http://arxiv.org/abs/1805.00907}
\showURL{%
\tempurl}


\bibitem[\protect\citeauthoryear{Sandler, Howard, Zhu, Zhmoginov, and
  Chen}{Sandler et~al\mbox{.}}{2018}]%
        {MobileNetV2}
\bibfield{author}{\bibinfo{person}{Mark Sandler}, \bibinfo{person}{Andrew~G.
  Howard}, \bibinfo{person}{Menglong Zhu}, \bibinfo{person}{Andrey Zhmoginov},
  {and} \bibinfo{person}{Liang{-}Chieh Chen}.} \bibinfo{year}{2018}\natexlab{}.
\newblock \showarticletitle{Inverted Residuals and Linear Bottlenecks: Mobile
  Networks for Classification, Detection and Segmentation}.
\newblock \bibinfo{journal}{\emph{CoRR}}  \bibinfo{volume}{abs/1801.04381}
  (\bibinfo{year}{2018}).
\newblock
\showeprint[arXiv]{1801.04381}
\urldef\tempurl%
\url{http://arxiv.org/abs/1801.04381}
\showURL{%
\tempurl}


\bibitem[\protect\citeauthoryear{Schrijver}{Schrijver}{1986}]%
        {schrijver86}
\bibfield{author}{\bibinfo{person}{Alexander Schrijver}.}
  \bibinfo{year}{1986}\natexlab{}.
\newblock \bibinfo{booktitle}{\emph{Theory of Linear and Integer Programming}}.
\newblock \bibinfo{publisher}{John Wiley \& Sons}.
\newblock


\bibitem[\protect\citeauthoryear{Smith, Liu, Lyubomirsky, Davidson, McMahan,
  Taylor, Ceze, and Tatlock}{Smith et~al\mbox{.}}{2021}]%
        {Glenside}
\bibfield{author}{\bibinfo{person}{Gus~Henry Smith}, \bibinfo{person}{Andrew
  Liu}, \bibinfo{person}{Steven Lyubomirsky}, \bibinfo{person}{Scott Davidson},
  \bibinfo{person}{Joseph McMahan}, \bibinfo{person}{Michael~B. Taylor},
  \bibinfo{person}{Luis Ceze}, {and} \bibinfo{person}{Zachary Tatlock}.}
  \bibinfo{year}{2021}\natexlab{}.
\newblock \showarticletitle{Pure Tensor Program Rewriting via Access Patterns
  (Representation Pearl)}.
\newblock \bibinfo{journal}{\emph{CoRR}}  \bibinfo{volume}{abs/2105.09377}
  (\bibinfo{year}{2021}).
\newblock
\showeprint[arXiv]{2105.09377}
\urldef\tempurl%
\url{https://arxiv.org/abs/2105.09377}
\showURL{%
\tempurl}


\bibitem[\protect\citeauthoryear{Steuwer, Remmelg, and Dubach}{Steuwer
  et~al\mbox{.}}{2017}]%
        {LIFT}
\bibfield{author}{\bibinfo{person}{Michel Steuwer}, \bibinfo{person}{Toomas
  Remmelg}, {and} \bibinfo{person}{Christophe Dubach}.}
  \bibinfo{year}{2017}\natexlab{}.
\newblock \showarticletitle{LIFT: A functional data-parallel IR for
  high-performance GPU code generation}. In \bibinfo{booktitle}{\emph{2017
  IEEE/ACM International Symposium on Code Generation and Optimization (CGO)}}.
  \bibinfo{publisher}{IEEE/ACM}, \bibinfo{pages}{74--85}.
\newblock
\urldef\tempurl%
\url{https://doi.org/10.1109/CGO.2017.7863730}
\showDOI{\tempurl}


\bibitem[\protect\citeauthoryear{team~within Google}{team~within
  Google}{2017}]%
        {XLA}
\bibfield{author}{\bibinfo{person}{XLA team~within Google}.}
  \bibinfo{year}{2017}\natexlab{}.
\newblock \bibinfo{title}{XLA: TensorFlow, Compiled.}
\newblock \bibinfo{howpublished}{Google Developers Blog}.
\newblock
\urldef\tempurl%
\url{https://developers.googleblog.com/2017/03/xla-tensorflow-compiled.html}
\showURL{%
\tempurl}


\bibitem[\protect\citeauthoryear{Tian, Saito, Preis, Garcia, Kozhukhov, Masten,
  Cherkasov, and Panchenko}{Tian et~al\mbox{.}}{2013}]%
        {IPDPS2013}
\bibfield{author}{\bibinfo{person}{Xinmin Tian}, \bibinfo{person}{Hideki
  Saito}, \bibinfo{person}{Serguei~V. Preis}, \bibinfo{person}{Eric~N. Garcia},
  \bibinfo{person}{Sergey~S. Kozhukhov}, \bibinfo{person}{Matt Masten},
  \bibinfo{person}{Aleksei~G. Cherkasov}, {and} \bibinfo{person}{Nikolay
  Panchenko}.} \bibinfo{year}{2013}\natexlab{}.
\newblock \showarticletitle{Practical SIMD Vectorization Techniques for Intel®
  Xeon Phi Coprocessors}. In \bibinfo{booktitle}{\emph{2013 IEEE International
  Symposium on Parallel Distributed Processing, Workshops and Phd Forum}}.
  \bibinfo{publisher}{IEEE}, \bibinfo{pages}{1149--1158}.
\newblock
\urldef\tempurl%
\url{https://doi.org/10.1109/IPDPSW.2013.245}
\showDOI{\tempurl}


\bibitem[\protect\citeauthoryear{Trifunovic, Cohen, Edelsohn, Li, Grosser,
  Jagasia, Ladelsky, Pop, Sj{\"o}din, and Upadrasta}{Trifunovic
  et~al\mbox{.}}{2010}]%
        {GRAPHITE}
\bibfield{author}{\bibinfo{person}{Konrad Trifunovic}, \bibinfo{person}{Albert
  Cohen}, \bibinfo{person}{David Edelsohn}, \bibinfo{person}{Feng Li},
  \bibinfo{person}{Tobias Grosser}, \bibinfo{person}{Harsha Jagasia},
  \bibinfo{person}{Razya Ladelsky}, \bibinfo{person}{Sebastian Pop},
  \bibinfo{person}{Jan Sj{\"o}din}, {and} \bibinfo{person}{Ramakrishna
  Upadrasta}.} \bibinfo{year}{2010}\natexlab{}.
\newblock \showarticletitle{{GRAPHITE Two Years After: First Lessons Learned
  From Real-World Polyhedral Compilation}}. In \bibinfo{booktitle}{\emph{{GCC
  Research Opportunities Workshop (GROW'10)}}}. \bibinfo{publisher}{HAL}.
\newblock
\urldef\tempurl%
\url{https://hal.inria.fr/inria-00551516}
\showURL{%
\tempurl}


\bibitem[\protect\citeauthoryear{Vasilache}{Vasilache}{2021}]%
        {VBlendPsLimitation}
\bibfield{author}{\bibinfo{person}{Nicolas Vasilache}.}
  \bibinfo{year}{2021}\natexlab{}.
\newblock \bibinfo{title}{Understanding and controlling some of the AVX shuffle
  emission paths}.
\newblock
  \bibinfo{howpublished}{\url{https://groups.google.com/g/llvm-dev/c/P_CXf5hPro8/m/RI6Fm3bzAQAJ}}.
\newblock


\bibitem[\protect\citeauthoryear{Vasilache, Meister, Baskaran, and
  Lethin}{Vasilache et~al\mbox{.}}{2012}]%
        {Vas12}
\bibfield{author}{\bibinfo{person}{Nicolas Vasilache}, \bibinfo{person}{Benoit
  Meister}, \bibinfo{person}{Muthu Baskaran}, {and} \bibinfo{person}{Richard
  Lethin}.} \bibinfo{year}{2012}\natexlab{}.
\newblock \showarticletitle{Joint scheduling and layout optimization to enable
  multi-level vectorization}. In \bibinfo{booktitle}{\emph{In Second
  International Workshop on Polyhedral Compilation Techniques (IMPACT)}}.
  \bibinfo{publisher}{Informal proceedings}.
\newblock


\bibitem[\protect\citeauthoryear{Vasilache, Zinenko, Theodoridis, Goyal,
  Devito, Moses, Verdoolaege, Adams, and Cohen}{Vasilache
  et~al\mbox{.}}{2019}]%
        {vasilache2019next}
\bibfield{author}{\bibinfo{person}{Nicolas Vasilache},
  \bibinfo{person}{Oleksandr Zinenko}, \bibinfo{person}{Theodoros Theodoridis},
  \bibinfo{person}{Priya Goyal}, \bibinfo{person}{Zachary Devito},
  \bibinfo{person}{William~S Moses}, \bibinfo{person}{Sven Verdoolaege},
  \bibinfo{person}{Andrew Adams}, {and} \bibinfo{person}{Albert Cohen}.}
  \bibinfo{year}{2019}\natexlab{}.
\newblock \showarticletitle{The next 700 accelerated layers: From mathematical
  expressions of network computation graphs to accelerated GPU kernels,
  automatically}.
\newblock \bibinfo{journal}{\emph{ACM Transactions on Architecture and Code
  Optimization (TACO)}} \bibinfo{volume}{16}, \bibinfo{number}{4}
  (\bibinfo{year}{2019}), \bibinfo{pages}{1--26}.
\newblock
\urldef\tempurl%
\url{https://doi.org/10.1145/3355606}
\showURL{%
\tempurl}


\bibitem[\protect\citeauthoryear{Verdoolaege}{Verdoolaege}{2010}]%
        {isl}
\bibfield{author}{\bibinfo{person}{Sven Verdoolaege}.}
  \bibinfo{year}{2010}\natexlab{}.
\newblock \showarticletitle{\emph{isl}: An Integer Set Library for the
  Polyhedral Model}. In \bibinfo{booktitle}{\emph{Mathematical Software --
  {ICMS} 2010, Third International Congress on Mathematical Software}}
  \emph{(\bibinfo{series}{Lecture Notes in Computer Science},
  Vol.~\bibinfo{volume}{6327})}, \bibfield{editor}{\bibinfo{person}{Komei
  Fukuda}, \bibinfo{person}{Joris van~der Hoeven}, \bibinfo{person}{Michael
  Joswig}, {and} \bibinfo{person}{Nobuki Takayama}} (Eds.).
  \bibinfo{publisher}{Springer}, \bibinfo{pages}{299--302}.
\newblock
\urldef\tempurl%
\url{https://doi.org/10.1007/978-3-642-15582-6\_49}
\showDOI{\tempurl}


\bibitem[\protect\citeauthoryear{Williams, Waterman, and Patterson}{Williams
  et~al\mbox{.}}{2009}]%
        {roofline}
\bibfield{author}{\bibinfo{person}{Samuel Williams}, \bibinfo{person}{Andrew
  Waterman}, {and} \bibinfo{person}{David~A. Patterson}.}
  \bibinfo{year}{2009}\natexlab{}.
\newblock \showarticletitle{Roofline: an insightful visual performance model
  for multicore architectures}.
\newblock \bibinfo{journal}{\emph{Commun. {ACM}}} \bibinfo{volume}{52},
  \bibinfo{number}{4} (\bibinfo{year}{2009}), \bibinfo{pages}{65--76}.
\newblock
\urldef\tempurl%
\url{https://doi.org/10.1145/1498765.1498785}
\showDOI{\tempurl}


\bibitem[\protect\citeauthoryear{Zhao, Kruse, and Cohen}{Zhao
  et~al\mbox{.}}{2018}]%
        {Zha18}
\bibfield{author}{\bibinfo{person}{Jie Zhao}, \bibinfo{person}{Michael Kruse},
  {and} \bibinfo{person}{Albert Cohen}.} \bibinfo{year}{2018}\natexlab{}.
\newblock \showarticletitle{A polyhedral compilation framework for loops with
  dynamic data-dependent bounds}. In \bibinfo{booktitle}{\emph{Proceedings of
  the 27th International Conference on Compiler Construction, {CC} 2018,
  February 24-25, 2018, Vienna, Austria}},
  \bibfield{editor}{\bibinfo{person}{Christophe Dubach} {and}
  \bibinfo{person}{Jingling Xue}} (Eds.). \bibinfo{publisher}{{ACM}},
  \bibinfo{pages}{14--24}.
\newblock
\urldef\tempurl%
\url{https://doi.org/10.1145/3178372.3179509}
\showDOI{\tempurl}


\bibitem[\protect\citeauthoryear{Zheng, Jia, Sun, Wu, Yu, Haj{-}Ali, Wang,
  Yang, Zhuo, Sen, Gonzalez, and Stoica}{Zheng et~al\mbox{.}}{2020}]%
        {zheng2020ansor}
\bibfield{author}{\bibinfo{person}{Lianmin Zheng}, \bibinfo{person}{Chengfan
  Jia}, \bibinfo{person}{Minmin Sun}, \bibinfo{person}{Zhao Wu},
  \bibinfo{person}{Cody~Hao Yu}, \bibinfo{person}{Ameer Haj{-}Ali},
  \bibinfo{person}{Yida Wang}, \bibinfo{person}{Jun Yang},
  \bibinfo{person}{Danyang Zhuo}, \bibinfo{person}{Koushik Sen},
  \bibinfo{person}{Joseph~E. Gonzalez}, {and} \bibinfo{person}{Ion Stoica}.}
  \bibinfo{year}{2020}\natexlab{}.
\newblock \showarticletitle{Ansor: Generating High-Performance Tensor Programs
  for Deep Learning}. In \bibinfo{booktitle}{\emph{14th {USENIX} Symposium on
  Operating Systems Design and Implementation, {OSDI} 2020, Virtual Event,
  November 4-6, 2020}}. \bibinfo{publisher}{{USENIX} Association},
  \bibinfo{pages}{863--879}.
\newblock
\urldef\tempurl%
\url{https://www.usenix.org/conference/osdi20/presentation/zheng}
\showURL{%
\tempurl}


\bibitem[\protect\citeauthoryear{Zinenko, Verdoolaege, Reddy, Shirako, Grosser,
  Sarkar, and Cohen}{Zinenko et~al\mbox{.}}{2018}]%
        {Zin18}
\bibfield{author}{\bibinfo{person}{Oleksandr Zinenko}, \bibinfo{person}{Sven
  Verdoolaege}, \bibinfo{person}{Chandan Reddy}, \bibinfo{person}{Jun Shirako},
  \bibinfo{person}{Tobias Grosser}, \bibinfo{person}{Vivek Sarkar}, {and}
  \bibinfo{person}{Albert Cohen}.} \bibinfo{year}{2018}\natexlab{}.
\newblock \showarticletitle{Modeling the conflicting demands of parallelism and
  Temporal/Spatial locality in affine scheduling}. In
  \bibinfo{booktitle}{\emph{Proceedings of the 27th International Conference on
  Compiler Construction, {CC} 2018, February 24-25, 2018, Vienna, Austria}},
  \bibfield{editor}{\bibinfo{person}{Christophe Dubach} {and}
  \bibinfo{person}{Jingling Xue}} (Eds.). \bibinfo{publisher}{{ACM}},
  \bibinfo{pages}{3--13}.
\newblock
\urldef\tempurl%
\url{https://doi.org/10.1145/3178372.3179507}
\showDOI{\tempurl}


\end{thebibliography}

\newpage
\appendix
\section{Annotated IR Examples}
\label{sec:annotated_examples}

\subsection{LLVM Dialect and Mix of Dialects}
\label{sec:appendix-llvm-mix-example}

The snippet in Figure~\ref{fig:llvm_dialect_example} (function \lstinline{llvm_example}) mixes operations and types within the declaration of a function in the \lstinline{llvm} dialect.

\begin{figure}[h!tb]
\begin{lstlisting}[language=mlir]
llvm.func @llvm_example(%v: i32, %cond: i1) -> !llvm.struct<(f32, vector<2xf32>, i32)> {
  %0 = llvm.mlir.undef : !llvm.struct<(f32, vector<2xf32>, i32)>
  %vf32 = arith.uitofp %v : i32 to f32
  %1 = llvm.insertvalue %vf32, %0[0] : !llvm.struct<(f32, vector<2xf32>, i32)>
  %f123 = arith.constant 123.0 : f32
  cond_br %cond, ^bb1(%f123: f32), ^bb2(%f123: f32)
  
^bb1(%fa: f32):
  %at123 = math.atan2 %fa, %fa : f32
  br ^bb2(%at123: f32)
  
^bb2(%fb: f32):
  %b = vector.broadcast %fb: f32 to vector<2xf32>
  %2 = llvm.insertvalue %b, %1[1] : !llvm.struct<(f32, vector<2xf32>, i32)>
  %c1024 = arith.constant 1024 : i32
  %3 = llvm.insertvalue %c1024, %2[2] : !llvm.struct<(f32, vector<2xf32>, i32)>
  llvm.return %3 : !llvm.struct<(f32, vector<2xf32>, i32)>
}
\end{lstlisting}
\caption{Illustration of an LLVM dialect function that takes an i32 and a bool and returns a struct. The struct is filled with different values depending on the bool function argument.}
\label{fig:llvm_dialect_example}
\end{figure}

\lstinline|llvm.undef| creates a \lstinline|!llvm.struct<(f32, vector<2xf32, i32)>| structure that is filled thanks to \lstinline|llvm.insertvalue|.
The values used to fill the struct are created with operations from the \lstinline|arith|, 
\lstinline|math| and \lstinline|vector| dialect. 
Builtin types are provided that are visible to all dialects (e.g. \lstinline|f32|, \lstinline|i32|, \lstinline|vector<2xf32>|). 
Dialects may introduce their own types (e.g. \lstinline|!llvm.struct<(f32, f32, i32)>| 
has no builtin representation and is not understood by ops in the \lstinline|arith| and \lstinline|math| dialects).

This simple example already provides a key insight about SSA form that we leverage in other dialects.
The instruction
\begin{lstlisting}[language=mlir]
%1 = llvm.insertvalue %vf32, %0[0] : !llvm.struct<(f32, vector<2xf32>, i32)>
\end{lstlisting}
inserts the value \lstinline|%vf32| at position \lstinline|0| in the 
\lstinline|!llvm.struct| named \lstinline|%0|. The instruction produces a new \lstinline|!llvm.struct| named \lstinline|%1| that contains a copy of all entries of \lstinline|%0| except for the \lstinline|f32| at position \lstinline|0| that takes the value \lstinline|%vf32|.

In a classical compilation flow to \LLVM IR, all operations and types must be converted to the \lstinline|llvm| dialect. This process is called \emph{dialect conversion}.
To make this process easier and improve interoperability between dialects, the \lstinline|llvm| dialect use the same representation for builtin types. 

\subsection{ArmNEON dialect example}
\label{sec:appendix-arm-neon-example}

\MLIR has a native higher-dimensional vector type that conveniently mixes with target-specific operations (effectively mirroring actual instructions) of the \lstinline|arm_neon| dialect.

\begin{lstlisting}[language=mlir]
// High-level 2-d vector flavor of the sdot operation.
func @sdot2d_4x4_i8i8(%a: vector<4xi32>, %b: vector<4x4xi8>, %c: vector<4x4xi8>) 
    -> vector<4xi32> {
  %0 = arm_neon.2d.sdot %a, %b, %c : vector<4x4xi8>, vector<4x4xi8> to vector<4xi32>
  return %0 : vector<4xi32>
}
// Lowers to 1-d intrinsic form.
llvm.func @lowered_sdot2d_4x4_i8i8(%a: vector<4xi32>, %b: vector<16xi8>, %c: vector<16xi8>) 
    -> vector<4xi32> {
  %0 = arm_neon.intr.sdot %a, %b, %c : vector<16xi8>, vector<16xi8> to vector<4xi32>
  return %0 : vector<4xi32>
}
// And further translates to llvm IR.
define <4xi32> @llvm_sdot2d_4x4_i8i8(<4xi32> %0, <16xi8> %1, <16xi8> %2) {
  %4 = call <4xi32> @llvm.aarch64.neon.sdot.v4i32.v16i8(<4xi32> %0, <16xi8> %1, <16xi8> %2)
  ret <4xi32> %4
}
\end{lstlisting}

The \lstinline|llvm.aarch64.neon.sdot.v4i32.v16i8| intrinsic performs 4 parallel dot products
on subvectors of \lstinline|4xi8|, each contributing to one of the lanes of the resulting \lstinline|4xi32|. The \lstinline|arm_neon.intr.sdot| closely mirrors the intrinsic.
However, we also provide a \lstinline|arm_neon.2d.sdot| that exposes the operation semantics more clearly to the higher level of the infrastructure.

This simple example illustrates the semantic benefits of more structured, higher-dimensional abstractions.

\subsection{Target-Specific Vector IR}
\label{sec:appendix-lowered-vector}

\begin{lstlisting}[language=mlir]
func @contract(%m0: memref<2xf32>, %m1: memref<2x16xf32>, %m2: memref<2x16xf32>) {
    %0 = vector.load %m0[0] : memref<2xf32>, vector<2xf32>
    %1 = vector.load %m1[0, 0] : memref<2x16xf32>, vector<8xf32>
    %2 = vector.load %m1[1, 0] : memref<2x16xf32>, vector<8xf32>
    %3 = vector.load %m1[0, 8] : memref<2x16xf32>, vector<8xf32>
    %4 = vector.load %m1[1, 8] : memref<2x16xf32>, vector<8xf32>
    %5 = vector.extract %0[0] : vector<2xf32>
    %6 = splat %5 : vector<8xf32>
    %7 = vector.fma %6, %1, 0.0f : vector<8xf32>
    %8 = vector.extract %0[1] : vector<2xf32>
    %9 = splat %8 : vector<8xf32>
    %10 = vector.fma %9, %2, %7 : vector<8xf32>
    %11 = vector.fma %6, %3, 0.0f : vector<8xf32>
    %12 = vector.fma %9, %4, %11 : vector<8xf32>
    vector.store %10, %m2[0, 0] : memref<2x16xf32>, vector<8xf32>
    vector.store %12, %m2[0, 8] : memref<2x16xf32>, vector<8xf32>
    vector.store %10, %m2[1, 0] : memref<2x16xf32>, vector<8xf32>
    vector.store %12, %m2[1, 8] : memref<2x16xf32>, vector<8xf32>
  return
}
\end{lstlisting}

\subsection{Python Bindings and Interoperability}
\label{sec:appendix-python}

Structured tensor code generation strongly influenced the design of the MLIR Python bindings, in particular, the ``natural'' mapping between nested Python context managers and the nested IR structure is illustrated in Figure~\ref{fig:python-ir-nesting}.

\begin{figure}[h!tb]
\begin{lstlisting}[language=python]
from mlir.ir import Context, InsertionPoint, Module
from mlir.dialects import builtin, scf

# Constructs IR in a fresh MLIR context.
with Context():
  module = Module()
  
  # Construct the following operations within module.
  with InsertionPoint(Module.body):
    func = builtin.FuncOp('foo')
    
    # Construct the following operations within function.
    # Create the entry block to turn declaration into definition.
    with InsertionPoint(func.add_entry_block()):
      loop = scf.ForOp(...)
      
      # Construct the following operations with loop.
      with InsertionPoint(loop.body):
        # etc.
\end{lstlisting}
\caption{Illustration of the L1-resident 2-D copy benchmark.}
\label{fig:python-ir-nesting}
\end{figure}

Structured data objects processed by the flow are exposed in Python as objects compatible with the Python buffer protocol and can therefore be converted to and from NumPy arrays, which are further convertible to framework-specific data types.

Once the program is constructed, compiled using the structured tensor code generation flow and lowered, it can be JIT-compiled and functions from this program can be invoked from Python using its name. The code below illustrates the execution of the \lstinline{matmul} function from the module defined above.

\begin{lstlisting}[language=python]
from mlir.execution_engine import ExecutionEngine
from mlir.runtime import get_ranked_memref_descriptor
import numpy as np
import ctypes

# Create NumPy arrays.
np_a, np_b = np.ones(M, K), np.ones(K, N)
np_c = np.zeros(M, N)

# Transform them into MLIR-compatible objects that share underlying data.
a, b, c = [ctypes.pointer(ctypes.pointer(get_ranked_memref_descriptor(v)))
           for v in (np_a, np_b, np_c)]
           
# JIT-compile and invoke the function.
engine = ExecutionEngine(module, opt_level=3)
engine.invoke('matmul', a, b, c)

# Results are readily available in the NumPy array.
print(np_c)
\end{lstlisting}

\end{document}